\documentclass[fleqn,usenatbib]{mnras}

\usepackage{newtxtext,newtxmath}

\usepackage[T1]{fontenc}

\DeclareRobustCommand{\VAN}[3]{#2}
\let\VANthebibliography\thebibliography
\def\thebibliography{\DeclareRobustCommand{\VAN}[3]{##3}\VANthebibliography}

\usepackage{graphicx}	
\usepackage{amsmath}	

\usepackage{orcidlink}

\usepackage{threeparttablex}
\usepackage{nicefrac}
\usepackage{multirow}
\usepackage{multicol}
\usepackage{longtable}
\usepackage{tabularx}
\usepackage[normalem]{ulem}
\usepackage{colortbl}

\usepackage{xfrac}
\usepackage{multirow}
\usepackage{array}
\renewcommand{\arraystretch}{1.1}
\usepackage{booktabs,threeparttable}
\usepackage{comment}

\usepackage{xcolor}
\usepackage{subcaption}
\usepackage{pdflscape}
\usepackage{flushend}

\DeclareCaptionFormat{custom}{\textbf{#1.}{#3}}

\newcommand{\msun}{\mbox{M$_{\odot}$}}
\newcommand{\kms}{\mbox{$\rm{km}\,s^{-1}$}}
\newcommand{\tardis}{$\textsc{tardis}$}
\newcommand{\carsus}{$\textsc{carsus}$}
\newcommand{\tausob}{$\tau_{\mathrm{\textsc{s}}}$}

\newcommand{\I}{{\sc i}}
\newcommand{\II}{{\sc ii}}
\newcommand{\III}{{\sc iii}}
\newcommand{\IV}{{\sc iv}}

\newcommand{\SrII}{Sr\,{\sc ii}}
\newcommand{\CaII}{Ca\,{\sc ii}}
\newcommand{\forbSrII}{[Sr\,{\sc ii}]}
\newcommand{\MSrII}{$M_{\rm{Sr}\,\textsc{ii}}$}
\newcommand{\YII}{Y\,{\sc ii}}
\newcommand{\ZrII}{Zr\,{\sc ii}}
\newcommand{\rpro}{\mbox{$r$-process}}
\newcommand{\Aval}{\mbox{$A$-values}}
\newcommand{\gfo}{AT2017gfo}
\newcommand{\xsh}{\mbox{X-shooter}}
\newcommand{\hst}{\mbox{\textit{HST}}}
\newcommand{\spitzer}{\mbox{\textit{Spitzer}}}
\newcommand{\Xlanth}{$X_{\textsc{ln}}$}

\newcommand{\ergsA}{erg\,s$^{-1}$\,\AA$^{-1}$}
\newcommand{\pcyg}{\mbox{P-Cygni}}

\newcommand{\Ye}{$Y_e$}

\newcommand{\AngIII}{\Ye$- 0.29$a}

\newcommand{\AngIX}{\Ye$- 0.21$b}

\defcitealias{PaperI}{Paper~I}
\newcommand{\paperI}{\citetalias{PaperI}}

\title[
    Modelling the post-photospheric spectra of AT2017gfo
]{
    Modelling the spectra of the kilonova AT2017gfo -- II: Beyond the photospheric epochs
}

\author[
    J. H. Gillanders et al.
]{
    J. H. Gillanders\,\orcidlink{0000-0002-8094-6108}$^{1, 2, 3}$\thanks{
        E-mail: james.gillanders@physics.ox.ac.uk
    },
    S. A. Sim\,\orcidlink{0000-0002-9774-1192}$^{1}$,
    S. J. Smartt\,\orcidlink{0000-0002-8229-1731}$^{1, 3}$,
    S. Goriely\,\orcidlink{0000-0002-9110-941X}$^4$ and 
    A. Bauswein\,\orcidlink{0000-0001-6798-3572}$^{5, 6}$
    \\
    $^1$Astrophysics Research Centre, School of Mathematics and Physics, Queen's University Belfast, BT7 1NN, UK \\
    $^2$Department of Physics, University of Rome `Tor Vergata', via della Ricerca Scientifica 1, I-00133 Rome, Italy \\
    $^3$Astrophysics sub-Department, Department of Physics, University of Oxford, Keble Road, Oxford, OX1 3RH, UK \\
    $^4$Institut d'Astronomie et d'Astrophysique, CP-226, Universit\'{e} Libre de Bruxelles, 1050 Brussels, Belgium \\
    $^5$GSI Helmholtzzentrum f\"ur Schwerionenforschung, Planckstrasse 1, 64291 Darmstadt, Germany \\
    $^6$Helmholtz Research Academy Hesse for FAIR (HFHF), GSI Helmholtz Center for Heavy Ion Research, \\ \hspace{0.2em} Campus Darmstadt, Planckstrasse 1, 64291 Darmstadt, Germany
    }

\date{Accepted XXX. Received YYY; in original form ZZZ}

\pubyear{2023}

\begin{document}
\label{firstpage}
\pagerange{\pageref{firstpage}--\pageref{lastpage}}
\maketitle

\begin{abstract}
Binary neutron star mergers are the first confirmed site of element nucleosynthesis by the rapid neutron-capture process ($r$-process). The kilonova AT2017gfo is the only electromagnetic counterpart of a neutron star merger spectroscopically observed. We analyse the entire spectral sequence of AT2017gfo (from merger to +10.4 days) and identify seven emission-like features. We confirm that the prominent 1.08\,$\mu$m feature can be explained by the Sr\,\textsc{ii} near-infrared triplet evolving from a P-Cygni profile through to pure emission. We calculate the expected strength of the [Sr\,\textsc{ii}] doublet and show that its absence requires highly clumped ejecta. Near-infrared features at 1.58 and 2.07\,$\mu$m emerge after three days and become more prominent as the spectra evolve. We model these as optically thick P-Cygni profiles and alternatively as pure emission features \mbox{(with FWHM~$\simeq 35600 \pm 6600$\,km\,s$^{-1}$)}, and favour the latter interpretation. The profile of the strong 2.07\,$\mu$m emission feature is best reproduced with two lines, centred at 2.059 and 2.135\,$\mu$m. We search for candidate ions for all prominent features in the spectra. Strong, permitted transitions of La\,\textsc{iii}, Ce\,\textsc{iii}, Gd\,\textsc{iii}, Ra\,\textsc{ii} and Ac\,\textsc{i} are plausible candidates for the emission features. If any of these features are produced by intrinsically weak, forbidden transitions, we highlight candidate ions spanning the three $r$-process peaks. The second $r$-process peak elements Te and I have plausible matches to multiple features. We highlight the need for more detailed and quantitative atomic line transition data.
\end{abstract}



\section{Introduction} \label{sec:Introduction}

In \cite{PaperI}, hereafter referred to as \paperI, we presented analysis of the early, photospheric spectra of the kilonova (KN) \gfo, the electromagnetic (EM) counterpart to a binary neutron star merger \citep[][]{LigoVirgo2017, Andreoni2017, Arcavi2017, Chornock2017, Coulter2017, Cowperthwaite2017, Drout2017, Evans2017, Kasliwal2017, Lipunov2017, Nicholl2017, Pian2017, Shappee2017, SoaresSantos2017, Smartt2017, Tanvir2017, Troja2017, Utsumi2017, Valenti2017}. KNe are thought to be ideal locations for the synthesis of material by the \rpro, and many theoretical simulations have corroborated this \citep[][]{Lattimer1974, Eichler1989, Freiburghaus1999, Rosswog1999, Goriely2011, Goriely2013, Goriely2015, Korobkin2012, Perego2014, Wanajo2014, Just2015}. In \paperI, we focussed on identifying features that we can attribute to a single transition (or a set of transitions) belonging to a specific species. Through the use of a mix of publicly available atomic data sets, and \tardis\ \citep[a 1D Monte Carlo radiative transfer spectral synthesis code;][]{tardis}, we were able to generate a set of models that self-consistently matched the observations of \gfo, from +0.5\,d, through to $\sim 1$~week post-explosion. At this stage, the observed \xsh\ spectra of \gfo\ appear to transition into an optically thin regime, and the spectral features evolve into broad emission profiles. The photospheric approximation within \tardis\ thus breaks down, and so \tardis\ modelling is not likely to be valid beyond $\sim 6 - 7$ days post-merger.

Our focus in \paperI\ was on studying specific species (e.g. \SrII, through the identification of the near-infrared triplet producing the broad absorption feature between $\sim 0.7 - 1.0$\,\micron), or on identifying sets of elements (e.g. the necessary presence of the lanthanides in our models, combining to suppress the flux in the near-UV and optical parts of the spectra $\gtrsim 2$ days). In this paper, our work from \paperI\ is extended, but now we predominantly focus on the later phases. The project has two main goals. First, we examine the \pcyg\ feature that is present between $\sim 0.7 - 1.2$\,\micron. In early-phase modelling, this feature has been associated with \SrII, and it remains persistent in the observed spectra until at least the +7.4\,d spectrum. We aim to determine whether this feature is consistent with \SrII\ producing an early \pcyg\ feature, which evolves through to later phases, and into a pure emission feature, or whether we require contribution from additional species beyond the first few days. Second, we analyse the other spectral features, with the goal of determining ejecta properties, and constraining the composition of the ejecta. We model the two broad near-infrared (NIR) features, present in the intermediate- and late-phase \xsh\ spectra of \gfo\ ($+3.4$~days onward), which we did not successfully reproduce with our \tardis\ modelling in \paperI. We also perform empirical fitting of the late-time ($+7.4$~days onward) features to deduce ejecta properties, before performing a line identification study to shortlist the species that may be contributing to the observed data. Positive identification would allow us to constrain the composition of the ejecta.

The work presented here is a continuation of that presented in \paperI, and so the reader is directed to that manuscript for further details. Here we use the same atomic data set as \paperI, which made use of data from the Chianti \citep[][]{Chianti-OG, Chianti-v10}, Kurucz \citep[][\url{http://kurucz.harvard.edu/atoms.html}]{Kurucz2017}, and \textsc{dream} data bases \citep[][]{DREAM1, DREAM2}, as well as the Pt and Au data presented by \cite{Gillanders2021} and \cite{McCann2022}. In addition, we also make use of collisional data for \SrII, presented by \cite{Bautista2002}. We also use the same spectral data set, which we compiled from \cite{Pian2017}  and \cite{Smartt2017}, with several additions. \cite{Tanvir2017} and \cite{Troja2017} present \textit{Hubble Space Telescope} (\hst) spectra of \gfo, at four separate epochs (+4.9, +7.3, +9.4 and +10.6~days). The spectra obtained at +4.9 and +9.4\,d contain a broad, emission-like feature at $\sim 1.40$\,\micron, which corresponds to a telluric region. Hence, this feature is not seen in any of our \xsh\ spectra at similar epochs. In the case of the +9.4\,d spectrum, the epoch of observation agrees exactly with one of the \xsh\ spectra, and so we merged this \hst\ spectrum with the \xsh\ spectrum at this epoch, replacing the pixels of \xsh\ with the flux-calibrated \hst\ pixels, within the telluric region. The +0.5 day spectrum from \cite{Shappee2017} is included, but it is blue and featureless, and so does not contribute to our analysis in this paper; however, we include it to illustrate the lack of prominent spectral features at very early times.

The manuscript is structured in the following manner. In Section~\ref{sec:Spectroscopic evolution of AT2017gfo} we briefly discuss the different features that are present in the entire observed spectral sequence of \gfo, and their evolution. Section~\ref{sec:Modelling the early NIR features} contains our \tardis\ analysis of the two NIR features that are prominent in the intermediate-phase ($+3.4$\,d onward) \xsh\ spectra. In Section~\ref{sec:Analysing the presence of strontium at late times}, we investigate the presence of \SrII\ at late times, and explore the possibility of it being responsible for the strong emission feature we observe at $\sim 1.08$\,\micron\ in the late-phase spectra. In Section~\ref{sec:Modelling the late-time emission features} we model the most prominent of the features in the late-time spectra with Gaussians. In Sections~\ref{sec:Search for potential candidate species (strong transitions)}~and~\ref{sec:Search for potential candidate species (weak transitions)}, we perform searches for potential candidates that may be responsible for these strong emission features at late times. Finally, we summarise our results, and conclude in Section~\ref{sec:Conclusions}. Here we note that, unless otherwise stated in the text, all wavelengths are given as in vacuum.

\section{Spectroscopic evolution of AT2017gfo} \label{sec:Spectroscopic evolution of AT2017gfo}

The spectra of \gfo, across all epochs observed ($+0.5 - 10.4$\,d) exhibit remarkably rapid evolution. Rapid cooling is observed and the centroids of the prominent features shift with time. This evolution is unrivalled by any other spectroscopically-observed extragalactic transient. At early times, the majority of the ejecta material is optically thick, and the spectra are dominated by contributions from the small mass of rapidly-expanding ejecta material that is present above the effective photosphere (see \paperI). As the KN ejecta expand and cool however, the outer material becomes transparent and so we can see deeper into the ejecta, meaning our spectra can now contain contributions from a broad range of ejecta material, spanning a large radial (or velocity) region. Given the inferred high velocities and low ejecta mass, it is of interest to investigate if/when we begin to observe evidence of a nebular spectrum. Recently, \cite{Pognan2023} proposed that KN ejecta material does not become completely optically thin very quickly. The results of their simulations show that while the rapidly-expanding outer ejecta quickly do become optically thin, it is possible for the innermost ejecta to remain optically thick out to +20~days.

Here we perform a simple exercise of determining where the spectral features peak, and how they evolve across multiple epochs. We can then attempt identification of the transitions that are responsible and determine information about the distribution of this material within the ejecta. The evolution of the peak wavelengths of the features can give us information about the velocity of the ejecta material contributing to the feature.

\begin{figure}
    \centering
    \includegraphics[width=\linewidth]{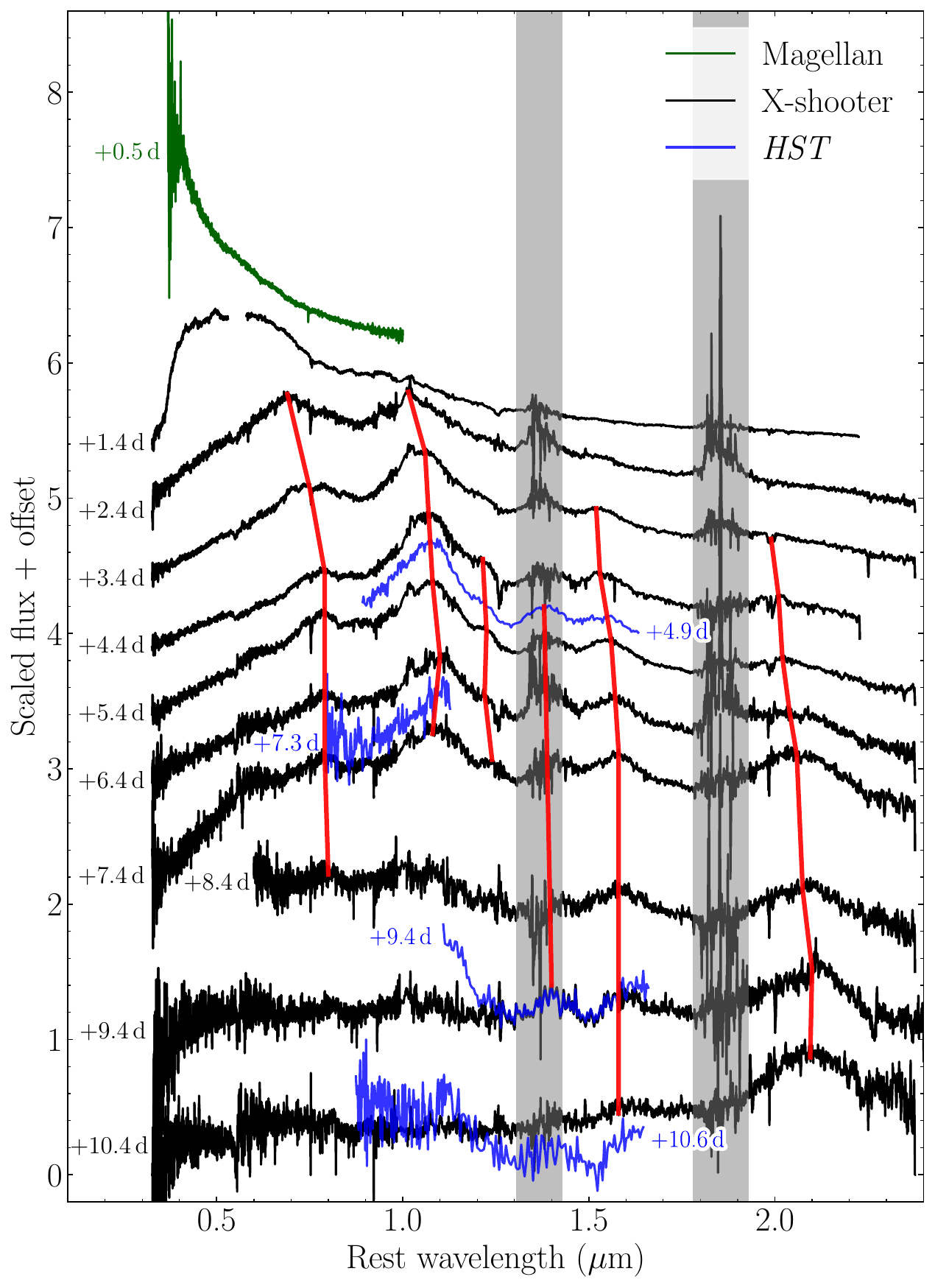}
    \caption{
        Sequence of \gfo\ spectra, scaled and offset for clarity. The early Magellan spectrum is plotted (green), as are the spectral sequences obtained with \xsh\ (black) and \hst\ (blue). Each observation has been annotated with its phase, relative to the gravitational wave (GW) trigger. The peaks of the most prominent emission-like features have been estimated, and joined (red lines) across epochs, to highlight their evolution. The vertical grey bands correspond to telluric regions.
    }
    \label{fig:Spectral evolution}
\end{figure}

\begin{figure}
    \centering
    \includegraphics[width=\linewidth]{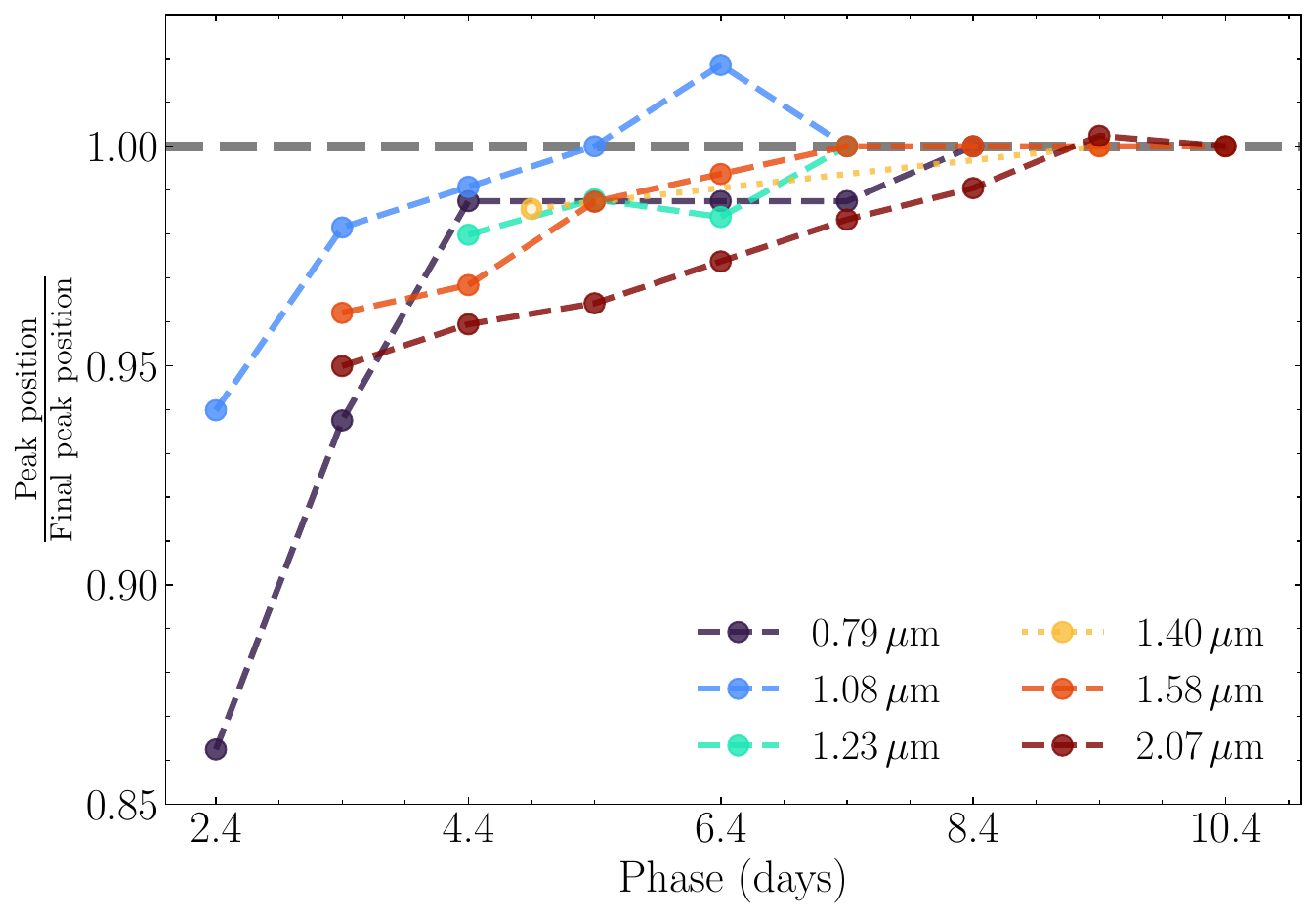}
    \caption{
        Ratio of the peak positions of the observed emission features and their final peak position. This is a visualisation of the data in Table~\ref{tab:Peak positions}, and helps illustrate the redward-shifting nature of the observed emission features with time. The 1.40\,\micron\ data point at +4.9\,d is included, although it does not appear in the main part of Table~\ref{tab:Peak positions} (but is clearly visible in the \hst\ spectrum presented in Figure~\ref{fig:Spectral evolution}). The horizontal grey dashed line marks a wavelength ratio of unity -- the ratios for all features should tend towards this value (by design).
    }
    \label{fig:Peak position vs. time}
\end{figure}

In Figure~\ref{fig:Spectral evolution}, we plot the sequence of \gfo\ spectra we study, in scaled flux. This is similar to the spectral data set presented in \paperI, with some additions. First we plot the ten \xsh\ spectra originally presented by \cite{Pian2017} and \cite{Smartt2017}, and the early Magellan spectrum from \cite{Shappee2017}. To this previous set, we have added the four \hst\ spectra from \cite{Tanvir2017}. Overlaid are lines to highlight the positions of the emission peaks in the spectra. Where features are present across multiple epochs, the lines are joined to illustrate how these positions evolve with time. Table~\ref{tab:Peak positions} provides the approximate peak positions for the features highlighted in Figure~\ref{fig:Spectral evolution}. These approximate peak positions have been deduced by visual inspection. In Figure~\ref{fig:Peak position vs. time}, we plot the peak positions of the features relative to their final peak position, to illustrate the evolution of the features and to visualise the data in Table~\ref{tab:Peak positions}. In general, the features shift redward with time, as we expect if the later spectra are dominated by contributions from material deeper within the ejecta (which would typically be expanding more slowly than the material observed at earlier times).

Previous works have drawn attention to some of these features. \cite{Smartt2017}, \cite{Tanvir2017}, \cite{Watson2019}, \cite{Domoto2021, Domoto2022}, \cite{Perego2022}, \cite{Sneppen2023_spherical_KN}, \cite{Sneppen2023_YII} and \cite{Tarumi2023} present analyses for some of them (and we highlight each work's specific contributions to the identification and study of these features below). However, no work to date has analysed all of the spectral emission-like features, and so here we present our own analysis and interpretation.

We identify six prominent spectral features, at \mbox{$\sim 0.79$\,\micron} \mbox{($+2.4 - 8.4$\,d)}, \mbox{$\sim 1.08$\,\micron} \mbox{($+2.4 - 7.4$\,d)}, \mbox{$\sim 1.23$\,\micron} \mbox{($+4.4 - 7.4$\,d)}, \mbox{$\sim 1.40$\,\micron} \mbox{($+4.9$ and $+9.4$\,d)}, \mbox{$\sim 1.58$\,\micron} \mbox{($+3.4 - 10.4$\,d)}, and \mbox{$\sim 2.07$\,\micron} \mbox{($+3.4 - 10.4$\,d)}. All features are present across a sequence of consecutive epochs, apart from the $\sim 1.40$\,\micron\ feature, which we only observe at +4.9 and +9.4\,d. It is likely that this feature exists at intermediate epochs too, but since this region of our \xsh\ spectra is affected by telluric absorption, we can only see it in the two \hst\ spectra.

\begin{table}
    \centering
    \caption{
        Approximate peak wavelengths of the prominent emission-like features in the observed \xsh\ spectra of \gfo. The peak locations of the features shift over time, indicating feature evolution.
    }
    \begin{threeparttable}
    \centering
    \begin{tabular}{ccccccc}
        \hline
        \hline
        Phase    &\multicolumn{6}{c}{Approximate peak position (\micron)}   \\
        \cline{2-7}
        (days)   &0.79    &1.08    &1.23   &1.40   &1.58   &2.07        \\
        \hline
        +0.5    &$-$    &$-$    &$-$     &$-$               &$-$     &$-$     \\
        +1.4    &$-$    &$-$    &$-$     &$-$               &$-$     &$-$     \\
        +2.4    &0.69   &1.015  &$-$     &$-$               &$-$     &$-$     \\
        +3.4    &0.75   &1.06   &$-$     &$-$               &1.52    &1.99    \\
        +4.4    &0.79   &1.07   &1.215   &$-$               &1.53    &2.01    \\
        +5.4    &0.79   &1.08   &1.225   &$-$               &1.56    &2.02    \\
        +6.4    &0.79   &1.10   &1.22    &$-$               &1.57    &2.04    \\
        +7.4    &0.79   &1.08   &1.24    &$-$               &1.58    &2.06    \\
        +8.4    &0.80   &$-$    &$-$     &$-$               &1.58    &2.075   \\
        +9.4    &$-$    &$-$    &$-$     &1.40\tnote{*}     &1.58    &2.10    \\
        +10.4   &$-$    &$-$    &$-$     &$-$               &1.58    &2.095   \\
        \hline
    \end{tabular}
    \begin{tablenotes}
        \item[*] Although this feature is only detectable in the \hst\ spectrum, we include it here since this \hst\ spectrum was merged with the \xsh\ spectrum taken \linebreak at the same epoch.
        \item \textbf{Note.} The emission feature with an approximate peak position of 1.40\,\micron\ peaks at 1.38\,\micron\ in the +4.9\,d \hst\ \linebreak spectrum (see Figure~\ref{fig:Spectral evolution}). 
    \end{tablenotes}
    \end{threeparttable}
    \label{tab:Peak positions}
\end{table}

First we look at the evolution of the 0.79\,\micron\ feature, which appears in the $+2.4 - 8.4$\,d spectra (previously noted by \citealt{Tanvir2017} and \citealt{Sneppen2023_spherical_KN}). We consider the most likely explanation of this feature as the edge of the blue wing of the strong \SrII\ NIR triplet \pcyg\ feature in the early phases \citep[identified by][see also \paperI]{Watson2019}. This point represents where the \SrII\ feature blends into the continuum, but since the continuum exhibits suppressed flux blueward of this, the location at which it joins the continuum becomes pronounced. Alternatively, it is possible that there is some as-yet unidentified species (unrelated to \SrII) producing a pronounced emission feature (see \citealt{Sneppen2023_YII}, where they propose the feature is produced by \YII). The peak of this feature shifts redward with time, indicating that the spectral features are being produced by sequentially slower-expanding material. If the material is expanding homologously, then this is simply a result of the outer regions of faster expanding ejecta becoming optically thin, allowing us to observe the effects of the inner regions of slower-expanding ejecta on the emergent spectrum. This leads to the blue wing not extending as far from the \SrII\ NIR triplet rest wavelengths, which agrees with our model results presented in \paperI. Alternatively, the observed redward-shifting of the feature could be explained by an unidentified emission line or blend, which evolves redward with time.

Next we analyse the 1.08\,\micron\ feature, which spans from \mbox{$+2.4 - 7.4$\,d} (noted by \citealt{Tanvir2017}; analysed by \citealt{Smartt2017, Watson2019, Domoto2021, Domoto2022, Perego2022, Sneppen2023_spherical_KN, Tarumi2023}). \cite{Watson2019} proposed that this feature, at early times, can be readily explained by emission from the \SrII\ NIR triplet. In \paperI, we confirmed that a composition extracted from a realistic hydrodynamical simulation of a binary neutron star (BNS) merger \citep[][]{Goriely2011, Goriely2013, Goriely2015, Bauswein2013} contained sufficient strontium for this feature to be produced by \SrII. As the ejecta material evolves from its photospheric phase, through to a quasi-nebular regime, the peak of the feature emission should broadly line up with that of the rest wavelength of the transition (or set of transitions) that produce it. The weighted average of the expected centroid of the \SrII\ NIR triplet blend ($\simeq 1.047$\,\micron), is blueward of the observed peak wavelength of the emission feature at late times (a point we will return to in Section~\ref{sec:Analysing the presence of strontium at late times}). We note that in our \tardis\ models presented in \paperI, we find generally good agreement between the peak positions of the emission components of the \SrII\ \pcyg\ profiles produced by our sequence of models, and the observed emission peaks. \cite{Perego2022} and \cite{Tarumi2023} have suggested the He\,\I\ $\lambda_{\rm air} = 10830$\,\AA\ line as a possible alternative to the \SrII\ NIR triplet (a point we return to in Section~\ref{sec:Analysing the presence of strontium at late times - Summary}).

We identify a weak emission component at $\sim 1.23$\,\micron, which appears at +4.4 days, and disappears beyond +7.4 days (also noted by \citealt{Tanvir2017} and \citealt{Sneppen2023_spherical_KN}). It shifts redward with time, as seen for the other features. It appears as a weak feature, that just rises above the continuum redward of the broad and pronounced \pcyg\ feature previously discussed.

As noted above, we find evidence for an emission feature at 1.40\,\micron, but due to the telluric region in the \xsh\ spectra, we cannot constrain its presence at any epochs apart from +4.9 and +9.4\,d, where we are able to supplement our \xsh\ observations with the \hst\ spectra presented by \cite{Tanvir2017}. This emission feature has been previously identified by \cite{Tanvir2017} and \cite{Sneppen2023_spherical_KN}. \cite{Domoto2022} model its absorption component (present at early times), and propose that this feature is produced by lanthanide absorption -- specifically, La\,\III\ (a point we return to in Section~\ref{sec:Search for potential candidate species (strong transitions)}).

Next we highlight the two emission features, present at $\sim 1.58$ and $\sim 2.07$\,\micron. These appear in the +3.4\,d spectrum, with both persisting until +10.4\,d. These features were noted in \paperI, as they rose significantly above the NIR continuum, although our modelling did not explore their properties (we return to the nature of these two features in Section~\ref{sec:Modelling the early NIR features}). The 1.58\,\micron\ feature has been previously noted by \cite{Smartt2017} and \cite{Tanvir2017}. \cite{Watson2019} and \cite{Sneppen2023_spherical_KN} note the presence of both the 1.58 and 2.07\,\micron\ features. None of these works identify the cause of these features. \cite{Domoto2022} model the $+1.4 - 3.4$~day \xsh\ spectra, and show that a Ce\,\III\ \pcyg\ feature can appear prominently at $\sim 14000 - 16000$\,\AA\ in their models at these phases (see Section~\ref{sec:Search for potential candidate species (strong transitions)}). These two emission features again follow the trend of shifting towards redder wavelengths as time evolves, for the reasons discussed for the previous features.

One of the key outstanding issues with interpreting the sources of these emission features is whether they originate from an optically thick or thin regime. To enable us to better understand these features, we consider the case that the emission features we have identified above are all the emission components of \pcyg\ profiles, as are routinely produced from photospheric regimes (e.g. the \SrII\ NIR triplet feature at $\sim 0.7 - 1.2$\,\micron). If this is indeed the case, then we can reasonably expect to see evidence for absorption troughs blueward of the respective emission components. We can make predictions for the wavelength ranges these proposed absorption features span by considering the \SrII\ feature.

\begin{figure}
    \centering
    \includegraphics[width=\linewidth]{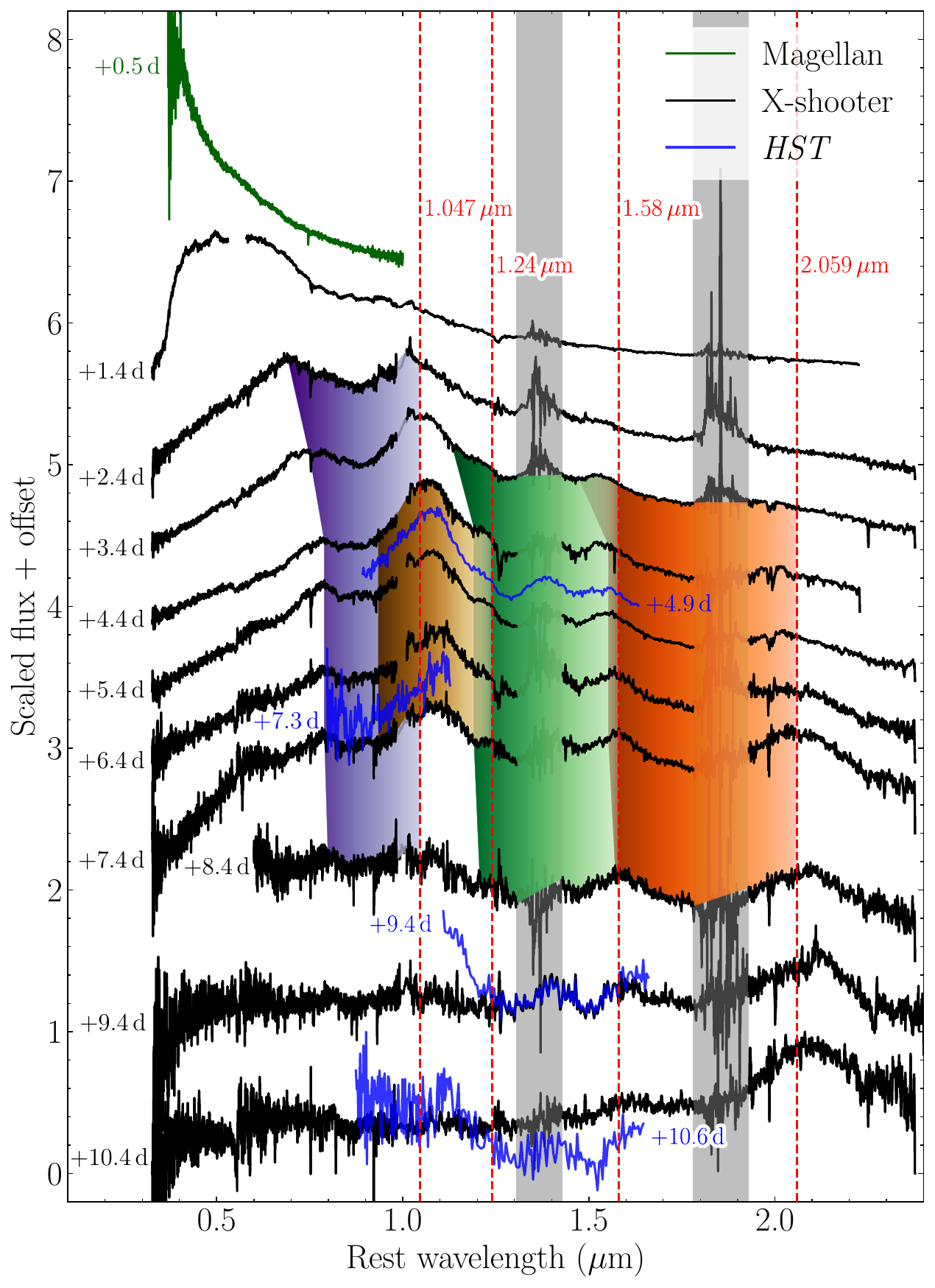}
    \caption{
    Sequence of \gfo\ spectra, scaled and offset for clarity. We plot the early Magellan spectrum (green), as well as the sequences of \xsh\ (black) and \hst\ (blue) spectra. All observations are labelled with their phase, relative to the GW trigger. We also plot shaded regions corresponding to the locations where we should see any absorption components of \pcyg\ features, with the intensity of the shading increasing with blueshift. The purple, brown, green and orange shaded regions correspond to the locations of proposed absorption for the $\sim 1.08$, 1.23, 1.58 and 2.07\,\micron\ features, respectively. The proposed rest wavelengths for these features are also plotted, as vertical dashed red lines. The vertical grey bands correspond to telluric regions.
    }
    \label{fig:Blueshifts}
\end{figure}

If the $\sim 1.08$\,\micron\ feature is produced by the \SrII\ NIR triplet (which has a weighted rest wavelength of $\simeq 1.047$\,\micron), and the $\sim 0.79$\,\micron\ feature corresponds to the edge of our blueshifted absorption feature from this \SrII\ \pcyg\ profile, we are able to estimate the maximum ejecta velocity of the \SrII\ material that is having an observable impact on the spectrum. Assuming that \SrII\ is uniformly distributed throughout the ejecta, we use the inferred velocities extracted from the \SrII\ \pcyg\ feature as representative velocities for the ejecta (at least the ejecta within the line-forming region). We then use these velocities to constrain the positions of the absorption features for the other proposed \pcyg\ profiles at each epoch the \SrII\ blue wing feature is prominent ($+2.4 - 8.4$\,d). From this, we can determine the expected positions of the absorption features (blue wings) of the other features in the spectra. For this calculation, we assume rest wavelengths of 1.24, 1.58 and 2.059\,\micron\ for the features at $\sim 1.23$, 1.58 and 2.07\,\micron, respectively (from Tables~\ref{tab:Peak positions} and \ref{tab:Gaussian fit parameters}). We present the results of this analysis in Figure~\ref{fig:Blueshifts}.

The purple shaded region in Figure~\ref{fig:Blueshifts} corresponds exactly to the proposed absorption region of the \SrII\ NIR triplet \pcyg\ profile (by design), and acts to highlight the spectral shape we should see for the other features, if they are also \pcyg\ features. The brown shaded region ($\sim 1.23$\,\micron\ feature) does not show any evidence for an absorption trough, while the green and orange shaded regions ($\sim 1.58$ and 2.07\,\micron\ features, respectively) are somewhat ambiguous. There does appear to be some flux deficit in these regions, but it is unclear whether this is a result of absorption, or whether these regions simply lie between prominent emission features, resulting in them appearing to have suppressed flux beneath some continuum.

One explanation for there being no convincing absorption components for these features is that emission from one (or many) other species is `filling in', or replenishing, the flux in these wavelength regions. Alternatively, it may indicate that these features are not associated with \pcyg\ profiles, and are indeed pure emission features, likely originating from some distinct component of optically thin ejecta material. Indeed, we proposed this may be the case in \paperI, since the shapes of our \tardis\ model continua seemed to reasonably match the slopes of the observed NIR spectral regions (aside from the prominent emission components at $\sim 1.58$ and 2.07\,\micron). From this analysis, it appears we cannot rule either scenario out (optically thick versus optically thin), and so we explore both cases in greater detail in Sections~\ref{sec:Modelling the early NIR features} and \ref{sec:Modelling the late-time emission features}, to better constrain which scenario best describes the data.

\section{Modelling the early phase features} \label{sec:Modelling the early NIR features}

There are two prominent emission-like features present in the NIR region, at $\lambda \simeq 1.58$ and 2.07\,\micron\ (see Section~\ref{sec:Spectroscopic evolution of AT2017gfo}). It is unclear what produces these features, and whether each is produced by a single transition, or a set of transitions, belonging to one or multiple species.

Given the possibility that the 1.08\,\micron\ feature might be (partly) due to the \mbox{He\,\I\ 2p~$^3$P -- 2s~$^3$S} $\lambda_{\rm air} = 10830$\,\AA\ transition (see Section~\ref{sec:Analysing the presence of strontium at late times}), it is particularly interesting to consider whether the observed 2.07\,\micron\ feature might be associated with the \mbox{He\,\I\ 2p~$^1$P -- 2s~$^1$S} $\lambda_{\rm air} = 20581$\,\AA\ transition (see also \citealt{Tarumi2023}, where they explore the presence of the 10830 and 20581\,\AA\ He\,\I\ transitions in KNe). Such an identification does provide a rather good wavelength match with the spectral feature (within 10\,\AA\ of the peak position of the 2.059\,\micron\ blue component of the blended 2.07\,\micron\ feature; see Section~\ref{sec:Modelling the late-time emission features}), making it an intriguing possibility. However, the apparent relative strengths of the observed 1.08 and 2.059\,\micron\ features does not seem to be compatible with expectations for these being produced consistently by He\,\II\ recombination. In particular, at +7.4 days, the approximate ratio of feature luminosities is $\sim 0.4$ (see Table~\ref{tab:Gaussian fit parameters}), while estimates for recombination line strengths suggest that the emissivity of the 10830\,\AA\ line should be at least an order of magnitude higher than the 20581\,\AA\ line \citep[][]{Benjamin1999}. Particularly in view of the fact that the observed 2.059\,\micron\ feature persists to later times after the 1.08\,\micron\ feature has faded, it therefore seems unlikely that the 2.059\,\micron\ spectral feature could be dominated by He\,\I.

A critical factor in identifying these NIR features is the lack of line information at these long wavelengths. As noted in \paperI, our atomic data set is incomplete for species with \mbox{$Z \gtrsim 40$}. Additionally, the heavy species for which we do have data typically only include transitions with wavelengths $\lesssim 1$\,\micron.\footnote{For example, only three of the species that we included from the \textsc{dream} data base \citep[][see Section~3 and Table~1 in \paperI\ for more on our atomic data set]{DREAM1, DREAM2} have data for transitions at $\lambda > 1$\,\micron\ (La\,\I, Pr\,\IV\ and Yb\,\II).} Therefore, it is likely that the currently available atomic data is insufficient to model these features. In Sections~\ref{sec:Search for potential candidate species (strong transitions)}~and~\ref{sec:Search for potential candidate species (weak transitions)}, we carry out a speculative search for candidate transitions -- however, here we first attempt to establish the best framework to use to interpret the data; i.e. can they be explained by optically thick material producing \pcyg\ features (similar to the 1.08\,\micron\ feature)? If so, what constraints can we place on ejecta properties (e.g. distribution of the species involved)?

\subsection{Motivation of parameters} \label{sec:Modelling the NIR emission features - Motivation of parameters}

We investigate whether these features can be formed in the same way as the \SrII\ feature at the same epochs; i.e. whether these features are produced by line scattering in the same region of the ejecta as \SrII. To constrain the nature of these features, we generate atomic data for two synthetic species. Both species are treated in a two-level atom approximation; i.e. we include a single excited level in each, where the energy gap between this level and the ground state corresponds to the wavelength of the observed features ($\lambda = 1.58$ and 2.07\,\micron). The two-level approximation means that \tardis\ will treat these as pure resonance scattering lines. We dub our mystery species X and Z, respectively.\footnote{We selected these symbols to avoid confusion with any real element, \linebreak e.g. Yttrium (Y).}

We adopt Einstein \Aval\ of $10^6$\,s$^{-1}$ for each of these features, which is characteristic of strong dipole transitions at these wavelengths. We then incorporated these synthetic species into our atomic data, using the \carsus\ package,\footnote{\url{https://github.com/tardis-sn/carsus}} and generated \tardis\ models to determine if we could reproduce these NIR features in this way. We note that the work presented in this section is based on the \tardis\ modelling we presented in \paperI, and where we refer to `our best-fitting \tardis\ models', we are referring to the models presented in \paperI\ that most closely matched the observational data.

\begin{figure}
    \centering
    \includegraphics[width=\linewidth]{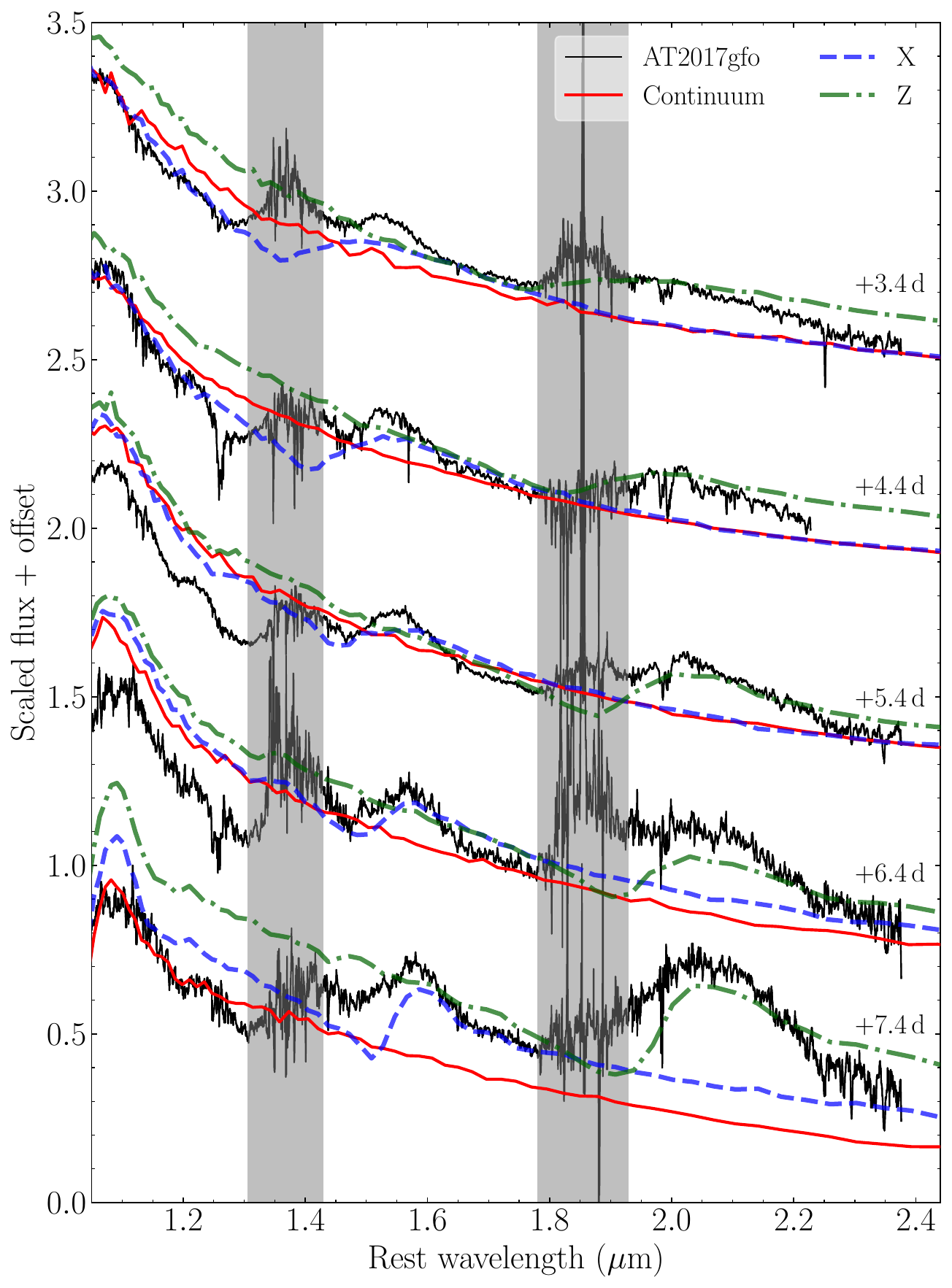}
    \caption{
        Sequence of \gfo\ spectra, from $+3.4 - 7.4$\,d, plotted alongside the continua from our best-fitting \tardis\ models (presented in \paperI; red). Also plotted are the best-fitting \tardis\ models with our mystery species X (blue dashed lines) and Z (green dash-dotted lines) included. All spectra have been plotted in scaled flux, and are offset for clarity (2.4, 1.8, 1.2 and 0.6 for the third, fourth, fifth and sixth epoch spectra, respectively). The models with our mystery species X and Z have in some instances an additional additive offset applied, to better match the local continua around the observed features (see Table~\ref{tab:Gillanderium mass fractions} for these model offsets). Telluric regions have been marked by grey shaded regions, and the observed flux in these regions are uncertain. The epochs have been annotated with their phase relative to GW~detection.
    }
    \label{fig:Gillanderium models 2022 models}
\end{figure}

We generate models for the $+3.4 - 7.4$\,d \gfo\ spectra, which correspond to the epochs modelled in \paperI\ that have prominent emission. The results of our modelling of these NIR features are shown in Figure~\ref{fig:Gillanderium models 2022 models}. We plot the observed spectra, as well as the continua from our best-fitting \tardis\ models presented in \paperI, for comparison. As noted in \paperI, our best-fitting models do not reproduce the flux level of the NIR continuum well. As a result, our model continua do not exactly match the observations at wavelengths $\gtrsim 1.2$\,\micron. As there is some systematic uncertainty in the continuum position, we give ourselves freedom to shift the continuum of our model to improve agreement between the feature profiles from our mystery species X and Z, and the observations.

The mass fractions of the two-level atoms X and Z needed to produce features with strength similar to observations are presented in Table~\ref{tab:Gillanderium mass fractions}.\footnote{The mass fractions presented were determined by assuming statistical weights for both the lower and upper level of 1.} Interpretation of these mass fractions requires care, owing to the simplicity of the two-level atom approach used. We summarise the dominant factors that are relevant with:
\begin{equation}
        \textrm{MF}_{\textsc{real}} \sim \textrm{MF}_{\textsc{tardis}} \left( \frac{1}{\chi_{\textsc{ion}}} \right) \left( \frac{p_{\textsc{tardis}}}{p_{\textsc{l}}} \right) \left( \frac{10^6\,{\rm s}^{-1}}{A_{\textsc{ul}}} \right) \left( \frac{A_{\textsc{ion}}}{100} \right) .
    \label{eqn:Species X and Z minimum masses}
\end{equation}

Here, MF$_{\textsc{tardis}}$ is the mass fraction of species X/Z in our \tardis\ models, whereas MF$_{\textsc{real}}$ represents the mass fraction of the elements that produce X/Z in a real astrophysical explosion. $\chi_{\textsc{ion}}$ is the fraction of the element that is ionised as species X/Z. $p_{\textsc{l}}$ is the fractional population of the ion in the lower level for the transition. $p_{\textsc{tardis}}$ is a variable that encapsulates the level populations as computed in our \tardis\ models (in our two-level atoms, typically $p_{\textsc{tardis}} \sim 1$, and in all our calculations $p_{\textsc{tardis}} > 0.86$; thus $p_{\textsc{tardis}}$ has only a modest impact on any interpretation of results). $A_{\textsc{ion}}$ is the atomic mass number of species X/Z (the calculations are performed assuming a mass number $A_{\textsc{ion}} = 100$, characteristic of elements between the first and second $r$-process peaks). Although all the parameters in this expression may vary, it is reasonable to expect that the largest potential deviations are in the ion fraction ($\chi_{\textsc{ion}}$) and lower level population fraction ($p_{\textsc{l}}$). Our two-level atomic models effectively assume both of these parameters are (close to) 1, but both can be significantly smaller (e.g. the transition of interest may originate from a sub-dominant ion, or from a low-population excited state). Thus we expect that MF$_{\textsc{tardis}}$ represents a \textit{minimum} mass fraction for the relevant species, since our two-level atom approach represents a \textit{best-case} scenario for line formation.

\begin{table}
    \renewcommand*{\arraystretch}{1.2}
    \centering
    \caption{
        \tardis\ mass fractions (MF$_{\textsc{tardis}}$) of mystery species X and Z needed to reproduce the features in the observed spectra of \gfo. These offset values correspond to the additive offsets that have been applied to our models to better match the local continua around the observed features (and are quoted in scaled flux).
    }
    \begin{tabular}{cccccc}
        \hline
        \hline
        Phase       &\multicolumn{2}{c}{X}      &       &\multicolumn{2}{c}{Z}  \\
        \cline{2-3}
        \cline{5-6}
        (days)  &MF$_{\textsc{tardis}}$ ($10^{-4}$)  &Offset   &   &MF$_{\textsc{tardis}}$ ($10^{-4}$)  &Offset \\
        \hline
        +3.4       &1.2    &0.0    &   &3.0   &0.1     \\
        +4.4       &1.2    &0.0    &   &3.0   &0.1     \\
        +5.4       &0.4    &0.0    &   &3.0   &0.05    \\
        +6.4       &0.4    &0.05   &   &1.8   &0.1     \\
        +7.4       &0.4    &0.1    &   &6.0   &0.25    \\
        \hline
    \end{tabular}
    \label{tab:Gillanderium mass fractions}
\end{table}

\subsection{Comparison between mystery species X and observations} \label{sec:Modelling the NIR emission features - Comparisons between species X and observations}

First, we look at the feature at $\sim 1.58$\,\micron, which is produced by our mystery species X. For the models at both +3.4 and +4.4 days, we find that no continuum offset is necessary to match the observations. The models have some small component of absorption blueward of the emission feature, that mostly lies in the telluric region of the observational data.\footnote{Here we include the data in the telluric regions, but stress that the flux within these regions is uncertain, and so should not be used for any inferences as to model agreement.} This prevents any definitive conclusion to be drawn with regards to the presence of a \pcyg\ feature, versus a pure emission feature, although we do note that the spectrum immediately blueward of the emission feature (at $\sim 1.45$\,\micron) appears to rise to the blue, in apparent contradiction with our models, which have absorption features at this wavelength. Our model emission components appear to be much broader than the observed features, and so we do not get good agreement between our models and the observed spectra. This indicates that the emission feature in the observed spectrum is produced by material that is expanding more slowly than the velocities required to reproduce the optical region of the spectra (the \SrII\ feature in particular). \cite{Domoto2022} analysed the  $+1.4 - 3.4$\,d spectra, assuming an optically thick photosphere, with a hybrid line list. They identify the same feature in the data, although they focus on modelling the wavelength position of the associated absorption component. \cite{Domoto2022} propose this feature is a blend of Ce\,\III\ lines, and their synthetic spectra produce a \pcyg\ profile similar to ours. However, it appears to peak at a redder wavelength, and is broader than the observed feature. The discrepancy in feature width between the synthetic and observed spectra is similar to what we find; hence, we reserve judgement here on whether or not mystery species X can be definitively explained by these Ce\,\III\ transitions.

Looking at later times, we find reasonable agreement with our model and the $\sim 1.58$\,\micron\ feature at +5.4 days. Again we require no continuum offset, but we do require a lower mass fraction of X than in the previous two epochs ($4 \times 10^{-5}$ at +5.4 days, versus $1.2 \times 10^{-4}$ for +3.4 and +4.4\,d). Again our emission feature is broader than the observation, but we get good agreement with the observed data at $\sim 1.45$\,\micron, directly between the emission component and the telluric region. Our +6.4\,d model requires a small continuum offset to better match observation, but the same mass fraction of X as needed at +5.4\,d. Again, we reasonably match the flux between the telluric and the emission feature. Our emission feature is still broader than the observations. At +7.4\,d, our model does not agree with the observations. Our model absorption feature has shifted to redder wavelengths (due to the receding photosphere within our \tardis\ models, leading to higher densities at lower velocities, and therefore most of our mystery species in this model has slower velocity than previous phases), and now does not agree with the observed spectrum. We apply an offset to the model to better match the continuum redward of the emission feature, but overall the fit is poor.

Our model emission features being broader than the observed features indicates our model expansion velocities are too high. This implies that our mystery species X is not present in the fastest-expanding material, which corresponds to the regions of ejecta that \SrII\ is present in (i.e. the regions that we modelled with \tardis\ in \paperI, where we obtained good agreement to the observations, up to $\lesssim 1$~week). This can be due to (i)~ionisation effects, where at higher velocities, our mystery species is not the dominant ion, or (ii)~ejecta stratification from different ejection mechanisms.

\subsection{Comparison between mystery species Z and observations} \label{sec:Modelling the NIR emission features - Comparisons between species Z and observations}

Now we summarise the results of modelling the second emission feature at $\sim 2.07$\,\micron, with our mystery species Z. Across all epochs, our model emission components are broader than the observations, similar to the behaviour of X. Our model for +3.4\,d requires a small offset to match the continuum blueward of the $\sim 2.07$\,\micron\ feature, but this leaves us with poor agreement to the data redward of the emission feature, which is perhaps evidence for needing a steeper continuum than our \tardis\ model produces. We see similar behaviour at +4.4\,d, where again we match the continuum at blue wavelengths, but not at redder wavelengths. The model emission component is still too broad to match the data.

At +5.4\,d, we get good agreement to the data, with a reasonable match to the continuum either side of the observed emission feature. Additionally, the shape of our emission feature appears to broadly agree with the observed one. Our model predicts a reasonably strong absorption feature, which cannot be compared with observation since it lies in a telluric region. For the final two phases (+6.4 and +7.4~days), the quality of our fits degrades. Now we cannot obtain a good match to the continuum at both sides of the emission feature. At +6.4\,d, the spectrum blueward of the emission feature appears to rise in flux (to the blue), at odds with our model, which predicts an absorption component at this position ($\sim 1.95$\,\micron). To reduce this discrepancy, we lowered our Z mass fraction (from $3 \times 10^{-4}$ at $+3.4 - 5.4$\,d, to $1.8 \times 10^{-4}$), but still cannot match the observed data well. At +7.4\,d, we match the continuum at wavelengths blueward of the emission feature, but not redward, and our model now requires a much higher mass fraction of Z ($6 \times 10^{-4}$) to produce an emission feature comparable in strength to the data. As a result, we now have a very pronounced absorption feature in our model that does not match the blue wing of the observed emission feature well. The fact our model requires such a large increase in mass fraction ($\gtrsim 3 \times$ increase) to reproduce the observed feature strength indicates that some additional species may be contributing to this feature at this epoch (see Section~\ref{sec:Modelling the late-time emission features}, although there we only see strong evidence for an additional emission component at +8.4\,d and later).

We have shown with our \tardis\ models that we cannot get good agreement with these features overall. Our model features are much broader than the equivalent features in the observed data, indicating that this $\sim 2.07$\,\micron\ feature originates from ejecta moving more slowly than that needed to reproduce the \SrII\ feature. This could be due to ionisation or stratification effects, where Z is only present at lower velocities.

\subsection{Summary}

The analysis above demonstrates that we are unable to convincingly reproduce the observations of the $\sim 1.58$ and 2.07\,\micron\ features from $+3.4 - 7.4$\,d under the assumption that these are produced by scattering lines forming in the same ejecta regions that are described by our \tardis\ models for the \SrII\ feature. Our model \pcyg\ profiles generally do not match the observations, especially the absorption components. Therefore, we ultimately disfavour this scenario (i.e. optically thick ejecta producing \pcyg\ features), and conclude that it is more likely that these features are in fact pure emission features, that are produced by either collisional excitation, recombination or fluorescence from some region of the ejecta, typically at lower velocity. In Section~\ref{sec:Modelling the late-time emission features}, we parameterise such a scenario by simple Gaussian fitting of the emission features.

\section{Analysing the presence of strontium at late times} \label{sec:Analysing the presence of strontium at late times}

\cite{Watson2019} first presented evidence for the presence of Sr in the early spectra of \gfo\ (later corroborated by \citealt{Domoto2021}, \citealt{Perego2022}, and \paperI). Specifically, they identified two features which they suggest are produced by \SrII\ absorption. These features lie at $\sim 3500$\,\AA\ (produced by the \SrII\ resonance lines from the ground state, with wavelengths, $\lambda_{\rm air} = 4077.7$ and 4215.5\,\AA) and \mbox{$\sim 7000 - 10000$\,\AA} (produced by the \SrII\ transitions from the metastable 4p$^6$4d levels, with wavelengths, $\lambda_{\rm air} = 10036.7$, 10327.3 and 10914.9\,\AA). However, they did not present detailed predictions for how the \SrII\ features may be expected to evolve at later times, beyond the early, photospheric phase.

In \paperI, we quantitatively confirmed that the \SrII\ NIR triplet can reproduce the strong 1.08\,\micron\ feature, but found that the effect of absorption from the resonance doublet was hidden by the strong line-blanketing in the blue ends of the spectra, dominated by \YII\ and \ZrII\ in our early models ($< 2$ days), and the lanthanides in our later models ($> 2$ days). Here we analyse whether the spectral feature at $\sim 1.08$\,\micron\ (prominent in the +7.4\,d \xsh\ spectrum) is consistent with the \SrII\ NIR triplet evolving from a \pcyg\ profile in the early spectra, to a pure emission feature at this position after $\sim 7$ days.

The \SrII\ NIR triplet transitions, and their connection to the ground state, are analogous to the well-known \CaII\ NIR triplet. The \CaII\ NIR triplet is commonly observed to transition from a \pcyg\ profile to a pure emission feature in supernova spectra \citep[see e.g.][for a recent example of such behaviour for SN\,2017iro, a well-sampled SN Ib]{Kumar2022}. As the density drops, the [\CaII] doublet (at $\lambda_{\rm air} = 7291.5$ and 7323.9\,\AA) becomes prominent as a strong emission feature through de-excitation processes. As a verification of our method (presented below), we are able to qualitatively reproduce this \CaII\ evolutionary behaviour. If the 1.08\,\micron\ feature observed in the photospheric spectra of \gfo\ is indeed produced by \SrII, as the density drops we expect to see this \pcyg\ profile transition into emission and for the equivalent [\SrII] doublet to appear, as we typically see for \CaII. The laboratory-confirmed line centroids for the \forbSrII\ doublet lie at wavelengths, $\lambda_{\rm air} = 6738.4$ and 6868.2\,\AA. Here we analyse the $+7.4 - 10.4$\,d \xsh\ spectra, and investigate if there is evidence for the \SrII\ NIR triplet evolving into a pure emission feature, and for the appearance of the expected \forbSrII\ emission.

\subsection{Method} \label{sec:Analysing the presence of strontium at late times - Method}

Once the ejecta become optically thin, the photospheric approximation used for modelling the early spectra breaks down, and can no longer be used. Additionally, we can no longer assume level populations remain close to their local thermal equilibrium (LTE) populations due to the fact that, in diffuse material, radiative effects will be comparable in strength to (or even dominate over) collisional effects. For example, \cite{Hotokezaka2021} and \cite{Pognan2022, Pognan2022_nlte} demonstrate that non-LTE effects may become important for determining accurate level populations after just a few days in KN ejecta. If the ejecta material is optically thin, then it will be dominated by emission, which may explain the presence of emission-like features in the spectra of \gfo\ taken after $\sim 1$ week.

For this study, we use the collisional data for \SrII, as presented by \cite{Bautista2002}.\footnote{This data was obtained via private communication.} With these data, we are able to estimate the level populations for the lowest few levels of \SrII, based on both radiative and collisional effects, which is more physically motivated than a simple LTE approximation. We do not discuss the details of calculating collision strengths or excitation and de-excitation rates here \citep[readers are instead directed to][for further information on the topic]{Burgess1992}. However, in brief, the thermally averaged collision strengths, $\Upsilon$, are used to derive the excitation ($q_{\textsc{lu}}$) and de-excitation ($q_{\textsc{ul}}$) rates for a level by collisions with free electrons. $q_{\textsc{lu}}$ and $q_{\textsc{ul}}$ both depend on temperature ($T$). For the following calculations, we assume a range of plausible temperatures, $T \in [2500, 3500, 4500]$\,K, for the late-time ejecta. From \paperI, our \tardis\ inner boundary temperatures and velocities are \mbox{$T = 3200$, 3100 and 2900\,K}, and $v_{\rm min} = 0.09$, 0.07 and $0.05 \, c$, for the +5.4, +6.4 and +7.4 day spectra, respectively. Accounting for Doppler corrections, these correspond to ejecta temperatures of $\sim 3000 - 3500$\,K. We propose that this range of temperatures explored by our models ($T \in [2500, 3500, 4500]$\,K) correspond to reasonable expectations for the ejecta temperature after $\sim 1$~week, as well as sampling slightly cooler or hotter ejecta than expected.

We also require an estimate for the electron density. \cite{Jerkstrand2017-HB} presents a review of supernovae in the nebular phase, which includes a useful parameterisation of the electron density of late-phase SN ejecta from a spherically symmetric distribution, given as:
\begin{multline}
    n_e \sim 2 \times 10^9 \left( \mu m_p \right)^{-1} \left( \frac{M}{1\,\msun} \right) \left( \frac{v}{3000\,\kms} \right)^{-3} \\ \times \left( \frac{x_e}{0.1} \right) \left( \frac{t}{200\,\rm{d}} \right)^{-3} \left( \frac{f}{0.1} \right)^{-1} \ \rm{cm}^{-3},
    \label{eqn:General electron density}
\end{multline}
where $\mu$ is the mean atomic weight, $m_p$ is the proton mass, $M$ is the ejecta mass of the system, $v$ is the typical velocity of the ejecta, $x_e$ is the electron fraction (\mbox{$x_e = n_e / n_{\rm nuclei}$}), $t$ is the expansion time, or time since explosion (assuming homologous expansion), and $f$ is the so-called filling factor (see the discussion below).

Here, we use this equation to estimate the electron density in KN ejecta at late times. Assuming $\mu = 100$, $M = 10^{-2}$\,\msun, $v = 0.1\,c$ \citep[see \paperI, or any of][]{Fernandez2013, Metzger2014, Perego2014, Just2015, Wu2016, Siegel2017, Fujibayashi2018, Miller2019, Curtis2021, Just2022} and that, on average, each atom present in the ejecta has been singly ionised ($x_e = 1$), we obtain:
\begin{equation}
    n_e \sim 1.6 \times 10^7 \left( \frac{t}{10 \, \rm{d}} \right)^{-3} \left( \frac{f}{0.1} \right)^{-1} \ \rm{cm}^{-3}.
    \label{eqn:Simplified electron density}
\end{equation}

Suitable values for $f$ in KN ejecta are unknown, and so we assume two different scenarios -- uniform density material ($f = 1$), and highly clumped material ($f = 10^{-2}$). We stress that when we refer to clumped material, we are not referring to large distinct clumps of ejecta with different compositions. Instead, we are referring to small ejecta clumps, which do not impact the column density along our line of sight, known as \textit{microclumping}. These clumps however do lead to regions of over- and under-dense ejecta compared to a homogeneous distribution, which impact the relative importance of collisional to radiative processes. Microclumping has been invoked to explain observations of stellar winds \citep[see e.g.][]{Hamann1998, Hillier1999, Puls2006}, and it has also been invoked empirically in the context of supernova ejecta \citep[see e.g.][]{Dessart2018, Wilk2020}. We therefore argue that it is reasonable to consider the potential impact of microclumping in the kilonova scenario,\footnote{The exact mechanisms that may lead to microclumped ejecta material in kilonovae are not yet understood.} and thus include the filling factor as a parameter in our modelling.\footnote{We highlight the strong velocity dependence in Equation~\ref{eqn:General electron density} ($n_e \propto v^{-3}$), which could lead to deviations in electron density from different velocity structures to be significant; a detailed investigation of this point lies outside the scope of the current project.} Finally, by considering the epochs of the late-time spectra, we can compute the electron densities at each epoch, and these are shown in Table~\ref{tab:electron densities}.

\begin{table}
    \centering
    \caption{
        Computed electron densities at times corresponding to the phases of the late-time \xsh\ spectra of \gfo.
    }
    \begin{tabular}{ccc}
        \hline
        \hline
        Phase       &$n_e$ ($10^{6}$\,cm$^{-3}$)   &$n_e$ ($10^{8}$\,cm$^{-3}$)     \\
        (days)      &[$f = 1$]                     &[$f = 10^{-2}$]                 \\
        \hline
        +7.4       &3.9    &3.9     \\
        +8.4       &2.7    &2.7     \\
        +9.4       &1.9    &1.9     \\
        +10.4      &1.4    &1.4     \\
        \hline
    \end{tabular}
    \label{tab:electron densities}
\end{table}

The thermally averaged collision strengths ($\Upsilon$), combined with the three $T$ values (2500, 3500 and 4500\,K) enable us to determine $q_{\textsc{lu}}$ and $q_{\textsc{ul}}$ values \citep{Burgess1992}. With these, the two sets of $n_e$ estimates (Table~\ref{tab:electron densities}), and the Einstein \Aval\ for all relevant transitions, we determine the level populations for the lowest five levels of \SrII\ \citep[as in, e.g.][]{Chianti-OG}. With quantitative estimates for level populations computed, we can then calculate the number of escaping photons, and the luminosities that we would expect for the emission lines ($L_{\rm em}$).

We need to account for the possibility of these emitted photons being scattered or re-absorbed by another \SrII\ ion before escaping. Therefore, we calculate the escape probability of the emitted photons, $\beta_{\rm esc}$, using:
\begin{equation}
    \beta_{\rm esc} = \frac{1 - e^{-\tau_{\textsc{s}}}}{\tau_{\textsc{s}}},
    \label{eqn:Escape probability}
\end{equation}
where \tausob\ is the Sobolev optical depth of the transition \citep[see e.g.][]{Noebauer2019}. We calculate \tausob\ from the level populations computed by assuming a uniform density sphere, with a radius derived from a constant expansion velocity, \mbox{$v = 0.1 \, c$} (assuming homologous expansion), and with the temperatures and masses presented in Table~\ref{tab:SrII model masses}. $\beta_{\rm esc}$ decreases with increasing \tausob, and so optically thick material will prevent photons from escaping. Since the escape probability of some of the transitions are not of order unity, we need to take this effect into consideration.\footnote{In our analysis, typically, $\tau_{\textsc{s}} \gg 1$ for the resonance doublet transitions, $\tau_{\textsc{s}} > 1$ for the NIR triplet transitions, and $\tau_{\textsc{s}} < 1$ for the forbidden \linebreak doublet transitions.} To do this, we computed \textit{effective} Einstein \Aval\ for all transitions ($\beta_{\rm esc} \times A_{\textsc{\textsc{ul}}}$). We then re-calculated the level populations with these effective \Aval, to obtain a truer representation of the level populations we would expect for \SrII. Line luminosities and profiles are calculated  \citep[as in][]{Gillanders2021} using:
\begin{equation}
    L_{\rm em} = N_{\textsc{u}} \left( \beta_{\rm esc} \times A_{\textsc{ul}} \right) \left( \frac{h c}{\lambda} \right),
    \label{eqn:Luminosity of gaussians}
\end{equation}
where $N_{\textsc{u}}$ is the number of atoms or ions in the excited state, and $\lambda$ is the wavelength corresponding to the transition. Gaussian line profiles for each \SrII\ transition were simulated with these $L_{\rm em}$ values and a full-width, half-maximum (FWHM) velocity of $0.1 \, c$ (see Section~\ref{sec:Modelling the late-time emission features}). We scale the mass of \SrII\ in our level population calculations, such that the strongest emission features in our +7.4\,d models have luminosities comparable to the observed features in the spectrum of \gfo\ at this epoch. These model masses remained fixed for the subsequent epochs, and are provided in Table~\ref{tab:SrII model masses}. To put these masses in context, our best-fitting models from \paperI\ require a range of \SrII\ masses (above the photosphere), from \MSrII~$= 10^{-7}$\,\msun\ for our +1.4\,d model, to \mbox{\MSrII~$= 2.25 \times 10^{-5}$\,\msun} for the model evolved to +7.4\,d. We expect our \MSrII\ values here to be larger than these, since we are sensitive to the entire ejecta in this regime (as opposed to just the line-forming region in the \tardis\ models). We impose an upper mass limit of \MSrII~$= 10^{-3}$\,\msun, to prevent our analysis exploring unphysically large mass estimates.

\begin{figure}
    \centering
    \includegraphics[width=\linewidth]{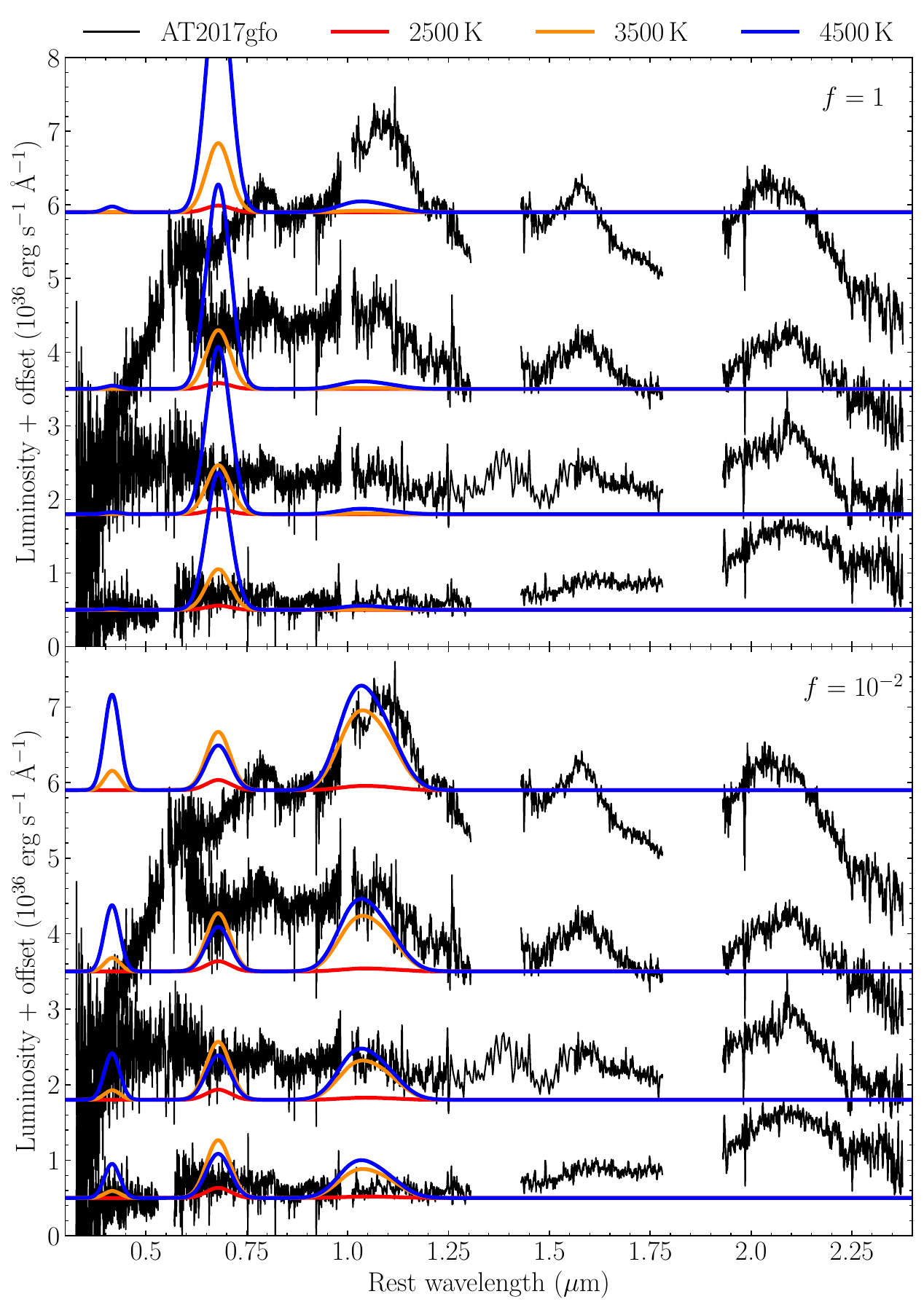}
    \caption{
        Comparison between the observed late-time spectra of \gfo\ (+7.4, +8.4, +9.4 and +10.4\,d, top -- bottom), and our \SrII\ synthetic emission spectra. The observed spectra have been offset for clarity (\mbox{$3 \times 10^{36}$}, \mbox{$2 \times 10^{36}$} and \mbox{$10^{36}$\,\ergsA}, for +7.4, +8.4 and +9.4\,d, respectively). \textit{Upper panel:} Here we present our uniform density material ($f = 1$) models, which are also offset to match the observations, to aid with comparisons. The different colours correspond to different temperatures (2500, 3500 and 4500\,K spectra are red, orange and blue, respectively). \textit{Lower panel:} Same as the upper panel, but with the synthetic emission spectra computed assuming clumped ejecta material ($f = 10^{-2}$).
    }
    \label{fig:sr2_emission_spectra_vs_late-time_17gfo}
\end{figure}

The individual Gaussian profiles were co-added to create a composite emission spectrum. Figure~\ref{fig:sr2_emission_spectra_vs_late-time_17gfo} shows the emission spectra we generated for our range of temperatures and electron densities. In the upper panel, we plot the emission spectra we generated, assuming uniform density ejecta material ($f = 1$), whereas the lower panel contains the models we generated assuming clumped ejecta material ($f = 10^{-2}$). The models in both panels have been over-plotted onto the observed late-phase spectra of \gfo\ ($+7.4 - 10.4$\,d), such that the models and observations that have the same phase are overlaid, to aid comparison.

\begin{table}
    \centering
    \caption{
        Masses of \SrII\ included in our synthetic emission spectra calculations, for both our uniform density ($f = 1$) and clumped ($f = 10^{-2}$) ejecta material.
    }
    \begin{tabular}{ccc}
        \hline
        \hline
        \multirow{2}{*}{$T$ (K)}    &$M_{\rm ej}$ (\msun)   &$M_{\rm ej}$ (\msun)   \\
                                    &[$f = 1$]              &[$f = 10^{-2}$]        \\
        \hline
        2500       &$10^{-3}$    &$10^{-3}$             \\
        3500       &$10^{-3}$    &$5 \times 10^{-4}$    \\
        4500       &$10^{-3}$    &$10^{-4}$             \\
        \hline
    \end{tabular}
    \label{tab:SrII model masses}
\end{table}

\subsection{Uniform density ejecta material ($\boldsymbol{f = 1}$)} \label{sec:Analysing the presence of strontium at late times - Uniform density material (f=1)}

Our 3500 and 4500\,K models, presented in the upper panel of Figure~\ref{fig:sr2_emission_spectra_vs_late-time_17gfo}, produce a strong emission feature at $\sim 6800$\,\AA\ that is not present in the observed spectra. This feature is produced by the \forbSrII\ 6738.4, 6868.2\,\AA\ lines. Although it may be possible for line-blanketing from other species to suppress some of the flux from this feature, the likelihood that a feature of this strength could be so heavily negated, such that it has no observational trace, seems low. Alternatively, it may be possible to explain the lack of this feature by invoking most of the \SrII\ material in the ejecta remains beneath the photosphere. However, this explanation becomes weaker for the later epochs, by which time the photosphere will have receded (or disappeared completely). Our model continues to predict this feature is prominent at these late times, and so this solution cannot reasonably explain the discrepancy between the models and the data. At these temperatures, electron densities and \SrII\ masses, the NIR triplet feature (which we conclude produces the $\sim 0.7 - 1.2$\,\micron\ \pcyg\ feature at early times; see \paperI), is much weaker than the data necessitate. Our 2500\,K model does not produce a strong \forbSrII\ doublet, but also does not reproduce the NIR triplet. In fact, at this temperature, no prominent features are produced, even though we have included our maximum \SrII\ mass (\MSrII~$= 10^{-3}$\,\msun). As the models evolve with time, the strong \forbSrII\ feature weakens, but still remains discrepant with the observations.

Our results indicate that, for the parameter space we have explored with this set of models, we cannot reproduce the observed feature at \mbox{$\sim 1.08$\,\micron} with \SrII. We would need significantly more \SrII\ material to increase the luminosity of the NIR triplet feature, but this would also lead to an increase in the luminosity of the [\SrII] doublet (and exceed our \SrII\ ejecta mass upper limit). Given there is already a discrepancy between our models and the observations at this wavelength, this suggests that a larger mass would not improve the agreement to the data (as well as being unphysical). The strong [\SrII] feature, for which we see no observational signature, strongly indicates that these models already possess too much \SrII\ material. If the ejecta of \gfo\ is well-represented by the range of temperatures and electron densities encompassed by this (i.e. $f = 1$) set of models, then the feature at $\sim 1.08$\,\micron\ cannot be produced by instantaneous \SrII\ emission.

\subsection{Clumped ejecta material ($\boldsymbol{f = 10^{-2}}$)} \label{sec:Analysing the presence of strontium at late times - Clumped ejecta material (f=0.01)}

For the set of models generated assuming $f = 10^{-2}$, we find that the 2500\,K model is too cool to produce any observable features, even when we include our maximum \SrII\ mass (\MSrII~$= 10^{-3}$\,\msun). For the hotter models (3500 and 4500\,K), we find that we need less \SrII\ mass overall (relative to the $f = 1$ models previously presented in Section~\ref{sec:Analysing the presence of strontium at late times - Uniform density material (f=1)}) to produce features with luminosities comparable to the observations, since for these models we have higher $n_e$, which leads to more highly populated excited levels, which in turn leads to stronger emitted luminosities. Another effect of this increase in $n_e$ is reflected in the relative strengths of the emission features present in the synthetic spectra we generate. In the previous set of models, the dominant feature was produced by the \forbSrII\ doublet. However, for $f = 10^{-2}$, the luminosity of the NIR triplet is comparable to (or exceeds) the luminosity of the \forbSrII\ doublet. For the 3500 and 4500\,K models, we have emission features that are comparable in strength to the observed $\sim 1.08$\,\micron\ feature.

However, these $f = 10^{-2}$ models highlight a concerning discrepancy between the synthetic spectra and observations. Here, the emission feature produced by the NIR triplet is too blue to reproduce the observations. The NIR triplet emission feature peaks at $\sim 10450$\,\AA, whereas the observed emission feature peaks at $\sim 10790$\,\AA\ (from our analysis of the late-time emission features; see Section~\ref{sec:Modelling the late-time emission features}). This corresponds to an offset of $\sim 340$\,\AA. To reconcile these features, we would require the bulk of the ejecta to be redshifted by $\sim 9800$\,\kms. This velocity is easily reconcilable with typical KN ejecta velocities (since we see evidence for material at velocities $\gtrsim 0.1 \, c$; see \paperI\ and Section~\ref{sec:Modelling the late-time emission features}), but the notion that the bulk of the ejecta will be redshifted by this amount is unexpected.

Although a bulk offset of $\sim 10000$\,\kms\ seems unusual, without other identifiable features to independently constrain ejecta properties, it is impossible to conclude if this is indeed redshifted \SrII\ emission, or some other effect. An alternative explanation for this discrepancy is that the feature is being produced by a \pcyg\ profile, with an absorption feature to the blue that affects the perceived peak wavelength of the emission. However, there exists no evidence for such absorption in the observed spectrum, although we note that this could be concealed if there is some other species `filling in' the absorption trough with its own emission. Another explanation is that there is some other contributing species redward of the \SrII\ NIR triplet, which blends to produce a single emission feature that peaks redward of the weighted peak wavelength of the \SrII\ NIR triplet. Other explanations include asymmetrical ejecta structures (which could lead to a disproportionate amount of \SrII\ material moving away from us), or observational effects arising from $\nu$-dependent arrival times from a rapidly-expanding medium. Finally, it is also possible that this feature is in fact not produced by \SrII\ at all, although this seems unlikely, given the agreement with \SrII\ at early phases (see \citealt{Watson2019}, \citealt{Perego2022} and \paperI). \cite{Tarumi2023} have proposed that the He\,\I\ $\lambda_{\rm air} = 10830$\,\AA\ line could be the dominant source, which is a much closer match to the line centroid of the observed emission feature at 10790\,\AA\ (see Section~\ref{sec:Analysing the presence of strontium at late times - Summary} for more details).

The \forbSrII\ feature is also present in these models, although not as prominently as in the $f = 1$ models. These model features again do not agree with the observations. As previously discussed, it may be possible that another species is contributing to the same region of the spectrum, and its absorption is removing any trace of this emission feature. A similar scenario may also explain the discrepancy between the 4500\,K model spectra and the observed data at $\sim 4150$\,\AA, which is produced by the \SrII\ $\lambda_{\rm air} = 4077.7$, 4215.5\,\AA\ resonance lines. Without some mechanism to destroy the emission from these features (both the resonance lines and the \forbSrII\ doublet feature for the 4500\,K model, or just the \forbSrII\ doublet for the 3500\,K model), these models cannot reproduce the data. Therefore, we either need to invoke the presence of line-blanketing in the spectrum at wavelengths $\lesssim 7500$\,\AA, or the ejecta of \gfo\ do not contain this amount of \SrII\ ($5 \times 10^{-4}$\,\msun\ for the 3500\,K model, and $10^{-4}$\,\msun\ for the 4500\,K model).

So far, we have not considered how our clumped ejecta models agree with the observations beyond the +7.4\,d spectrum. Our good models (3500 and 4500\,K) become less luminous with time, as the expanding material leads to lower electron densities. The strong emission feature at $\sim 1.08$\,\micron\ in the observed spectra also decays with time, and the rate at which it fades seems to roughly match the fading of the model features at the same position (assuming our continuum is placed correctly at +8.4\,d, which is not clear). The resonance feature and the forbidden doublet remain reasonably persistent across all epochs, and so if it is the case that line-blanketing (or some other effect) is acting to mask their presence in the observed data, it must be the case that it continues to do so across all epochs that we model here ($+7.4 - 10.4$\,d). Alternatively, these features could be removed at later epochs by recombination (or ionisation) effects; i.e. most of the \SrII\ recombines to Sr\,\I\ (or ionises to Sr\,\III), which would also accelerate the rate of fading of these \SrII\ features, improving the agreement to the data.

\begin{figure}
    \centering
    \includegraphics[width=\linewidth]{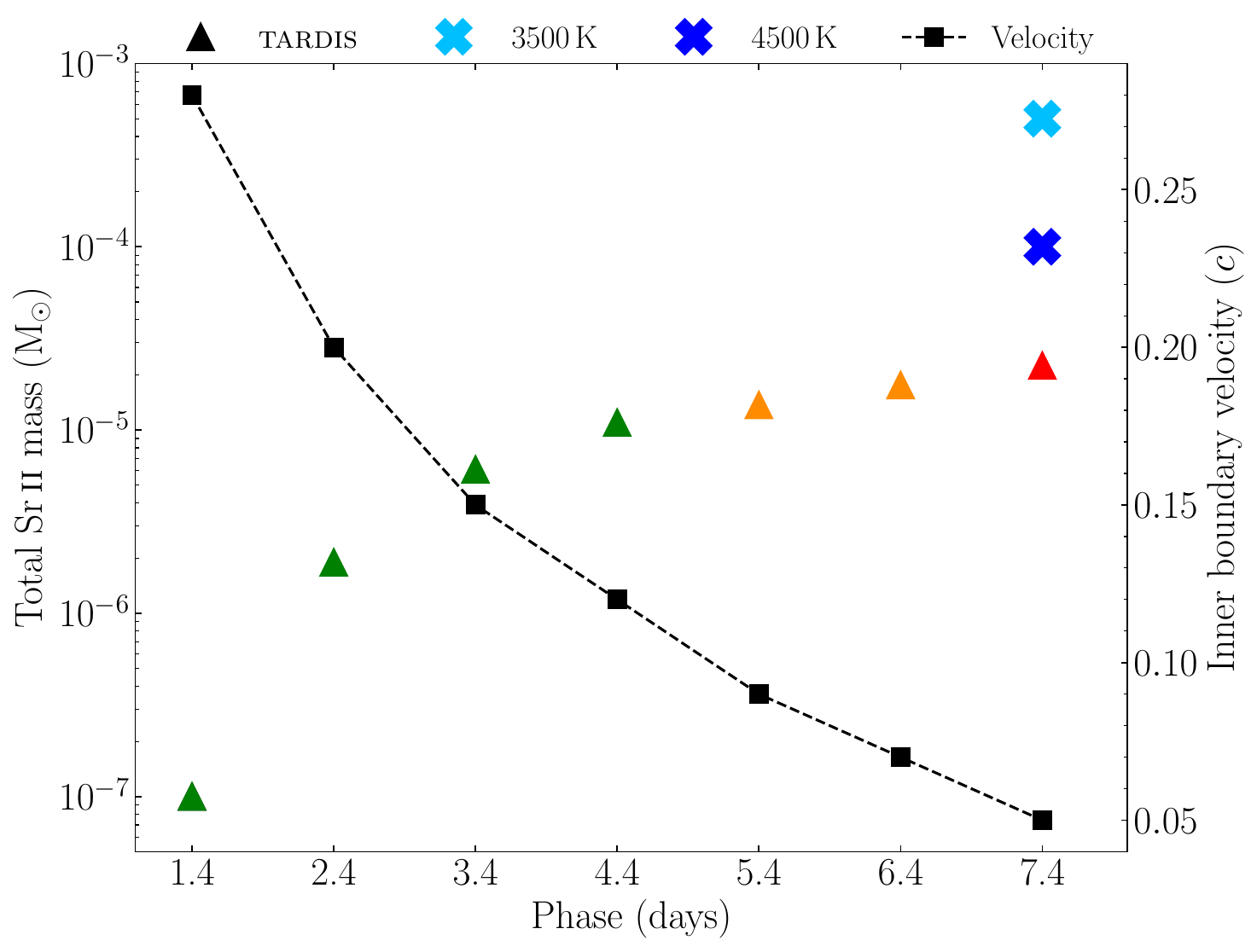}
    \caption{
        Total \SrII\ masses inferred from our \tardis\ modelling in \paperI\ (lower limits; triangles). The reliability of these masses is indicated by marker colour, with the green, orange and red triangles indicating good, medium and poor reliability, respectively. Also plotted are the \SrII\ masses inferred from our modelling here, with the points for the $T = 3500$ and 4500\,K models included. We also plot the evolution of the inner boundary velocities from our sequence of \tardis\ models, to indicate how much of the total velocity range of the system our \tardis\ lower limit masses represent at each epoch.
    }
    \label{fig:SrII_mass_evolution}
\end{figure}

In \paperI, we presented the inferred masses of \SrII\ needed to model the observed sequence of \xsh\ spectra of \gfo, up to +7.4\,d. Due to the approximations made within \tardis, we argued that our models are capable of accurately representing the ejecta properties of \gfo\ at $+0.5 - 4.4$~days. The models for the +5.4 and +6.4\,d spectra do an adequate job of reproducing the observations, with some caveats, as the observed spectra appear to be transitioning away from a purely photospheric regime. The +7.4\,d model is a poor reproduction of the data as, at this phase, \tardis\ is no longer suitable for modelling the ejecta properties. Therefore, we expect our model parameters (including inferred \SrII\ mass) to be most reliable up to +4.4 days, be reasonably reliable at +5.4 and +6.4 days, and not reliable at +7.4 days and beyond. In Figure~\ref{fig:SrII_mass_evolution}, we present the total \SrII\ mass in each of our \tardis\ models. The points are qualitatively colour-coded to indicate the reliability of the \tardis\ \SrII\ masses presented at each phase (with the green, orange and red points indicating good, medium and poor reliability, respectively). Since \tardis\ only simulates the line-forming region of the ejecta, our \tardis\ masses are actually \textit{lower limits} for the total \SrII\ mass at each phase (hence they are marked as lower limits on the plot; see \paperI\ for more on these models).

Also plotted in Figure~\ref{fig:SrII_mass_evolution} are the inferred \SrII\ masses from our modelling efforts here, invoking $T = 3500$ and 4500\,K. Clearly, we need more \SrII\ material to model the +7.4\,d spectrum in this way than we did using \tardis\ for the same epoch. This is not too informative, as the reliability of this \tardis\ model is poor. Note that the total Sr mass included in our \tardis\ model compositions (from $+2.4$\,d onward) does not change, but since the inner boundary is receding with time, we are modelling a greater portion of the total system, and so we infer a larger \SrII\ mass. Therefore, we also expect our modelling work here to require more \SrII\ mass than our \tardis\ models, which they do. This is consistent with most (or all) of the ejecta becoming optically thin, allowing us to see the effects of all of the \SrII\ material in the ejecta. We note that the inferred masses for \SrII\ from this analysis are not unphysically high either. Our best-fitting composition profile for the early spectra (+2.4\,d onward) of \gfo\ contains a Sr mass fraction of 0.0325 (see \paperI). Assuming a total ejecta mass for the system, $M \simeq 10^{-2}$\,\msun, this implies a total Sr mass, $M_{\rm Sr} = 3.3 \times 10^{-4}$\,\msun, which is close to the inferred mass of \SrII\ for our 3500\,K model ($5 \times 10^{-4}$\,\msun), and comfortably higher than the \SrII\ mass inferred for the 4500\,K model ($10^{-4}$\,\msun).\footnote{Note that we have not made any assumptions regarding the ionisation state of Sr at late times, and so this mass comparison between Sr from the early phases, and \SrII\ from the present work, although informative, will be somewhat uncertain. A more complete understanding of both non-LTE and non-thermal effects is needed to enable a sensible comparison.}

\cite{Watson2019} invoke Sr masses of $1 - 5 \times 10^{-5}$\,\msun\ (in the line-forming region) to reproduce the feature between $7000 - 10000$\,\AA\ in the $+1.4 - 4.4$\,d spectra, while \cite{Perego2022} suggest that $\sim 8 - 16 \times 10^{-5}$\,\msun\ of Sr was ejected by \gfo. Both of these ranges are typically larger than the range of values we present for \SrII\ masses from our \tardis\ modelling, which is somewhat expected, since we are comparing Sr masses with \SrII\ masses.

\subsection{Summary and discussion} \label{sec:Analysing the presence of strontium at late times - Summary}

Although this modelling approach seems quite simplistic, we can deduce some useful constraints for the ejecta of \gfo. First, we find that the identification of the \SrII\ NIR triplet transitions as the dominant source of the $\sim 1.08$\,\micron\ feature is problematic for uniform density material, with reasonable estimates for ejecta temperature, velocity and ionisation. This is because the [\SrII] doublet transitions are the dominant transitions, leading to strong emission at $\sim 6800$\,\AA, for which we see no observational trace (see the upper panel of Figure~\ref{fig:sr2_emission_spectra_vs_late-time_17gfo}). Therefore, if the ejecta is smooth (i.e. $f = 1$), then the \mbox{$\sim 1.08$\,\micron} feature cannot be produced by instantaneous \SrII\ emission.

Second, we show that, by invoking significant clumping of the ejecta material, we can reach regimes of high enough electron density such that the NIR triplet emission is stronger (or at least comparably as strong) as the [\SrII] doublet (see the lower panel of Figure~\ref{fig:sr2_emission_spectra_vs_late-time_17gfo}). Clumped ejecta material has been invoked to interpret SN ejecta \citep[see e.g.][]{Dessart2018, Wilk2020}, and so it seems reasonable that we invoke it here. If the $\sim 1.08$\,\micron\ feature is produced by \SrII, then we require clumped ejecta material to boost the relative strength of the NIR triplet feature over the [\SrII] doublet and the resonance lines. Either the ejecta material was always clumped, or became clumped as it expanded and cooled.

Third, even accounting for clumped ejecta material, which enables our models to produce a feature comparable in strength to observations at $\sim 1.08$\,\micron, we still have an issue with feature offset. Our model emission peaks at $\sim 10450$\,\AA, which is $\sim 340$\,\AA\ blueward of the observed feature peak at +7.4\,d, corresponding to a $\sim 9800$\,\kms\ bulk redshift in the observed feature. The source of this discrepancy may be the result of a number of factors, the most likely of which is our assumption that these late-time spectra originate from a fully nebular regime, which likely does not completely encapsulate the properties of the system. Alternatively, the difference could be evidence of a contribution from some (one or multiple) unidentified species that contributes to (or completely dominates) the emission immediately redward of the \SrII\ NIR feature. Other sources for the discrepancy could be some asymmetrical ejecta structure, or observational effects arising from $\nu$-dependent arrival times from a rapidly-expanding medium.

In light of these issues, it is worth considering the alternative possibility as investigated by \cite{Perego2022} and \cite{Tarumi2023}, namely that He\,\I\ can produce the $\sim 1.08$\,\micron\ feature. As noted above, the He\,\I\ $\lambda_{\rm air} = 10830$\,\AA\ line closely matches the peak wavelength of the observed feature at late times. There are no other contaminating lines that obviously do not match the data, which could be used to rule out the presence of He\,\I. We note in particular that the suggestion by \cite{Tarumi2023} -- that the feature could be dominated by Sr\,\II\ at early times, with He\,\I\ becoming dominant at later times -- could alleviate the issues identified above. Specifically, late-time contribution from He\,\I\ would help explain the apparent wavelength shift, as well as being able to reproduce the 1.08\,\micron\ feature without the need to invoke a large \SrII\ mass that leads to strong \forbSrII\ emission (in the $f = 1$ case).

The case for He\,\I\ is also strengthened when considering nucleosynthesis arguments. He is among the most abundant elements in many of the composition profiles we present in \paperI. These composition profiles were extracted from nuclear network calculations based on a realistic hydrodynamical simulation of a BNS merger \citep{Goriely2011, Goriely2013, Goriely2015, Bauswein2013}, with some modification. The data have a prescribed distribution of \Ye, from which we extract `representative' composition profiles, effectively binned by \Ye. This approach was taken to obtain a diverse range of feasible composition profiles for the ejecta from BNS mergers. These composition profiles span a range of \Ye\ values ($0.44 - 0.05$), which correspond to compositions dominated by the light iron-group and first \rpro\ peak elements ($Y_e = 0.44$), through to compositions dominated by the heavy lanthanides and third \rpro\ peak elements ($Y_e = 0.05$). Consulting these composition profiles, we see that He makes up 15 per cent of all material (by mass) in our \AngIX\ composition profile (see \paperI\ for more details). This is in agreement with other studies; for example, \cite{Just2023} show BNS mergers are expected to produce significant quantities of He (mass fractions of $\sim 10$~per~cent). As the analysis by \cite{Tarumi2023} and the calculations presented here show, a full understanding of this (and other) features will depend on further studies, in which full non-thermal and non-LTE excitation and ionisation effects are incorporated for realistic ejecta models.

\section{Modelling the late phase features} \label{sec:Modelling the late-time emission features}

As previously highlighted in \paperI, the evolution of the spectral energy distribution (SED) of \gfo\ is unlike any previously observed extragalactic transient. After $\sim 1$ week, the ejecta has expanded and cooled sufficiently such that the spectra appear to be originating from an optically thin regime. As such, the photospheric approximation used with our \tardis\ modelling in \paperI\ can no longer be applied to model the spectra. Hence, we need an alternative method to analyse the late-time observations. If the spectra are truly originating from an optically thin region, then the features observed will be produced predominantly by collisionally-excited emission and/or recombination. Here we fit Gaussian profiles to the most prominent emission-like features in the late-time spectra to constrain the peak positions, which we then use in our attempts to identify candidate species for each of the features (see Sections~\ref{sec:Search for potential candidate species (strong transitions)}~and~\ref{sec:Search for potential candidate species (weak transitions)}). In addition, measured line widths and centroids can be used to search for evolution of the ejecta properties, and fluxes can be used to constrain the properties of the emitting lines. Note that this approach represents an alternative interpretation for the features at $\sim 1.58$ and 2.07\,\micron\ than that presented in Section~\ref{sec:Modelling the early NIR features}, and we note that both features become more prominent in the $+7.4 - 10.4$\,d spectra.

\begin{figure}
    \centering
    \includegraphics[width=\linewidth]{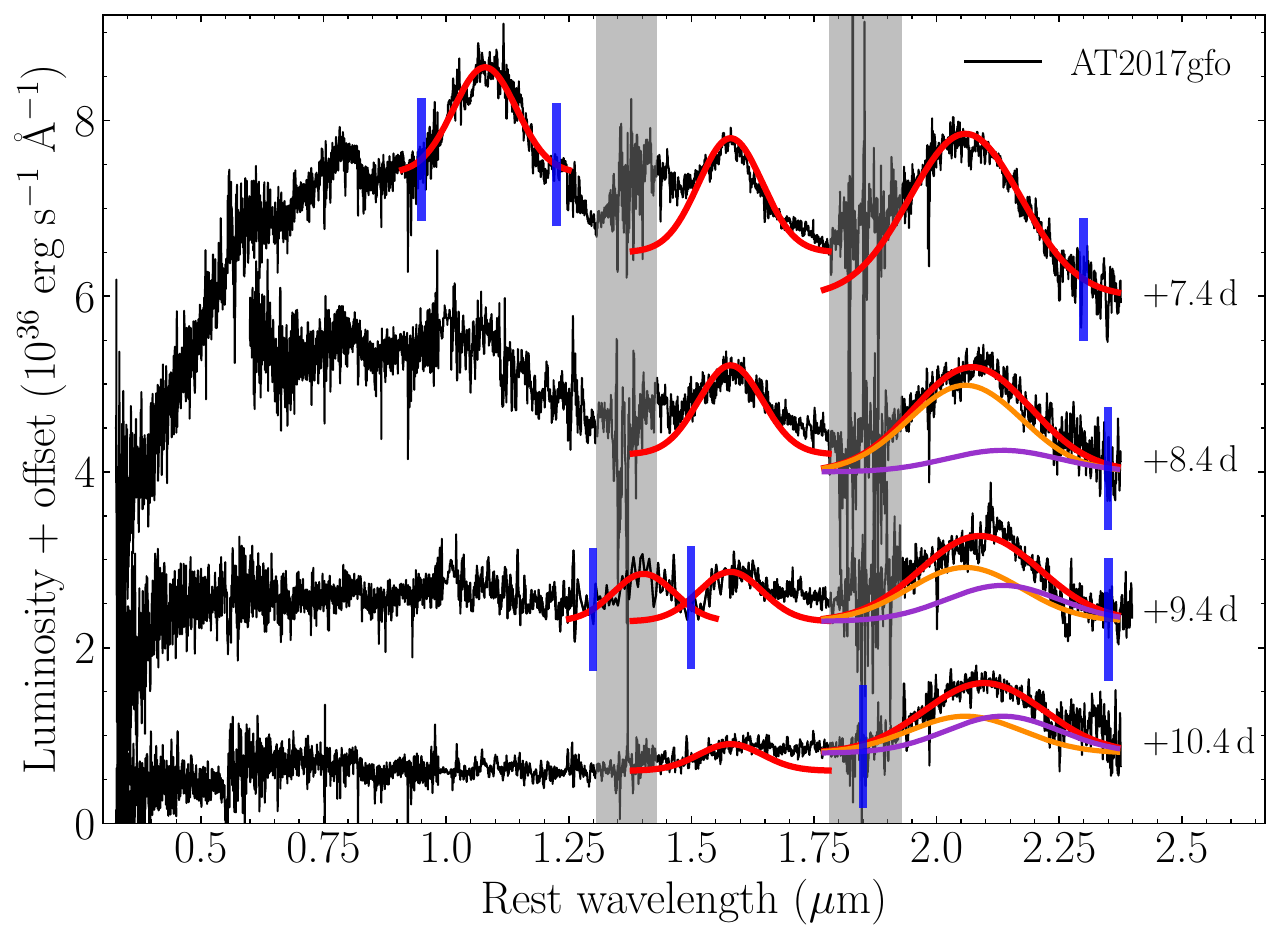}
    \caption{
        Best-fitting Gaussians overlaid on the late-time spectra of \gfo. The observed spectra are offset for clarity (\mbox{$4.5 \times 10^{36}$}, \mbox{$3.0 \times 10^{36}$} and \mbox{$1.5 \times 10^{36}$\,\ergsA}, for +7.4, +8.4 and +9.4\,d, respectively), and are annotated with their phase, relative to the GW trigger. The solid red lines represent our best-fitting Gaussians for the most prominent emission features. The feature at 2.07\,\micron\ in the $+8.4 - 10.4$\,d spectra can be reproduced well by the composite of two Gaussians, which are plotted in orange and purple, while the co-added resultant Gaussian is plotted in red. The vertical blue lines mark where we estimate the emission features blend into the continuum. As the position of the continuum is quite uncertain, we do not make predictions for these positions at the blue and red wings of all features. The vertical grey bands correspond to telluric regions.
    }
    \label{fig:Gaussian fits}
\end{figure}

We focussed on fitting the most prominent of these emission features -- specifically, we fit the $\sim 1.08$, 1.40, 1.58 and 2.07\,\micron\ features. We used a $\chi$-squared analysis to determine the best-fitting Gaussians for these features, and these are shown in Figure~\ref{fig:Gaussian fits}. The parameters corresponding to the fits are presented in Table~\ref{tab:Gaussian fit parameters}.

\begin{table*}
    \renewcommand*{\arraystretch}{1.2}
    \centering
    \caption{
    Parameters of the best-fitting Gaussians presented in this work.
    }
    \begin{threeparttable}
        \centering
        \begin{tabular}{cccccccc}
            \hline
            \hline
            \multirow{2}{*}{\begin{tabular}[c]{@{}c@{}}Approx.\\$\lambda_{\rm peak}$ (\micron)\end{tabular}}
            &\multirow{2}{*}{\begin{tabular}[c]{@{}c@{}}Peak\\(10$^{35}$\,erg\,s$^{-1}$\,\AA$^{-1}$)\end{tabular}}
            &\multirow{2}{*}{\begin{tabular}[c]{@{}c@{}}Luminosity\\(10$^{38}$\,erg\,s$^{-1}$)\end{tabular}}
            &\multirow{2}{*}{\begin{tabular}[c]{@{}c@{}}$v_{\textsc{fwhm}}$\\(\kms)\end{tabular}}
            &\multicolumn{2}{c}{$v_{\rm edge}$ (\kms)}
            &\multirow{2}{*}{\begin{tabular}[c]{@{}c@{}}$\lambda_{\rm peak}$\\(\AA)\end{tabular}}
            &\multirow{2}{*}{\begin{tabular}[c]{@{}c@{}}Continuum\\(10$^{36}$\,erg\,s$^{-1}$\,\AA$^{-1}$)\end{tabular}}    \\
            \cline{5-6}
                &    &    &     &Blue    &Red    &  &     \\
            \hline
            \multicolumn{8}{c}{+7.4\,d} \\
            \hline
            1.08     &12.0    &19.5    &42190    &35840  &40620      &10790   &2.9   \\
            1.58     &13.0    &21.1    &29050    &$-$    &$-$        &15800   &2.0   \\
            2.07     &18.5    &52.6    &38910    &$-$    &35160      &20590   &1.5   \\
            \hline
            \multicolumn{8}{c}{+8.4\,d} \\
            \hline
            1.58     &10.1      &16.5       &29050   &$-$   &$-$        &15800    &1.2   \\
            2.07     &9.87      &28.1       &38900   &$-$   &42400      &20590    &1.0   \\
            2.07     &2.47      &7.28       &38900   &$-$   &30210      &21350    &1.0   \\
            \hline
            \multicolumn{8}{c}{+9.4\,d} \\
            \hline
            1.40    &5.40       &8.46       &31420      &21980\tnote{*}     &20790\tnote{*}    &14030     &0.8   \\
            1.58    &5.63       &9.18       &29050      &15230\tnote{*}     &$-$               &15800     &0.8   \\
            2.07    &6.14       &17.5       &38900      &$-$                &42400             &20590     &0.8   \\
            2.07    &4.09       &12.1       &38900      &$-$                &30210             &21350     &0.8   \\
            \hline
            \multicolumn{8}{c}{+10.4\,d} \\
            \hline
            1.58    &3.04       &4.95       &29050      &$-$        &$-$    &15800      &0.6   \\
            2.07    &4.21       &12.0       &38900      &30450      &$-$    &20590      &0.8   \\
            2.07    &4.21       &12.4       &38900      &40050      &$-$    &21350      &0.8   \\
            \hline
        \end{tabular}
        \begin{tablenotes}
              \small
              \item[*] These $v_{\rm edge}$ values possess some associated error from spectral splicing, and so we expect them to be very uncertain.
        \end{tablenotes}
    \end{threeparttable}
\label{tab:Gaussian fit parameters}
\end{table*}

The 1.08\,\micron\ feature is only prominent in the +7.4\,d spectrum, and so we were only able to fit this feature at this single epoch. This is the feature that we have concluded is likely produced by \SrII\ at early times (see \paperI), and late times (see Section~\ref{sec:Analysing the presence of strontium at late times}).

The telluric region at 1.40\,\micron\ prevents identification of any emission feature at this wavelength in the \xsh\ data. However, the +9.4\,d spectrum, with the merged \hst\ spectrum covering this wavelength range, clearly shows an emission feature. The presence of this feature in the +9.4\,d spectrum should not be completely unexpected, since we can see it is prominent in the earlier +4.9\,d \hst\ spectrum (see Figure~\ref{fig:Spectral evolution}).

The feature at 1.58\,\micron\ is prominent in the $+7.4 - 10.4$\,d spectra (although by +10.4\,d it has weakened significantly, and appears to barely rise above the continuum to the red). The blue wing of this feature in the +9.4\,d spectrum contains data from the \hst\ spectrum that has been merged into the \xsh\ spectrum here to demonstrate the emission feature at 1.40\,\micron. This join in the two spectra may have introduced some error in this region of the spectrum. For these reasons, we trust the results of our fit to the earlier two epochs at this wavelength range more. Hence, we freely fit the 1.58\,\micron\ feature in the +7.4 and +8.4\,d spectra, and then took an average of these values and refit. We found that the average of these fits was able to broadly reproduce the feature across both epochs, thus motivating no additional analysis; i.e. the data can be adequately modelled with a single emission feature with identical width and peak wavelength across both epochs. We then took the peak wavelength and width from the +7.4 and +8.4\,d fits, and imposed those on the +9.4 and +10.4\,d spectra, allowing only the line luminosity to vary. For +9.4\,d, this fit does not replicate the blue wing of the feature, but this is not surprising, based on the discussion above. For +10.4\,d, this fit matches the data reasonably well, although we again note the continuum to the red is higher than the blue, which our Gaussian does not match.

The 2.07\,\micron\ feature exhibits the most evolution across these four epochs. In the +7.4\,d spectrum, it peaks at $\sim 2.06$\,\micron, but by +10.4\,d, this has shifted by about 400\,\AA, peaking at $\sim 2.10$\,\micron\ (see Section~\ref{sec:Spectroscopic evolution of AT2017gfo}). We find that the evolution of this feature can be easily modelled by the presence of two distinct emission features,\footnote{We find that the data cannot be adequately reproduced with a single Gaussian with fixed width and peak wavelength, hence we invoke two Gaussian components for our analysis for this specific feature.} the relative strengths of which evolve across the multiple observations. By assuming that the feature in the +7.4\,d spectrum is composed primarily of a single emission feature, and by fixing the properties of this best-fitting Gaussian for all subsequent epochs, we are able to satisfactorily reproduce the emission feature across all subsequent features by introducing a second Gaussian emission feature redward of the first. We also fixed the properties of this Gaussian across the \mbox{$+8.4 - 10.4$\,d} spectra. The only changes to either of these Gaussians are the strengths of the features -- both absolute luminosity from the transitions, and the relative strengths of the features. By transitioning from a regime dominated by the bluer Gaussian, to a blend between the two, with increasing contribution from the redder Gaussian as time evolves, we can reproduce the feature across all four spectra. We find that this feature can be well-reproduced by these two Gaussians, and there is no requirement from the data to have different FWHM velocities, indicating they are both originating from similar regions of the ejecta, which seems to be a physically motivated explanation for their source. These individual Gaussians are plotted in orange and purple in Figure~\ref{fig:Gaussian fits}, while the co-added resultant Gaussian is plotted in red.

The maximum velocity of the emitting region can be estimated by considering where the emission features blend into the continuum ($v_{\rm edge}$). However, this can be difficult to determine accurately due to the uncertainty in continuum location. Nevertheless, we manage to estimate some of these positions (see Table~\ref{tab:Gaussian fit parameters} and Figure~\ref{fig:Gaussian fits}).

For cases when identifying where the feature blends into the continuum is more difficult, we adopt the FWHM velocity as a proxy for the velocity extent of the emitting region, noting that FWHM estimates are argued to be comparable to half-width zero-intensity measurements in a study of nebular phase supernova spectra \citep{Nicholl2019}. Our FWHM velocities are generally comparable to our $v_{\rm edge}$ values, indicating the validity of this approach. From our best-fitting Gaussians, we conclude that a FWHM ejecta velocity, \mbox{$v_{\textsc{fwhm}} = 35600 \pm 6600$\,\kms}, is capable of reproducing all of the prominent emission features in the late-time spectra of \gfo. Hence, for simplicity in all of our analyses \citep[e.g. Section~\ref{sec:Analysing the presence of strontium at late times}, Appendix~\ref{sec:Search for potential candidate species (weak transitions) - Kurucz analysis}, and previously in][]{Gillanders2021}, we adopt a FWHM velocity of $0.1\,c$ as a characteristic ejecta velocity.

We find that all emission features across all late phases \mbox{($+7.4 - 10.4$ days)} can be well-matched by invoking a velocity of $\sim 0.1 \, c$. We find no evidence for continued feature peak evolution at these late phases (i.e. the peak positions remain constant across the \mbox{$+7.4 - 10.4$\,d} spectra). We also found that the 2.07\,\micron\ feature evolution is best explained as a blend of two prominent components, which have fixed peak positions, but whose relative strengths vary. Finally, our feature luminosity measurements here can be used to constrain the masses needed to reproduce the observations. These luminosity values enable us to make quantitative statements regarding species we favour or disfavour, based on the validity of their derived mass estimates (see e.g. Section~\ref{sec:Modelling the early NIR features} and Appendix~\ref{sec:Search for potential candidate species (weak transitions) - Kurucz analysis}).

\section{Search for potential candidate species (strong transitions)} \label{sec:Search for potential candidate species (strong transitions)}

With estimates for the peak position of each emission feature from Sections~\ref{sec:Spectroscopic evolution of AT2017gfo}~and~\ref{sec:Modelling the late-time emission features}, we can now attempt to identify the transitions that are responsible for producing these features. Additionally, the luminosity required to produce these features can be used to infer the mass of the emitting species (if we know the intrinsic line strength). In this section, we consider permitted line transitions (which may produce \pcyg\ features as explored in Section~\ref{sec:Modelling the early NIR features}). In Section~\ref{sec:Search for potential candidate species (weak transitions)} we extend the search to weaker candidate transitions.

\subsection{Method} \label{sec:Appendix - Search for potential candidate species (strong transitions) - Method}

For our search for permitted transitions coincident with the prominent emission features in the spectra of \gfo, we generate synthetic line lists from the level information that is available for the heavy elements. Specifically, we extracted the level information from the National Institute of Standards and Technology Atomic Spectra Data base \citep[NIST ASD;][]{NIST2020} for all neutral, singly and doubly ionised species, with $38 \leq Z \leq 92$.\footnote{Although the Kurucz data base has atomic data for $Z = 38 - 46$, we analyse these species again via this method as a check to see if we recover the \SrII\ NIR triplet, which we have shown is capable of reproducing the $\sim 1.08$\,\micron\ emission feature reasonably well at both early and late times.} \cite{Tanaka2020} find that for KN ejecta at times, \mbox{$t \gtrsim 1$\,d}, and temperatures, \mbox{$T \lesssim 20000$\,K}, the dominant ionisation stages of heavy elements are neutral up to triply ionised (\mbox{$\textsc{i} - \textsc{iv}$}). Since our \tardis\ models for the third epoch (and all later epochs) have characteristic temperatures much lower than this (3400\,K for the +3.4\,d best-fitting \tardis\ model), we assume that the dominant ionisation stages will be neutral up to doubly ionised (\mbox{$\textsc{i} - \textsc{iii}$}). We can validate this choice by analysing the ionisation of the species in our best-fitting \tardis\ models, presented in \paperI. As an example, for each of the lanthanides, the mass fraction of triply ionised material, $M_{\textrm{X}\,\textsc{iv}} \lesssim 10^{-14}$, and so they can be safely excluded from consideration. Therefore, we focus only on extracting the first three ion stages from the NIST ASD.

With this level information, we perform a rudimentary set of calculations, where we determine the energy differences between all levels. This gives us the wavelengths for any theoretical transitions between all known levels within the NIST ASD. We note that all energy levels within the NIST ASD are critically evaluated, and so our transition wavelengths that we compute are sufficiently accurate for line identification studies. However, this approach lacks Einstein~$A$-coefficients, and so we use a different method to estimate line strength (see below). Since we are considering permitted transitions as the cause of the features, we can make some further cuts on our line list. First, permitted, or electric dipole, transitions (traditionally labelled E1) require a parity change between levels. Therefore, we can discard any lines from our list that do not satisfy this clause. Additionally, E1 transitions can only have variations in the quantum J number, $| \Delta \rm{J} | = 0$ or 1, so we excluded all lines from our list that do not satisfy this rule too. For the instances where the configurations and terms for levels can be easily expressed by the LS-coupling formalism, we also eliminate lines that do not obey the LS-coupling rules.\footnote{We note that the LS-coupling rules do not capture the complete line information for heavy species, and so we do not fully recover all possible lines for the heavy species here. However, as discussed in Appendix~\ref{sec:Appendix - Search for potential candidate species (weak transitions) - NIST - Verification of analysis}, we recover many of the relevant lines for the cases of [Pt\,\I] and [Au\,\I]. We expect this issue to be less pronounced for the lighter species in our study.} Finally, we also discard any lines from these lists that have an upper level with a higher excitation energy than the ionisation energy for the species under consideration (also extracted from the NIST ASD).

We lack any measure of intrinsic line strengths from this approach, and so, in an effort to begin to constrain whether the transitions we propose are likely to be prominent, we calculate relative level populations for the upper levels of each of the transitions. We assume LTE, and so the level populations will be governed simply by the Boltzmann equation:
\begin{equation} \label{eqn:no_ions_upper_lvl}
    N_{\textsc{u}} = N_{\textsc{t}} \left( \frac{g_{\textsc{u}}}{Z} \right) e^{-\frac{E_{\textsc{u}}}{k_{\textsc{b}} T}},
\end{equation}
where $N_{\textsc{u}}$ is the number of atoms or ions in the upper level, $N_{\textsc{t}}$ is the total number of atoms or ions, $g_{\textsc{u}}$ is the statistical weight of the upper level, $Z$ is the LTE partition function, $E_{\textsc{u}}$ is the energy of the upper level, and $T$ is the temperature. We assume that LTE level populations should be a reasonable approximation, at least for the earlier epochs. We cannot compute total numbers of populated levels without an accurate constraint on the ejecta composition and mass, but we can compute \textit{relative} level populations using the statistical weights and energies of the levels, and some estimate for the ejecta temperature. From \paperI, our \tardis\ models for the $+3.4 - 7.4$\,d spectra have ejecta temperatures ranging from 3400\,K at +3.4\,d, to 2900\,K at +7.4\,d. For simplicity, we assume $T = 3000$\,K for the calculations here.

This approach allows us to better understand the relative strengths of these transitions, since, if we assume they are all equally strong lines (identical \Aval), then larger level populations directly correspond to more flux. Hence, we use upper level population as a proxy for line strength, and subsequently filter our data, to identify what we consider to be the most prominent, and therefore most relevant, lines in our analysis.

We normalise these level populations to the upper level that has the strongest potential E1 transition with a wavelength that lies within the range covered by the \xsh\ spectra of \gfo\ \mbox{($0.5 \leq \lambda \leq 2.5$\,\micron)}. As an example for how to interpret these relative upper level population values, consider La\,\III\ (see Table~\ref{tab:NIST - Candidate transitions (E1)}). This ion has two candidate transitions for the 1.40\,\micron\ feature. Based on the relative level population calculation, the 13898\,\AA\ transition comes from an excited level that is $\sim 1.5$ times more populous than the excited level that produces the 14100\,\AA\ transition. This indicates that, if all other parameters remain identical, the number of photons emitted through the 13898\,\AA\ candidate transition will be approximately 1.5 times that emitted through the 14100\,\AA\ candidate transition.

With our line list and associated estimates for line strengths, we make comparisons to the observed spectra of \gfo. Species that do not produce any strong emission features coincident with the observed features of interest ($\sim 0.79$, 1.08, 1.23, 1.40, 1.58 and 2.07\,\micron) are discarded. Additionally, species that produce features coincident with one (or more) emission feature, but also produce comparably prominent emission at wavelengths where we can quantitatively rule out significant flux are also discarded. However, we do not rule out species that only have a few lines in the observable wavelength range of \gfo\ on this basis. Due to uncertainties associated with our estimates, we retain the cases with a small number of lines, as the contaminant line may be weaker than our analysis predicts, and thus the model could potentially match the observations. When we consider species with many lines with broad flux contamination, we discard the species, since even if the relative strengths are uncertain, we still expect the effect of many blended lines on the spectra to be prominent.

We are interested in the lines that may be producing the observed features, but the locations of these peak positions are somewhat uncertain. Therefore, we adopted an arbitrary wavelength range that corresponds to a $0.05 \, c$ Doppler shift tolerance for the peak wavelength of each feature, within which lines are considered viable candidate transitions.

\subsection{Results} \label{sec:Search for potential candidate species (strong transitions) - Results}

We recover two shortlists of candidate species that contain transitions that match the observational data: (i) a viable list of candidate ions that match the data with no perceived limitations, and (ii) a list of ions that may contribute to the data, with some caveat.\footnote{An example of a caveat is an ion that possesses a line in agreement with an observed feature that is expected to have another comparably strong line at a position where we do not see any observational evidence for emission (see Section~\ref{sec:Appendix - Search for potential candidate species (strong transitions) - Method}).} Our list of viable candidate lines for each of the observed features is given in Table~\ref{tab:NIST - Candidate transitions (E1)}.\footnote{For completeness, we present the wavelengths of the candidate transitions in Table~\ref{tab:NIST - Candidate transitions (E1)} both in vacuum and air.} Our viable list of candidate species include \SrII, La\,\III, Ce\,\III, Gd\,\III, Ra\,\II\ and Ac\,\I\ (see Table~\ref{tab:Candidate ions}).

We find candidate species for all prominent emission features, but some are less plausible than others. Additional points of consideration are the number of total contaminating lines, and the upper level energy.\footnote{Our analysis computes a \textit{relative} level population, so if only highly excited transitions are available, they will appear prominently in our analysis, despite the fact they originate from energy levels beyond what we expect to be significantly populated in a 3000\,K plasma (ignoring any effects that can alter the level populations significantly from LTE). Candidate lines originating from highly-excited levels ($E_{\textsc{u}} > 4$\,eV) are discarded.} Note that where we refer to the number of lines obtained from our analysis (to indicate the numbers of contaminant lines), we are referring to all lines that lie within the wavelength range, $0.5 \leq \lambda \leq 2.5$\,\micron, that have strengths $> 10^{-3}$ the strength of the strongest line for that species in our analysis (within the same wavelength range). Our most viable candidate species are presented below.

\begin{itemize}

    \item \textbf{Sr\,$\boldsymbol{\textsc{ii}}$:}
    In our analysis, this ion yields two lines coincident with the 1.08\,\micron\ feature, providing a consistency check for our method (see Appendix~\ref{sec:Appendix - Search for potential candidate species (strong transitions) - Verification of analysis}).

    \item \textbf{La\,$\boldsymbol{\textsc{iii}}$:}
    This species returned only three candidate lines in our wavelength range of interest. Two of these lie in the wavelength range of the 1.40\,\micron\ feature (13898 and 14100\,\AA). The third lies in a telluric region (17883\,\AA), and so we cannot determine whether it is present in the observational data.

    \item \textbf{Ce\,$\boldsymbol{\textsc{iii}}$:}
    This species returned 30 lines in our analysis, the strongest of which are coincident with many of the observed features. For instance, there is good agreement between our shortlist and the observed features, with the 10723\,\AA\ ($\sim 1.08$\,\micron), 12760 and 12825\,\AA\ ($\sim 1.23$\,\micron), 14659\,\AA\ ($\sim 1.40$\,\micron), 15720, 15852, 15961 and 16133\,\AA\ ($\sim 1.58$\,\micron), and 20691\,\AA\ (coincident with the blended $\sim 2.07$\,\micron\ feature) lines all matching the observations.

    \item \textbf{Gd\,$\boldsymbol{\textsc{iii}}$:}
    Our analysis returned nine lines in our wavelength range of interest. The strongest of these is coincident with the $\sim 1.40$\,\micron\ feature (14337\,\AA). There are other non-negligible lines that lie at wavelengths $\gtrsim 17000$\,\AA. Some of these are coincident with the blended $\sim 2.07$\,\micron\ feature (20002 and 21265\,\AA), while there is no observational evidence for the others. Gd\,\III\ is a viable candidate for the 1.40\,\micron\ feature, but to be able to reproduce the blended 2.07\,\micron\ feature, the relative strengths of the contaminant lines need to be weaker than what we have estimated here.

    \item \textbf{Ra\,$\boldsymbol{\textsc{ii}}$:}
    There are only three lines extracted by our analysis; two of these are coincident with the 0.79 and 1.08\,\micron\ features (8021.9 and 10791\,\AA).

    \item \textbf{Ac\,$\boldsymbol{\textsc{i}}$:}
    This species returned 27 lines in our analysis. The strongest line is coincident with the 1.40\,\micron\ feature (13374\,\AA), although it lies at the extreme blue end of our permitted wavelength range. Ac\,\I\ also has a strong line coincident with the 0.79\,\micron\ emission feature (8145.6\,\AA).

\end{itemize}

Apart from the above six species, the full list of candidate species that may contribute to the data, albeit with some caveats, are summarised and presented in Appendix~\ref{sec:Appendix - Search for potential candidate species (strong transitions) - Shortlisted species} -- these include In\,\I, Ba\,\I, Eu\,\II, Gd\,\III, Tb\,\II, Ho\,\I, Er\,\I$-$\II, Tm\,\I$-$\II, Yb\,\I, Lu\,\I, Hf\,\I$-$\II, Fr\,\I\ and Th\,\III\ (see also Table~\ref{tab:Candidate ions}).

\subsection{Discussion} \label{sec:Search for potential candidate species (strong transitions) - Comparison to previous works and discussion}

We present a summary of our shortlisted viable and potential contributing species for each of the observed features in Table~\ref{tab:Candidate ions} (for individual discussion on each shortlisted species, see Section~\ref{sec:Search for potential candidate species (strong transitions) - Results} and Appendix~\ref{sec:Appendix - Search for potential candidate species (strong transitions) - Shortlisted species}). We also present some verification of our analysis by comparing with previous works in Appendix~\ref{sec:Appendix - Search for potential candidate species (strong transitions) - Verification of analysis}.

We note that many of our shortlisted species that have satisfied our various cuts belong to the lanthanides. This is not too unexpected, since the valence $f$-shell electron structure leads to many low-lying levels, between which many lines exist, some of which we expect to coincidentally line up with our observed features. From \paperI, lanthanides are expected to make up around 5 per cent (by mass)  of the total ejecta material (\Xlanth~$ = 0.05^{+0.05}_{-0.02}$).

If the assumption that the observed spectral features are produced by permitted transitions is valid, then we can speculate that our mystery species X (see the analysis in Section~\ref{sec:Modelling the early NIR features}) is most likely to be Ce\,\III, but could possibly be Eu\,\II, Tb\,\II, Yb\,\I\ or Th\,\III. Mystery species Z is also most likely to be Ce\,\III, but could also be any of Ba\,\I, Gd\,\III, Yb\,\I, Hf\,\I\ and Th\,\III. For both cases, the most likely candidate is Ce\,\III, a lanthanide. It is also worth noting again that these features may not be made up of emission from a unique species -- at the high velocities characteristic for KN ejecta, line blending is also plausible. If this is indeed the case, we would not be able to obtain a unique identification for either mystery species X or Z.

Identification of features produced by multiple ions of the same element are intriguing, as it would lend additional credence to the presence of that element in the ejecta. We find multiple candidate species for the same element: Er\,\I$-$\II, Tm\,\I$-$\II\ and Hf\,\I$-$\,\II, although we note that none of these are a convincing viable candidate species for any of the features under investigation. Further analysis of these elements in particular would be worthwhile to see if they do contribute significantly to the spectra -- we note that having positive identifications of multiple ions for the same element in the ejecta would be extremely useful for constraining the ionisation. Finally, the specific cases of In\,\I, Ce\,\III, Eu\,\II, Gd\,\III, Ho\,\I, Er\,\I, Yb\,\I, Hf\,\I$-$\II, Ra\,\II, Ac\,\I\ and Th\,\III\ are intriguing, since we find each of these species are capable of contributing to multiple observed features. In particular, Ce\,\III\ shows remarkable agreement with the observations, as it is a viable candidate for five of the emission features observed at late times ($\sim 1.08$, 1.23, 1.40, 1.58 and 2.07\,\micron). Not only is Ce\,\III\ the most promising candidate, but it is also supported by nucleosynthesis arguments, since it is among the most abundant lanthanides produced in our realistic composition profiles (see \paperI). We note however, that since the lanthanide elements are expected to be co-produced, it may be difficult to disentangle evidence for a single lanthanide species from the contribution expected from all the other lanthanides (although particularly strong lines belonging to the more abundant lanthanide species may be identifiable).

Our rudimentary line search here successfully recovers the candidate La\,\III\ and Ce\,\III\ lines presented by \cite{Domoto2022} \linebreak (see Appendix~\ref{sec:Appendix - Search for potential candidate species (strong transitions) - Verification of analysis}). Specifically, they propose that La\,\III\ is responsible for absorption at $\sim 12000$\,\AA, and that Ce\,\III\ is responsible for absorption at $\sim 14000$\,\AA\ (these absorption features are linked to the emission features we analyse here, at $\sim 1.40$ and 1.58\,\micron, respectively). While the synthetic KN spectral models presented by \cite{Domoto2022} show that Ce\,\III\ is capable of producing a prominent absorption feature at $\sim 1.40$\,\micron, their models do not produce any other prominent absorption due to Ce\,\III. The absence of absorption for the other Ce\,\III\ transitions we shortlist here as candidates for the other observed features does not rule them out from contributing to the observed emission. Since we are exploring the emission components, we are affected by the upper level populations, whereas the study of absorption features will be dependent on lower level populations. Additionally, the upper level populations can be significantly affected by both non-LTE and non-thermal effects, and so these could act to boost the level populations needed for our shortlisted transitions to produce prominent emission.

We do not recover \YII\ as a candidate for the 0.79\,\micron\ feature, as proposed by \cite{Sneppen2023_YII}. Their models assume a single blackbody fit to the data, and P-Cygni profiles for the strongest \YII\ lines. They show that strong \YII\ transitions can produce a prominent P-Cygni profile that matches the data at intermediate phases. The presence of strong absorption from \YII\ has been noted in other works (e.g. \paperI, \citealt{Shingles2023}). Those studies predict broad absorption due to \YII\ across blue wavelengths, but do not produce a prominent and detectable feature due to the \mbox{\YII\ 4d5p -- 4d$^{2}$} family of transitions (with $\lambda_{\rm air} = 7264.2 - 7881.9$\,\AA), as proposed by \cite{Sneppen2023_YII}. However, the strength of this feature may be dependent on details of the velocity evolution in the models, and further work is required. For our work in this section, we are considering the source of the emission at later times ($> 7$~days), and so we are focussed on species that can produce prominent emission. In our analysis for \YII, we recover many contaminant lines that we also expect to be producing prominent emission, for which we see no evidence. Hence, we rule it out as a viable candidate for dominating the 0.79\,\micron\ emission feature (at +7.4~days and beyond). \cite{Pognan2023} found that the \YII\ candidate lines from \cite{Sneppen2023_YII} are expected to be weak in their KN ejecta simulations, and instead propose Rb\,\I\ as an alternative candidate for the same 0.79\,\micron\ P-Cygni feature.

We stress that without estimates for line strengths (i.e. \Aval) for the lines, it is impossible to definitively determine whether any of our shortlisted lines are the cause of the features we observe in the \gfo\ spectra. However, this type of analysis does highlight which atoms or ions could be likely contributors, considering our atomic data restrictions. Despite the uncertainties associated with our analysis, we have obtained a list of candidate transitions for the observed features (see Table~\ref{tab:NIST - Candidate transitions (E1)}). We present a summary of our shortlisted species for each prominent feature in the observed spectra in Table~\ref{tab:Candidate ions}. This table contains a summary of all shortlisted species discussed above, and we argue that this shortlist encapsulates the species that most pertinently need better atomic data. This list can act as a `priority list' for future atomic data studies.

\begin{table*}
    \renewcommand*{\arraystretch}{1.2}
    \centering
    \caption{
    Summary of the candidate species we identify in our analysis for each of the prominent emission features in the late-time spectra of \gfo. The ions we refer to as viable candidate species are those which we propose are capable of reproducing the observations, with no perceived limitations. The ions we propose as potential candidate species are those that may contribute to the observations, but with some caveat (discussed in the main text).
    }
    \begin{threeparttable}
        \centering
        \begin{tabular}{cccccc}
            \hline
            \hline
            \multirow{2}{*}{\begin{tabular}[c]{@{}c@{}}Approx.\\$\lambda_{\rm peak}$ (\micron)\end{tabular}}    &\multicolumn{2}{c}{Candidate permitted (i.e. E1) species}    &    &\multicolumn{2}{c}{Candidate forbidden (i.e. M1 \& E2) species}        \\
            \cline{2-3}
            \cline{5-6}
                    &Viable         &Potential          &       &Viable         &Potential      \\
            \hline

            \addlinespace[1ex]

            \multirow{2}{*}{0.79}    &\multirow{2}{*}{\begin{tabular}[c]{@{}c@{}} Ra\,\II, Ac\,\I \end{tabular}}    &\multirow{2}{*}{\begin{tabular}[c]{@{}c@{}} Ho\,\I, Tm\,\I, \\ Hf\,\I, Fr\,\I \end{tabular}}    &    &\multirow{2}{*}{\begin{tabular}[c]{@{}c@{}} [I\,\III] \end{tabular}}    &\multirow{2}{*}{\begin{tabular}[c]{@{}c@{}} [Os\,\III], [Au\,\II] \end{tabular}}    \\
                    &       &       &       &       \\

            \addlinespace[1ex]
            \hline
            \addlinespace[1ex]

            \multirow{2}{*}{1.08}    &\multirow{2}{*}{\begin{tabular}[c]{@{}c@{}} \SrII, Ce\,\III, \\ Ra\,\II \end{tabular}}    &\multirow{2}{*}{\begin{tabular}[c]{@{}c@{}} Eu\,\II, Ho\,\I, \\ Tm\,\II, Hf\,\II \end{tabular}}    &    &\multirow{2}{*}{\begin{tabular}[c]{@{}c@{}} [I\,\III], [Au\,\I], \\\ [Ac\,\II] \end{tabular}}    &\multirow{2}{*}{\begin{tabular}[c]{@{}c@{}} [Tc\,\II], [Rh\,\I], [Gd\,\III], \\\ [Yb\,\II], [Ta\,\I], [Pt\,\I] \end{tabular}}    \\
                    &       &       &       &       \\

            \addlinespace[1ex]
            \hline
            \addlinespace[1ex]
            
            \multirow{2}{*}{1.23}    &\multirow{2}{*}{\begin{tabular}[c]{@{}c@{}} Ce\,\III \end{tabular}}    &\multirow{2}{*}{\begin{tabular}[c]{@{}c@{}} In\,\I, Ho\,\I, \\ Er\,\I, Hf\,\II \end{tabular}}    &    &\multirow{2}{*}{\begin{tabular}[c]{@{}c@{}} [Rh\,\II], [Te\,\III], \\\ [Ir\,\II] \end{tabular}}    &\multirow{2}{*}{\begin{tabular}[c]{@{}c@{}} [Te\,\II], [Xe\,\III], \\\ [Os\,\III] \end{tabular}}    \\
                    &       &       &       &       \\

            \addlinespace[1ex]
            \hline
            \addlinespace[1ex]
            
            \multirow{2}{*}{1.40}    &\multirow{2}{*}{\begin{tabular}[c]{@{}c@{}} La\,\III, Ce\,\III, \\ Gd\,\III, Ac\,\I \end{tabular}}    &\multirow{2}{*}{\begin{tabular}[c]{@{}c@{}} In\,\I, Eu\,\II, Ho\,\I, Er\,\I, \\ Er\,\II, Yb\,\I, Lu\,\I, Th\,\III \end{tabular}}    &    &\multirow{2}{*}{\begin{tabular}[c]{@{}c@{}} [Y\,\II], [I\,\II], \\\ [Dy\,\I], [Ir\,\I] \end{tabular}}    &\multirow{2}{*}{\begin{tabular}[c]{@{}c@{}} [Mo\,\III], [Pd\,\III], \\\ [Er\,\II], [Ta\,\I] \end{tabular}}     \\
                    &       &       &       &       \\
    
            \addlinespace[1ex]
            \hline
            \addlinespace[1ex]
            
            \multirow{2}{*}{1.58}    &\multirow{2}{*}{Ce\,\III}    &\multirow{2}{*}{\begin{tabular}[c]{@{}c@{}} Eu\,\II, Tb\,\II, \\ Yb\,\I, Th\,\III \end{tabular}}    &    &\multirow{2}{*}{[I\,\II], [Ir\,\I]}    &\multirow{2}{*}{$-$}    \\
                    &       &       &       &       \\

            \addlinespace[1ex]
            \hline
            \addlinespace[1ex]
            
            \multirow{2}{*}{2.059}     &\multirow{2}{*}{Ce\,\III}    &\multirow{2}{*}{\begin{tabular}[c]{@{}c@{}} Gd\,\III, Yb\,\I, \\ Hf\,\I, Th\,\III \end{tabular}}    &    &\multirow{2}{*}{\begin{tabular}[c]{@{}c@{}} [Ba\,\II], [Tb\,\III], \\\ [Ir\,\II], [Ac\,\I] \end{tabular}}    &\multirow{2}{*}{[Ce\,\III]}    \\
                    &       &       &       &       \\

            \addlinespace[1ex]
            \hline
            \addlinespace[1ex]

            \multirow{2}{*}{2.135}    &\multirow{2}{*}{Ce\,\III}    &\multirow{2}{*}{\begin{tabular}[c]{@{}c@{}} Ba\,\I, Gd\,\III, Yb\,\I, \\ Hf\,\I, Th\,\III \end{tabular}}    &    &\multirow{2}{*}{\begin{tabular}[c]{@{}c@{}} [Pd\,\III], [Ag\,\III], [Te\,\I], \\\ [Te\,\III], [Tb\,\III], [Ac\,\I] \end{tabular}}    &\multirow{2}{*}{\begin{tabular}[c]{@{}c@{}} [Ce\,\III], [Gd\,\II], \\\ [Dy\,\II], [Er\,\II] \end{tabular}}    \\
                    &       &       &       &       \\
                    
            \addlinespace[1ex]
            
            \hline
        \end{tabular}
    \end{threeparttable}
\label{tab:Candidate ions}
\end{table*}

\section{Search for potential candidate species (weak transitions)} \label{sec:Search for potential candidate species (weak transitions)}

In Section~\ref{sec:Modelling the late-time emission features} we showed that the observed features in the late-time spectra of \gfo\ can be reproduced by pure emission features. We proposed that these may be produced by intrinsically weak transitions originating from some optically thin region of ejecta. Therefore, here we perform a similar search as in Section~\ref{sec:Search for potential candidate species (strong transitions)}, but this time searching for intrinsically weak transitions that we expect to be prominent. \cite{Hotokezaka2021} and \cite{Pognan2022, Pognan2022_nlte} suggest that forbidden lines are important for understanding the late phases of KNe, and so here we consider that they may have an effect on the late-phase spectra of \gfo.

\subsection{Method} \label{sec:Appendix - Search for potential candidate species (weak transitions) - NIST - Method}

Here we follow broadly the same approach as outlined in Section~\ref{sec:Search for potential candidate species (strong transitions)}, but now consider only forbidden magnetic dipole and electric quadrupole transitions, typically denoted as M1 and E2, respectively. The main steps of the analysis are summarised here, with emphasis on the differences with the approach presented in Section~\ref{sec:Search for potential candidate species (strong transitions)}.

As in Section~\ref{sec:Search for potential candidate species (strong transitions)}, we ingest the level information from the NIST ASD for all species with $Z \geq 38$. We again focus our efforts on the lowest three ion stages of these elements (for the same reason discussed in Section~\ref{sec:Search for potential candidate species (strong transitions)}). We calculate energy differences between all the levels, which gives us a list of all possible transitions. As highlighted in Section~\ref{sec:Search for potential candidate species (strong transitions)}, the levels in the NIST ASD \citep[][]{NIST2020} have been critically evaluated, and so these computed transition wavelengths will be accurate and reliable for line identification studies. For both M1 and E2 transitions, parity must be conserved. Additionally, only specific variations in the quantum J number are allowed. For M1 transitions, $| \Delta \rm{J} | = 0$ or 1, whereas for \mbox{E2, $| \Delta \rm{J} | = 0$, 1 or 2}. As before, where the levels have configurations and terms that can be easily expressed by the LS-coupling formalism, we consider the LS-coupling rules. Consulting all of the above, we discard any lines that do not satisfy these rules. Finally, we truncate all lines that have upper levels above the ionisation threshold of the species under investigation.

In the earlier analysis in Section~\ref{sec:Search for potential candidate species (strong transitions)}, we considered a wavelength range corresponding to a $0.05 \, c$ Doppler shift, to set our tolerated wavelength range for each of the emission features explored. We deliberately set a broad tolerable threshold for the optically thick case, as we wanted to accommodate the effect that any potential \pcyg\ absorption might have on the inferred emission feature peak location. However, this effect is not of concern in this optically thin regime, and so we set a smaller tolerable range (Doppler velocity of $0.02 \, c$).

We again compute the upper level population, and use these values as a proxy for intrinsic line strength (as in Section~\ref{sec:Search for potential candidate species (strong transitions)}), still assuming $T = 3000$\,K, for convenience. These are normalised to the upper level with the strongest potential M1 or E2 transition within some observable wavelength range \mbox{($0.5 \leq \lambda \leq 2.5$\,\micron)}.

In addition to the observed spectra of \gfo\ presented throughout this work, there are other observations that may help to constrain potential candidate species. \cite{Villar2018} and \cite{Kasliwal2022} present \textit{Spitzer Space Telescope} (\spitzer) observations of \gfo\ at +43 and +74 days. There is a significant flux detection at 4.5\,\micron\ (in the IRAC band, which covers the $3.96 - 5.02$\,\micron\ wavelength range), whereas there is no such signal present at 3.6\,\micron\ (which samples the $3.18 - 3.96$\,\micron\ wavelength range). We can use the detections from \spitzer\ to additionally favour or disfavour shortlisted candidate ions, although we note that these observations are much later than the spectra we analyse, and so there may be significant evolution in the ejecta.

We perform a complementary analysis to that presented in this section using the Kurucz atomic data \citep[][see Appendix~\ref{sec:Search for potential candidate species (weak transitions) - Kurucz analysis}]{Kurucz2017}. This analysis includes E1, M1 and E2 transitions, and possesses estimates for intrinsic line strengths. However, the Kurucz data base is incomplete for heavy elements. However, utilising this data set can still provide useful results, and as we highlight below, we recover [\YII] as a viable candidate ion from our analysis with the Kurucz data (see Appendix~\ref{sec:Search for potential candidate species (weak transitions) - Kurucz analysis} for details).

\subsection{Results} \label{sec:Search for potential candidate species (weak transitions) - NIST analysis}

Our list of forbidden candidate lines that we extract from our analysis are presented in Table~\ref{tab:Forbidden candidate transitions}.\footnote{For completeness, we present the wavelengths of the candidate transitions in Table~\ref{tab:Forbidden candidate transitions} both in vacuum and air.} Our list of viable candidate species for each of the observed features include [Y\,\II], [Rh\,\II], [Pd\,\III], [Ag\,\III], [Te\,\I], [Te\,\III], [I\,\II], [I\,\III], [Ba\,\II], [Tb\,\III], [Dy\,\I], [Ir\,\I], [Ir\,\II], [Au\,\I], [Ac\,\I] and [Ac\,\II] (see Table~\ref{tab:Candidate ions}). Our analysis yields multiple candidate transitions for each of the features identified in the late-time spectra of \gfo. As in Section~\ref{sec:Search for potential candidate species (strong transitions)}, we discard any candidate species that have highly excited upper levels. We also disfavour species with many lines, since these produce an intolerable amount of flux, for which we have no observational evidence. We summarise our most viable candidate species below.

\begin{itemize}

    \item \textbf{[Y\,$\boldsymbol{\textsc{ii}}$]:}
    We recover 24 lines in our analysis. The strongest lines (12990, 13355, 13711, 13731, 13961, 14372, 14539 and 15259\,\AA) produce a candidate blend, which we expect to produce an emission feature that peaks slightly blueward of the 1.40\,\micron\ feature (note that many of these lines are the same lines identified in our line identification search using the Kurucz data; see Appendix~\ref{sec:Search for potential candidate species (weak transitions) - Kurucz analysis}). There are other emission features produced, but at $T = 3000$\,K, these are subdominant. This approach recovers the same strong lines we found in Appendix~\ref{sec:Search for potential candidate species (weak transitions) - Kurucz analysis}, although we do not obtain the same relative strengths for the features. Note that only one of these strong lines satisfies our tolerable wavelength range for the 1.40\,\micron\ feature (13961\,\AA). The recovery of the lines shortlisted from our more detailed analysis in Appendix~\ref{sec:Search for potential candidate species (weak transitions) - Kurucz analysis} acts to verify the validity of this rudimentary analysis (see Appendix~\ref{sec:Appendix - Search for potential candidate species (weak transitions) - NIST - Verification of analysis}.)

    \item \textbf{[Rh\,$\boldsymbol{\textsc{ii}}$]:}
    Our analysis recovers six lines for this species. Of these, two are coincident with the 1.23\,\micron\ feature (12248 and 12325\,\AA).

    \item \textbf{[Pd\,$\boldsymbol{\textsc{iii}}$]:}
    Our analysis recovers seven lines, the strongest of which (21338\,\AA) is coincident with the 2.135\,\micron\ feature. The next three strongest lines lie at 9775.2, 14284 and 18039\,\AA. The 14284\,\AA\ line is a potential candidate for the 1.40\,\micron\ feature, if it is stronger than these two contaminant lines.

    \item \textbf{[Ag\,$\boldsymbol{\textsc{iii}}$]:}
    Only one line is predicted from our analysis, which is coincident with the 2.135\,\micron\ feature (21696\,\AA).
    
    \item \textbf{[Te\,$\boldsymbol{\textsc{i}}$]:}
    Our analysis recovers two lines, both of which coincide with the 2.135\,\micron\ feature (21049 and 21247\,\AA).
   
    \item \textbf{[Te\,$\boldsymbol{\textsc{iii}}$]:}
    Our analysis returns three lines, the strongest of which is coincident with the 2.135\,\micron\ feature (21050\,\AA). Another strong line is predicted at 12248\,\AA, which is coincident with the small emission feature at 1.23\,\micron.

    \item \textbf{[I\,$\boldsymbol{\textsc{ii}}$]:}
    There are two lines recovered by our analysis. These are coincident with the 1.40 and 1.58\,\micron\ features (14111 and 15509\,\AA), although the redder of these is at the extreme blue edge of our wavelength cut.
    
    \item \textbf{[I\,$\boldsymbol{\textsc{iii}}$]:}
    We recover five lines from our analysis. The two most prominent lie at 7943.9 and 10641\,\AA, which are coincident with the 0.79\,\micron\ and 1.08\,\micron\ features (see Section~\ref{sec:Spectroscopic evolution of AT2017gfo}).
    
    \item \textbf{[Ba\,$\boldsymbol{\textsc{ii}}$]:}
    Only two lines recovered by our analysis -- the strongest two (17622 and 20518\,\AA) are coincident with a telluric region and the 2.059\,\micron\ feature.

    \item \textbf{[Tb\,$\boldsymbol{\textsc{iii}}$]:}
    Our analysis recovers 56 lines, but the three strongest coincide with the 2.059 and 2.135\,\micron\ features (20760 and 21121\,\AA), with the third lying just blueward of the 1.58\,\micron\ feature (15456\,\AA). Tb\,\III\ has a transition between two levels in the ground configuration that coincides exactly with the 3.6\,\micron\ \spitzer\ band non-detection (35655\,\AA). We predict this to be our strongest line, and so some explanation for not seeing this line at later times (e.g. recombination/ionisation effects, intrinsically weak transition, etc.) is needed.

    \item \textbf{[Dy\,$\boldsymbol{\textsc{i}}$]:}
    Our analysis returns 140 lines, but we predict most of these are weak, ineffectual transitions. The strongest line within our wavelength range is coincident with the 1.40\,\micron\ feature (14183\,\AA).

    \item \textbf{[Ir\,$\boldsymbol{\textsc{i}}$]:}
    There are 52 lines returned by our analysis, three of which we predict to be prominent. Two of these strongest lines are coincident with the 1.40\,\micron\ (14071\,\AA) and 1.58\,\micron\ (15813\,\AA) features, while we have no observational evidence for the third (17287\,\AA). If this contaminant line is weaker than we estimate, then we have very good agreement with the other two strong lines.  

    \item \textbf{[Ir\,$\boldsymbol{\textsc{ii}}$]:}
    In our analysis, this ion yields 30 lines, with the strongest within our observable wavelength range coinciding with the 2.059\,\micron\ feature (20886\,\AA). There is another weaker (but still detectable) line at 12215\,\AA, which is coincident with the 1.23\,\micron\ feature.
    
    \item \textbf{[Au\,$\boldsymbol{\textsc{i}}$]:}
    This species returned two lines in our analysis, only one of which is expected to be prominent. This line is coincident with the 1.08\,\micron\ feature (10916\,\AA), and is the same line as previously identified by \cite{Gillanders2021}.
    
    \item \textbf{[Ac\,$\boldsymbol{\textsc{i}}$]:}
    Our analysis identifies 50 lines. The strongest line in our analysis lies at 44814\,\AA, which is exactly coincident with the 4.5\,\micron\ \spitzer\ detection. The strongest lines within our observable wavelength range lie at 20837 and 21458\,\AA, coincident with the 2.059 and 2.135\,\micron\ features, respectively. 

    \item \textbf{[Ac\,$\boldsymbol{\textsc{ii}}$]:}
    This species has 23 lines in our analysis. The strongest lie at 11004, 37218 and 46310\,\AA, which correspond to the 1.08\,\micron\ feature, and the 3.6 and 4.5\,\micron\ \spitzer\ bands, respectively. Although we do not see strong evidence for flux in the 3.6\,\micron\ \spitzer\ band, we propose [Ac\,\II] is a viable candidate for the epochs of the \xsh\ spectra.

\end{itemize}

Our candidate species that may contribute emission to the observed spectra, albeit with some caveats, are summarised and presented in Appendix~\ref{sec:Appendix - Search for potential candidate species (weak transitions) - NIST - Shortlisted species} -- these include [Mo\,\III], [Tc\,\II], [Rh\,\I], [Pd\,\III], [Te\,\II], [Xe\,\III], [Ce\,\III], [Gd\,\II], [Gd\,\III], [Dy\,\II], [Er\,\II], [Yb\,\II], [Ta\,\I], [Os\,\III], [Pt\,\I] and [Au\,\II] (see Table~\ref{tab:Candidate ions}).

\subsection{Discussion} \label{sec:Search for potential candidate species (weak transitions) - Comparison to previous works and discussion}

We present a summary of our shortlisted viable and potential contributing species for each of the features in Table~\ref{tab:Candidate ions}. For individual discussion on each of the shortlisted species, the reader is referred to Section~\ref{sec:Search for potential candidate species (weak transitions) - NIST analysis} and Appendix~\ref{sec:Appendix - Search for potential candidate species (weak transitions) - NIST - Shortlisted species}.

For our modelling at early times (see \paperI), we favoured a composition with little material heavier than the lanthanides. However, as we probe later times, and become sensitive to different regions of ejecta, it is possible that the composition deviates from that inferred at earlier times, and so we cannot rule out the possibility of these features being produced by trans-lanthanide species, such as [Ta\,\I], [Ir\,\I], [Ir\,\II], [Os\,\III], [Pt\,\I], [Au\,\I], [Au\,\II], [Ac\,\I] and [Ac\,\II]. We note that many of our candidate species are lanthanides, as was also the case in Section~\ref{sec:Search for potential candidate species (strong transitions)}. This is again a result of the lanthanide species having valence $f$-shell electrons, and so there exist many low-lying levels, between which there are many lines.

We expect some of these candidate species to have coincidental agreement with at least one of the observed features in the spectra. As before, the best way to constrain the presence of a particular species is to search for evidence of it producing (or at least contributing to) a number of features. From our list, [Pd\,\III], [Te\,\III], [I\,\II], [I\,\III], [Ce\,\III] (although this is due to a single line that lies within the tolerable wavelength range of both the 2.059 and 2.135\,\micron\ features), [Tb\,\III], [Er\,\II], [Ta\,\I], [Os\,\III], [Ir\,\I], [Ir\,\II] and [Ac\,\I] all potentially contribute to more than one emission feature.

Elements for which we shortlist multiple ion stages include [Rh\,\I] \& [Rh\,\II], [Te\,\I], [Te\,\II] \& [Te\,\III], [I\,\II] \& [I\,\III], [Gd\,\II] \& [Gd\,\III], [Dy\,\I] \& [Dy\,\II], [Ir\,\I] \& [Ir\,\II], [Au\,\I] \& [Au\,\II] and [Ac\,\I] \& [Ac\,\II]. Positive identification of any of these multiple ion stages would be extremely useful for constraining the ionisation of the ejecta material.

Both I and Te possess multiple candidate transitions that match the observed emission features, across multiple ion stages. In addition, they are both predicted to be abundant by nucleosynthesis calculations (see \paperI). I (which has atomic mass, $A = 127$ and 129) corresponds to the top of the second \rpro\ peak, and is expected to be quite abundant,\footnote{In \paperI, we found that the photospheric spectra, from $+2.4 - 7.4$\,d, were best reproduced by the \AngIII\ composition (see \paperI, Table~2). This best-fitting composition contained a 4.4 per cent mass fraction of I, making it the seventh most abundant element (by mass) in our best-fitting composition profile.} and so I having a detectable effect on the observed spectra seems plausible. We find agreement with four of the emission features in the observed spectra (0.79, 1.08, 1.40 and 1.58\,\micron) with two ion stages of I \mbox{([I\,\II] \& [I\,\III])} -- see Table~\ref{tab:Candidate ions}.

For the specific case of Te, we shortlist both [Te\,\I] and [Te\,\III] as viable candidates for the 2.135\,\micron\ feature, as well as finding that [Te\,\II] is capable of producing a prominent feature coincident with the 4.5\,\micron\ \spitzer\ detection. Also, [Te\,\II] and [Te\,\III] have lines that coincide with the emission feature at 1.23\,\micron. Te (which has atomic mass, $A = 128$ and 130) corresponds to the top of the second \rpro\ peak, due to the $N = 82$ magic number. This implies that we can expect Te to be abundant in many different \rpro\ scenarios. In fact, Te is among the most abundant elements in many of our composition profiles presented in \paperI\ (and has a mass fraction of $\approx 10$ per cent in our best-fitting \AngIII\ composition profile).

We conclude that there are a reasonable number of potential candidate ions for the features in the late phase spectra of \gfo\ (see Table~\ref{tab:Forbidden candidate transitions}), and we propose that our rudimentary study here has produced a reasonable shortlist of ions that warrant further analysis. As in Section~\ref{sec:Search for potential candidate species (strong transitions)}, we again present a summary of the candidate species we identify from our analysis in Table~\ref{tab:Candidate ions}. This table contains a summary of all shortlisted species discussed above, and we again argue that this shortlist contains the species that most pertinently need better atomic data.

Although our approach here is quite rudimentary, we are performing as detailed an analysis as is currently possible with the existing atomic data. Despite the obvious limitations encountered by not having access to complete line lists, we still manage to identify candidate transitions for each of the observed emission features observed in the spectra of \gfo. While our shortlist is likely incomplete, we propose that it is still useful, as it demonstrates the species that we think are most likely to contribute to the observed data. These species (see Table~\ref{tab:Candidate ions}) are ideal candidates for future atomic data studies, which would allow us to build on our conclusions here. Improved atomic data would enable us to quantitatively determine whether our proposed candidate ions contribute to the observations of \gfo. We note that it is important that future atomic data studies not only expand to heavier elements and to longer wavelengths, but also to encompass line information for forbidden transitions, since these are key to interpreting the spectra of kilonovae at late times.

\section{Conclusions} \label{sec:Conclusions}

The spectra of the kilonova \gfo\ exhibit remarkable evolution. Due to the high expansion velocities and low ejecta mass, the transient evolves extremely rapidly, as is evidenced by the extreme spectral evolution. The spectra quickly reach a quasi-nebular regime, with prominent emission features rising above the continuum. To date, no study has modelled the late-phase spectra in detail, and so here we present an empirical approach to aid interpretation of these spectra. First we show that the spectral feature peaks evolve redward with time. We find that there are seven perceived features in the ejecta of \gfo\ (at $\sim 0.79$, 1.08, 1.23, 1.40, 1.58, 2.059 and 2.135\,\micron), most of which appear prominently across multiple epochs. We have mapped the evolution of these features and performed a line identification study.

We modelled the  1.58 and 2.07\,\micron\ NIR features at intermediate epochs ($+3.4 - 7.4$ days) with \tardis\ as single strong permitted transitions produced by some mystery species X and Z. We find that such an interpretation can only achieve agreement with the data at a few epochs (+5.4 and +6.4\,d for X, and +5.4\,d for Z), and that generally these models produce features that are broader than the observations. This indicates that mystery species X and Z exist at lower velocities than the \SrII\ material producing the \mbox{$\sim 1.08$\,\micron} \pcyg\ feature at these intermediate phases. This could be a result of ionisation (where X and Z are not the dominant ions at higher velocities), or ejecta stratification arising from different ejection mechanisms.

Next we analysed the $\sim 1.08$\,\micron\ feature evolution. The \pcyg\ feature at $\sim 0.7 - 1.2$\,\micron\ is thought to be produced by the \SrII\ NIR triplet in the first few days. The \SrII\ NIR triplet feature plausibly evolves into a pure emission feature at later times (+7.4\,d), but the emission component for \SrII\ would require a $\sim 9800$\,\kms\ bulk redshift to match observations. It is possible that this feature is not originating from a fully nebular regime at +7.4\,d, or it may be contaminated by another emerging species at redder wavelengths. Our calculations of the emission expected from \SrII\ show that if this emission profile is due to the NIR triplet, the \forbSrII\ doublet should be detectable. There is no detection of this feature in the observed data, implying the ejecta is highly clumped ($f = 0.01$). It may also cast doubt on the \SrII\ NIR triplet emission persisting to the later epochs. We explore the possibility of He\,\I\ emission  contributing to the data. While the He\,\I\ $\lambda_{\rm air} = 10830$ and 20581\,\AA\ lines are remarkably good wavelength matches for the observed 1.08 and 2.059\,\micron\ features, we conclude that He\,\I\ emission cannot reproduce the observed relative strengths of these features. While He\,\I\ could contribute to the 1.08\,\micron\ feature, it cannot be responsible for the 2.059\,\micron\ feature.

We next modelled all emission features present in the late-time (+7.4~days onward) spectra as Gaussian-shaped emission lines arising from a nebular regime, with characteristic FWHM ejecta velocity, \mbox{$v_{\textsc{fwhm}} = 35600 \pm 6600$\,\kms}. We favour this scenario for line formation and perform a search for candidate permitted transitions of ions within the NIST ASD. We find that almost all of our candidate species are lanthanides, and that the 1.58 and 2.07\,\micron\ features (mystery species X and Z) are both likely to be Ce\,\III. Although \SrII\ is still the favoured explanation for the  1.08\,\micron\ emission profile, our search returned alternative identifications including Ce\,\III\ or Ra\,\II. While we propose the observed feature at 0.79\,\micron\ likely corresponds to the point at which the \SrII\ NIR triplet \pcyg\ rejoins the continuum, our search returned Ra\,\II\ and Ac\,\I\ as alternative explanations. We also shortlist Ce\,\III\ as the species most likely responsible for the 1.23\,\micron\ feature. The shortlisted species responsible for the $\sim 1.40$\,\micron\ feature include La\,\III, Ce\,\III, Gd\,\III\ and Ac\,\I. We note it is likely that line-blending is commonplace in KN ejecta -- even at later times -- and so it is possible that none of these emission features have unique and identifiable contributions from a single species. 

An alternative interpretation of the emission features in the spectra $\gtrsim 7$ days is that they are produced by intrinsically weak lines in optically thin emitting material, and so we perform a similar search for possible forbidden transitions. Candidate species include neutral, singly and doubly ionised ions from the light \rpro\ elements \mbox{(Y -- Ba)}, lanthanides (Tb \& Dy), and very heavy trans-lanthanide elements (Ir -- Ac). We propose that Te and I are the most promising elements that warrant further investigation, as they possess multiple candidate transitions that match the observed emission features, across multiple ion stages, in addition to lying close to one of the \rpro\ abundance peaks. However, with the existing atomic data available, we cannot make unique and definitive identifications of forbidden lines for the observed emission profiles. 

The major limitation throughout this work is a lack of atomic data. We lack estimates for intrinsic line strengths for many of our shortlisted species for the observed features, which prevents a more in-depth analysis. Not only does the community need access to an accurate and experimentally-calibrated atomic line list, but it needs atomic data that also encompasses the NIR (and MIR),  that also contains intrinsically weak (i.e. forbidden) transitions. Having these would make line modelling and identification studies much more accessible. Finally, although line blending makes unique identifications difficult, future observations extending into the MIR will be invaluable, since the density of transitions decreases with increasing wavelength, making individual line identifications more feasible.

\section*{Acknowledgements}

The authors thank the anonymous referee for useful comments that helped improve the final version of the manuscript.
%
The authors thank Manuel Bautista for providing collisional information for the \SrII\ ion.
%
JHG thanks Nial Tanvir and Connor Ballance for helpful comments on this analysis when first presented in his PhD thesis.
JHG thanks Brendan O'Connor and Eleonora Troja for comments on the formatting of the manuscript.
%
SAS and SJS acknowledge funding from STFC Grants ST/P000312/1, ST/T000198/1, ST/X00094X/1 and ST/X006506/1.
SG acknowledges support by the Fonds de la Recherche Scientifique (F.R.S.-FNRS) and the Fonds Wetenschappelijk Onderzoek -- Vlaanderen (FWO) under the EOS Projects nr O022818F and O000422F.
AB acknowledges support by the European Research Council (ERC) under the European Union’s Horizon 2020 research and innovation programme under grant agreement No. 759253, support by Deutsche Forschungsgemeinschaft (DFG, German Research Foundation) through Project-ID 279384907 -- SFB 1245 and through Project-ID 138713538 -- SFB 881 (``The Milky Way System'', subproject A10) and support by the State of Hesse within the Cluster Project ELEMENTS.
AB and SAS acknowledge support by the ERC under the European Union’s research and innovation program (ERC Grant HEAVYMETAL No. 101071865).
%
CHIANTI is a collaborative project involving George Mason University, the University of Michigan (USA), University of Cambridge (UK) and NASA Goddard Space Flight Center (USA).
%
This research made use of \tardis, a community-developed software package for spectral synthesis in SNe. The development of \tardis\ received support from GitHub, the Google Summer of Code initiative, and from ESA's Summer of Code in Space program. \tardis\ is a fiscally sponsored project of NumFOCUS. \tardis\ makes extensive use of Astropy and PyNE.
%
We are grateful for use of the computing resources from the Northern Ireland High Performance Computing (NI-HPC) service funded by EPSRC (EP/T022175).
%
We made use of the flux-calibrated versions of the \xsh\ spectra publicly available through ENGRAVE, which are based on observations collected at the European Southern Observatory (ESO), available through the ESO Science Archive Facility.

\section*{Data Availability}

The data underlying this article will be shared on reasonable request to the corresponding author.



\bibliographystyle{mnras}
\bibliography{ref} 

\appendix

\section{Search for potential candidate species (strong transitions)} \label{sec:Appendix - Search for potential candidate species (strong transitions)}

\subsection{Shortlisted species} \label{sec:Appendix - Search for potential candidate species (strong transitions) - Shortlisted species}

Our candidate species that may contribute to the data, albeit with some caveats, are summarised and presented below (see Table~\ref{tab:NIST - Candidate transitions (E1)}).

\begin{itemize}

    \item \textbf{In\,$\boldsymbol{\textsc{i}}$:}
    Our analysis recovers 15 lines. Two of the strongest lines are coincident with the 1.23 and 1.40\,\micron\ features (12916 and 13434\,\AA). Although the upper level energies are quite high for these transitions, they do lie below the upper level energy limit imposed for this analysis.

    \item \textbf{Ba\,$\boldsymbol{\textsc{i}}$:}
    We obtain 35 lines for this species in our analysis. The three strongest lines lie at 22318, 23260 and 25522\,\AA. The bluest of these lies at the red end of our tolerable wavelength range for the 2.135\,\micron\ feature. Although this line is likely too red to be the main source of the flux for the 2.135\,\micron\ feature, it is capable of contributing to the data (although we would need the two contaminating lines to be weaker than what we have estimated).

    \item \textbf{Eu\,$\boldsymbol{\textsc{ii}}$:}
    Our analysis returns 81 lines, although the majority of these are clustered at the blue end of the observations ($< 8000$\,\AA). Of the redder lines in our analysis, many are coincident with the 1.08\,\micron\ (10312, 10739 and 10907\,\AA), 1.40\,\micron\ (13611, 13882 and 14150\,\AA) and 1.58\,\micron\ (15075 and 15322\,\AA) features. The only contaminant lines at wavelengths $> 8000$\,\AA\ lie immediately blueward of the 1.08\,\micron\ feature, which produce a candidate blend for a feature at $\sim 1$\,\micron, for which we see no evidence. We need line blanketing for wavelengths $< 8000$\,\AA, as well as the contaminant lines at $\sim 1$\,\micron\ to be relatively weaker than what we have estimated here for Eu\,\II\ to be a viable candidate.
    
    \item \textbf{Gd\,$\boldsymbol{\textsc{iii}}$:}
    See the main text (Section~\ref{sec:Search for potential candidate species (strong transitions) - Results}).

    \item \textbf{Tb\,$\boldsymbol{\textsc{ii}}$:}
    There are 162 lines in our wavelength range of interest. Some of the strongest lines are coincident with the 1.58\,\micron\ feature (15967, 16068 and 16204\,\AA). However, there are multiple weaker (but non-negligible) lines in our observed wavelength range, for which we see no evidence.

    \item \textbf{Ho\,$\boldsymbol{\textsc{i}}$:}
    Our analysis recovered 262 lines for this species. Of these, many are expected to be weak. Many of the most prominent transitions are coincident with the observed features; specifically, the 0.79\,\micron\ (8100.7 and 8104.4\,\AA), 1.08\,\micron\ (10932\,\AA), 1.23\,\micron\ (11935 and 11866\,\AA) and 1.40\,\micron\ (14441 and 14452\,\AA) features. The relative strengths of these lines do not match the observations, and so we need these strongest transitions to have different relative strengths for Ho\,\I\ to be considered a viable species.

    \item \textbf{Er\,$\boldsymbol{\textsc{i}}$:}
    This ion returns 161 lines in our analysis, although many of these are negligibly weak. The strongest two lines are coincident with the 1.23 and 1.40\,\micron\ features (12992 and 13934\,\AA). Er\,\I\ also has a small cluster of weak, but still detectable lines at $\sim 8500$\,\AA, which we see no strong evidence for, and so we need these lines to be weaker than what we estimate (or we need to invoke line-blanketing) for Er\,\I\ to be a viable candidate.

    \item \textbf{Er\,$\boldsymbol{\textsc{ii}}$:}
    There are 96 lines in our wavelength range of interest. Er\,\II\ has its most prominent line coincident with the 1.40\,\micron\ feature, although we note that it lies at the very edge of our permitted wavelength range for this feature (14653\,\AA). There are many weaker (but non-negligible) lines between $\sim 7500 - 10000$\,\AA, which may blend to produce some detectable flux that we do not see evidence for in the data.
    
    \item \textbf{Tm\,$\boldsymbol{\textsc{i}}$:}
    Our analysis returned 242 lines for this species. Of these, the strongest transition lies at 7622.2\,\AA, coincident with the 0.79\,\micron\ feature. We need estimates for the strengths of the many other lines in our analysis to verify that they are much weaker than this prominent line, for Tm\,\I\ to be a viable species.
    
    \item \textbf{Tm\,$\boldsymbol{\textsc{ii}}$:}
    This species returned 168 lines in our analysis. We find very strong transitions in the near-UV and blue end of the visible spectrum ($\lambda \leq 6000$\,\AA), which corresponds with the heavily line-blanketed region of the \gfo\ spectra. Two of its strongest lines at redder wavelengths are coincident with the 1.08\,\micron\ feature (10714 and 11090\,\AA).

    \item \textbf{Yb\,$\boldsymbol{\textsc{i}}$:}
    Our analysis returned 60 lines for this species. Many of the strongest lines are coincident with the 1.40\,\micron\ (13888\,\AA), 1.58\,\micron\ (15391\,\AA) and blended 2.07\,\micron\ feature (19835, 20926, 22202 and 22276\,\AA). We note the presence of two comparably strong lines, which lie at 17984 and 14793\,\AA\ -- these lines are coincident with a telluric region, and the gap between the 1.40 and 1.58\,\micron\ features, respectively. While telluric absorption can explain away the contaminant 17984\,\AA\ line, we see no evidence for the 14793\,\AA\ line in the observed data. We need the relative strength of this line to be much weaker than our estimate for Yb\,\I\ to be a viable candidate.

    \item \textbf{Lu\,$\boldsymbol{\textsc{i}}$:}
    Our analysis shortlists only three transitions (13375, 18240 and 24177\,\AA). The strongest is coincident with the 1.40\,\micron\ feature (although it lies at the extreme blue end of permitted wavelengths), one lies in a telluric region, and the third lies just beyond the wavelengths covered by our \xsh\ spectra. The 13375\,\AA\ line is too blue to be the sole producer of the 1.40\,\micron\ feature, but it may contribute.

    \item \textbf{Hf\,$\boldsymbol{\textsc{i}}$:}
    Our analysis finds 68 lines within our wavelength range of interest. Many of these lines lie at the blue end of the spectrum ($\lesssim 7500$\,\AA), where the data may be significantly impacted by line blanketing effects. Two of the strongest lines recovered are coincident with the blended 2.07\,\micron\ feature (19865 and 20533\,\AA). Other weaker, but still prominent lines are coincident with the 0.79\,\micron\ feature (8206.8 and 8279.2\,\AA). There are other detectable lines across the spectral range, with a clustering around $\sim 1$\,\micron, which are too blue to reproduce the 1.08\,\micron\ feature. Some explanation for not seeing evidence for prominent flux from Hf\,\I\ at wavelengths $\lesssim 7500$\,\AA\ is needed before Hf\,\I\ can be considered a viable candidate.

    \item \textbf{Hf\,$\boldsymbol{\textsc{ii}}$:}
    Our analysis returns 63 lines for this ion, many of which lie at wavelengths $\lesssim 7000$\,\AA. There are a small number of  strong lines at redder wavelengths -- 10787 and 10904\,\AA\ (which are coincident with the 1.08\,\micron\ feature), 12604 and 12875\,\AA\ (which are coincident with the emission feature at $\sim 1.23$\,\micron), and a small grouping of lines at $\sim 9500$\,\AA, for which we see no evidence in the observations.

    \item \textbf{Fr\,$\boldsymbol{\textsc{i}}$:}
    Our analysis recovered 19 lines, two of which are expected to be much more prominent than the rest (7181.8 and 8171.7\,\AA). The redder of these two lines is coincident with the 0.79\,\micron\ feature, while there is no observational evidence for the bluer line -- if this bluer transition is relatively weaker than we predict here, then Fr\,\I\ can be a viable candidate for the 0.79\,\micron\ feature.

    \item   \textbf{Th\,$\boldsymbol{\textsc{iii}}$:}
    We obtained 63 lines from our analysis. The strongest lines generally agree with the observed peak positions, with the 13446\,\AA\ (1.40\,\micron), 16064\,\AA\ (1.58\,\micron), 19948, 20011, 20306, 20993 and 21510\,\AA\ (2.07\,\micron) lines all broadly agreeing with observation. However, there are a number of weaker (but non-negligible) lines -- the most prominent of these lie at 12726, 14767 and 14781\,\AA. We see no evidence for any of these three lines in the observed data.
    
\end{itemize}

\subsection{Verification of analysis} \label{sec:Appendix - Search for potential candidate species (strong transitions) - Verification of analysis}

Here we present some discussion to highlight the validity of our rudimentary analysis. By comparing to previous results, we can verify whether our analysis is returning sensible results.

As a sanity check, and proof of the ability to recover sensible candidate transitions from this analysis, we note that we recover two candidate transitions from \SrII\ for the 1.08\,\micron\ feature. These are two of the three \SrII\ lines that are responsible for producing this feature at early times (i.e. the \SrII\ NIR triplet; see \paperI). It is worth noting however that only the reddest two of these three transitions satisfies our wavelength cut, indicating again that this feature does indeed lie redward of the weighted average of the expected centroid of the \SrII\ NIR triplet blend (see Section~\ref{sec:Analysing the presence of strontium at late times}).

As discussed in Section\,\ref{sec:Modelling the NIR emission features - Comparisons between species X and observations}, \cite{Domoto2022} focus on identifying the NIR features present in the same $+1.4 - 3.4$\,d \xsh\ spectra of \gfo\ as we analyse here. They assume the ejecta is optically thick, and attempt to identify the cause of the inferred \textit{absorption} features that have troughs at $\sim 12000$ and 14000\,\AA. \cite{Domoto2022} propose La\,\III\ as the cause of absorption at $\sim 1.20$\,\micron, and Ce\,\III\ for the absorption at $\sim 1.40$\,\micron. Specifically, they identify two strong transitions for La\,\III\ (with $\log \left[ gf \right] > -3$) at 13898 and 14100\,\AA, as the cause of the observed absorption trough at $\sim 1.20$\,\micron. These are the same transitions as we find in our search for candidates for the $\sim 1.40$\,\micron\ emission feature (see Table~\ref{tab:NIST - Candidate transitions (E1)}). Hence it is plausible that these strong La\,\III\ lines cause both shallow absorption in the early spectra and evolve into pure emission in the later spectra.

\cite{Domoto2022} further identify 38 strong transitions for Ce\,\III\ at NIR wavelengths, and find that the strongest of these could produce a P-Cygni feature in the first $\sim 3$ days, with emission peaking around 1.59\,\micron. We discussed this in Section\,\ref{sec:Modelling the NIR emission features - Comparisons between species X and observations} and could not quantitatively confirm that our synthetic, photospheric phase spectra matched the observations for these lines. However, these strong, permitted Ce\,\III\ lines could potentially be  capable of reproducing the 1.58\,\micron\ emission feature that develops later. In our line search, we recover many of the same lines as \cite{Domoto2022}. As well as being able to match the emission feature at $\sim 1.58$\,\micron, many of our strongest Ce\,\III\ lines also broadly match the emission features at $\sim 1.08$, 1.23, 1.40 and 2.07\,\micron.

Our agreement with many of the candidate lines proposed by \cite{Domoto2022}, belonging to both La\,\III\ and Ce\,\III, demonstrates that we are able to recover sensible shortlists of candidate transitions with our approach here.

\section{Search for potential candidate species (weak transitions)} \label{sec:Appendix - Search for potential candidate species (weak transitions) - NIST}

\subsection{Shortlisted species} \label{sec:Appendix - Search for potential candidate species (weak transitions) - NIST - Shortlisted species}

Our candidate species that may contribute to the data, albeit with some caveats, are summarised and presented below (see Table~\ref{tab:Forbidden candidate transitions}).

\begin{itemize}
    
    \item \textbf{[Mo\,$\boldsymbol{\textsc{iii}}$]:}
    Our analysis returns 12 lines within our wavelength range. Of these, we identify a strong line at 14152\,\AA, coincident with the 1.40\,\micron\ feature. We also find a number of comparably strong lines between $\sim 17500 - 19500$\,\AA, which do not match the data.
    
    \item \textbf{[Tc\,$\boldsymbol{\textsc{ii}}$]:}
    This ion has 15 lines in our analysis. The strongest of these are the 10922, 11905, 12582, 13062 and 13305\,\AA, with only the 10922\,\AA\ line matching the observed data (specifically the 1.08\,\micron\ feature).

    \item \textbf{[Rh\,$\boldsymbol{\textsc{i}}$]:}
    We recover 54 lines in our analysis. Among the few strong lines, we find that the 10845\,\AA\ line matches the 1.08\,\micron\ feature. We find comparably strong lines at 13002, 15098, 17396 and 22315\,\AA, which do not match the data.

    \item \textbf{[Pd\,$\boldsymbol{\textsc{iii}}$]:} 
    See the main text (Section~\ref{sec:Search for potential candidate species (weak transitions) - NIST analysis}).

    \item \textbf{[Te\,$\boldsymbol{\textsc{ii}}$]:}
    Our analysis returns four lines within our observable wavelength range. One of the strongest of these is coincident with the $\sim 1.23$\,\micron\ emission feature (12308\,\AA). We need the other contaminant lines to be weaker than our model predicts for Te\,\II\ to be a viable candidate for the 1.23\,\micron\ feature. Our analysis also predicts a prominent line coincident with the 4.5\,\micron\ \spitzer\ band (45466\,\AA), which we expect to be significantly stronger than any of the lines within $0.5 \leq \lambda \leq 2.5$\,\micron.

    \item \textbf{[Xe\,$\boldsymbol{\textsc{iii}}$]:}
    Our analysis recovers only two lines. These lie at 10210 and 12300\,\AA. The redder of these is coincident with the 1.23\,\micron\ feature. We see no evidence for the 10210\,\AA\ line in the observed data, and so we need this line to be relatively weaker than the 12300\,\AA\ line for Xe\,\III\ to be considered a viable candidate.
    
    \item \textbf{[Ce\,$\boldsymbol{\textsc{iii}}$]:}
    Our analysis recovers 25 lines. The strongest two lines lie at 20987\,\AA\ (matching the wavelength threshold for both the 2.059 and 2.135\,\micron\ features), and 19976\,\AA, which lies immediately blueward of the 2.059\,\micron\ feature. The contaminant 19976\,\AA\ line is the strongest line in our analysis, so we need it to be much weaker for Ce\,\III\ to be a viable species.

    \item \textbf{[Gd\,$\boldsymbol{\textsc{ii}}$]:}
    There are 281 shortlisted lines in our analysis. Among the strongest lines are the three lines at 21105, 21249 and 21414\,\AA, which all agree with the 2.135\,\micron\ feature. The number of contaminant lines is high, and some of these may produce detectable flux, and so Gd\,\II\ can only be a viable candidate species if the three lines highlighted are relatively stronger than all other transitions.
    
    \item \textbf{[Gd\,$\boldsymbol{\textsc{iii}}$]:}
    We recover five lines from our analysis -- all with comparable strength, at 10875, 11216, 11764, 12682 and 14467\,\AA. The 10875\,\AA\ line is coincident with the 1.08\,\micron\ feature, while we see no evidence for the other lines. For Gd\,\III\ to be a viable candidate, we need these other lines to be weaker than what we have estimated here.

    \item \textbf{[Dy\,$\boldsymbol{\textsc{ii}}$]:}
    We recover 124 lines in our analysis. The strongest few lines lie at 13360, 15022, 15070, 21028 and 23036\,\AA. While the 21028\,\AA\ line matches the 2.135\,\micron\ feature, these other lines do not match observation, and so they need to be weaker than estimated here for Dy\,\II\ to be a viable candidate.
  
    \item \textbf{[Er\,$\boldsymbol{\textsc{ii}}$]:}
    We recover 132 lines from our analysis, many of which have a negligible effect on our model. The strongest few produce emission features coincident with the 1.40\,\micron\ feature (due to the 13898 and 13987\,\AA\ lines), and the 2.135\,\micron\ feature (due to the 21312\,\AA\ line). There are contaminant lines that may be detectable, at 14804, 14904, 19483 and 20148\,\AA\ -- these need to be weaker than we estimate here for Er\,\II\ to be a viable candidate.

    \item \textbf{[Yb\,$\boldsymbol{\textsc{ii}}$]:}
    We recover 103 from our analysis. The strongest few lines lie at 10936\,\AA\ (coincident with the 1.08\,\micron), 11356\,\AA\ (redward of the 1.08\,\micron\ feature), 13625\,\AA\ (blueward of the 1.40\,\micron\ feature), and 18726\,\AA\ (coincident with a telluric region). Although there are some close matches to observed feature (i.e. the 11356 and 13625\,\AA\ lines) these do not satisfy our wavelength range thresholds, and so are considered contaminant lines. For Yb\,\II\ to be a viable candidate, these lines need to be weaker than we estimate here.
    
    \item \textbf{[Ta\,$\boldsymbol{\textsc{i}}$]:}
    We recover 64 lines from our analysis for this ion. Some of the strongest lines are coincident with the observed features; specifically, the 10807 and 13806\,\AA\ lines are coincident with the 1.08 and 1.40\,\micron\ features. Other equally strong lines lie at 16477, 16531 and 18905\,\AA, which do not agree with the data. We need these lines to be weaker than our analysis predicts for Ta\,\I\ to be a viable candidate ion.
    
    \item \textbf{[Os\,$\boldsymbol{\textsc{iii}}$]:}
    Our analysis recovers 57 lines for this species. Of these, only four are expected to be strong -- 7977.3, 11707, 11512 and 12567\,\AA. Two of these are coincident with the 0.79\,\micron\ (7977.3\,\AA) and 1.23\,\micron\ features (12567\,\AA), while we see no observational evidence to support the presence of the other two lines. We need these two contaminant lines to be weaker than our analysis suggests for Os\,\III\ to be a viable contributing species.
    
    \item \textbf{[Pt\,$\boldsymbol{\textsc{i}}$]:}
    Our analysis recovered 19 lines for this species. One of the strongest lines is coincident with the 1.08\,\micron\ feature (10761\,\AA), and we note that this line was previously identified by \cite{Gillanders2021}. However, there are comparably strong lines outside our wavelength ranges of interest, and so these need to be relatively weaker than we estimate here.

    \item \textbf{[Au\,$\boldsymbol{\textsc{ii}}$]:}
    We recover two lines for this species from our analysis (7857.8 and 9876.4\,\AA; previously identified by \citealt{Gillanders2021}). The bluer of these transitions is coincident with the 0.79\,\micron\ feature, while we see no evidence for the redder line in the observational data. For Au\,\II\ to be a viable candidate, we need the contaminant 9876.4\,\AA\ line to be weaker than the 7857.8\,\AA\ line.

\end{itemize}

\subsection{Verification of analysis} \label{sec:Appendix - Search for potential candidate species (weak transitions) - NIST - Verification of analysis}

Here we briefly summarise the comparisons to other results from previous analyses, which aid to verify the validity of our rudimentary analysis presented here.

In Section~\ref{sec:Search for potential candidate species (weak transitions) - NIST analysis}, we use upper level populations as a proxy for line strength, which results in relative emission strengths that differ somewhat from those computed using the Kurucz data (Appendix~\ref{sec:Search for potential candidate species (weak transitions) - Kurucz analysis}). For the Kurucz data, we used an atomic data set that possesses full line information, and so our shortlisted transitions will be more reliable using this approach, than that carried out in Section~\ref{sec:Search for potential candidate species (weak transitions) - NIST analysis}. Although our upper level populations as a proxy for line strength approximation is expected to be a source of uncertainty, which we expect impacts the inferences that can be made with our models, we still manage to shortlist the only potentially viable candidate species that we obtained using the Kurucz data -- [\YII]. 

As well as comparing our Kurucz and NIST analyses presented in this work, we can also compare with previous works. Specifically, we can compare our shortlisted transitions for [Pt] and [Au] in Section~\ref{sec:Search for potential candidate species (weak transitions) - NIST analysis} with our previous analysis, where we utilised newly-calculated Pt and Au data \citep[][]{Gillanders2021}, to show how our results vary when considering intrinsic line strengths. This previous study was more detailed, as it had full line information (i.e. intrinsic line strengths as well as transition wavelengths).

Our shortlisted 10761\,\AA\ transition here for [Pt\,\I] is the most prominent line we find in \cite{Gillanders2021}. For the [Au\,\I] case, we recover the same three most prominent transitions shortlisted in \cite{Gillanders2021}, albeit with different relative strengths (due to the effects of consulting \Aval). We note that for the [Pt\,\I] case, we fail to recover some of the weaker (but non-negligible) lines that we shortlist in \cite{Gillanders2021} in this analysis. This is due to our analysis discarding these lines as possibilities, due to them not satisfying our imposed LS-coupling rules. It is known that the LS-coupling formalism does not apply well to heavy ions, and so this result is somewhat expected. This effect will be more pronounced for heavier ions, and so we propose the possibility of missing transitions that are relevant to our inferences in this work is smaller for the lighter ions.

While expanding our analysis to include the intermediate coupling rules as well as the LS-coupling rules would assist our recovery of the missing [Pt\,\I] lines, we opted not to pursue a study of these additional lines, for the following reason. Throughout our analysis, we have found that there is no convincing candidate ion or element that can explain all (or many) of the observed features. We are already finding many possible candidate ions (even after enforcing the strict LS-coupling rules), and so adding additional lines to this analysis does nothing to improve our ability to shortlist a small number of candidate ions -- instead, it will likely only increase the numbers of possible line identifications.\footnote{We note that in some cases, the inclusion of the intermediate coupling rules would allow us to identify many more contaminant lines, which would enable us to rule out particular ions. However, this is not expected to be the case for all of our shortlist, and we expect it to introduce additional candidate lines and ions, on top of the ones already shortlisted in our analysis.} If all of the spectral features arise from different species, or if they arise from blended emission from multiple lines, then we really cannot advance with this type of line identification study without possessing Einstein \Aval.

Although this comparison highlights the uncertainty of our approach, it also demonstrates that level population as a proxy for line strength is a useful approximation, since we can use this approach to find the strongest few lines -- although the relative strengths of the strongest lines appear to be uncertain relative to one another (which is something we have considered while compiling our shortlist of viable candidates).

\section{Search for potential candidate species (Kurucz)} \label{sec:Appendix - Search for potential candidate species (weak transitions) - Kurucz}

The Kurucz data base \citep[][]{Kurucz2017}, from which we exported atomic data to include in our \tardis\ modelling (see \paperI), contains atomic data for most of the lowest few ionisation stages of species between $Z = 38 - 46$, as well as data for $Z = 56$. Specifically, it contains line lists that include forbidden M1 and E2 transitions in addition to the permitted E1 transitions that we needed for our earlier \tardis\ modelling. The Kurucz line lists contain a mix of theoretical and experimentally measured lines, but we note that our analysis favours the transitions between low-lying levels, which tend to be the ones that have been experimentally verified. With these line lists, we search for candidate transitions for each of the emission features present in the late-time spectra of \gfo.

Note that the Kurucz atomic data base presents all transition wavelengths $> 2000$\,\AA\ as air wavelengths, and so all wavelengths in this appendix are presented as in air.\footnote{We present the wavelengths of the candidate transitions in Table~\ref{tab:Kurucz candidates} in air and vacuum for completeness, and for ease of comparison with the lines discussed in the other sections.}

\subsection{Method} \label{sec:Appendix - Search for potential candidate species (weak transitions) - Kurucz - Method}

For our candidate transition search using the Kurucz atomic data, we follow the same method as outlined by \cite{Gillanders2021}, which we summarise below \citep[see also][for details]{Thesis_Gillanders}. In this analysis, we include all transition types (i.e. E1, M1 and E2), since there are some permitted transitions that are intrinsically weak (i.e. have very small Einstein \Aval; \mbox{$\ll 100$\,s$^{-1}$}), and so they may be relevant for understanding the dominant transitions at the late phases. We filter the line list based on whether we expect the lines to be strong at late times, which we can estimate by determining the upper level populations. First, we calculate the mean radiative lifetime ($\tau_{\rm rad}$) of all excited levels for the species under investigation.\footnote{In \cite{Gillanders2021}, we defined the mean radiative lifetime of a level, $\tau_{\rm rad}$, as the inverse of the sum of the $A$-values of all transitions that originate from that level.} We only consider lines that originate from upper levels that have \mbox{$\tau_{\rm rad} \geq 10^{-2}$\,s}, as the rate of emission from levels with shorter $\tau_{\rm rad}$ at late times is likely to be much lower than in LTE. For the remaining levels, we assume LTE to calculate the upper level populations.

Then, we generate synthetic emission spectra \citep[as in Section~\ref{sec:Analysing the presence of strontium at late times - Method}, and][]{Gillanders2021}. We initially assume $M = 10^{-3}$\,\msun\ for all species, but note that, if any provide promising fits to the data, the line luminosity scales linearly with mass (i.e. $L_{\rm em} \propto M$), and so we can easily determine how much mass we need in our models to replicate the data. From this approach, we can investigate how much material is needed for the transitions of each species to become prominent. We can also investigate how the relative feature strengths vary as we alter our ejecta temperature ($T \in [2000, 3500, 5000]$\,K, as used by \citealt{Gillanders2021}). This analysis is more accurate than that performed in Sections~\ref{sec:Search for potential candidate species (strong transitions)} and \ref{sec:Search for potential candidate species (weak transitions)} (since here we have estimates for intrinsic line strengths), although those other approaches are able to analyse the species for which we do not have NIR line lists ($Z \gtrsim 46$). With these simple emission spectra, we compare to the observed spectra of \gfo, and determine whether the prominent features are coincident with the emission features in the observed data.

Finally, we have one additional point of consideration, that may help to constrain potential candidate species -- namely, the \spitzer\ observations presented by \cite{Villar2018} and \cite{Kasliwal2022} (see Section~\ref{sec:Appendix - Search for potential candidate species (weak transitions) - NIST - Method}). \cite{Hotokezaka2022} present an analysis exploring the possibility of forbidden transitions from the lowest configuration of \rpro\ species producing transitions in agreement with these \spitzer\ observations -- specifically, they search for species that can produce strong emission at $\sim 4.5$\,\micron, while simultaneously not producing any detectable features at $\sim 3.6$\,\micron. We extend our analysis here to encompass these additional wavelengths, and further favour or disfavour candidate species, beyond what we could do considering only the features present in the \xsh\ spectra.

\subsection{Results} \label{sec:Search for potential candidate species (weak transitions) - Kurucz analysis}

Here we present our candidate transitions for the $\sim 0.79$, 1.08, 1.23, 1.40, 1.58, 2.059 and 2.135\,\micron\ features present in the late-phase spectra of \gfo. We also present the lines that are coincident with the 3.6 and 4.5\,\micron\ \spitzer\ bands -- see Table~\ref{tab:Kurucz candidates}.

From our analysis of the species in the Kurucz data base, we find that [\YII], [Zr\,\II], Nb\,\I, Mo\,\I, [Rh\,\II] and [Rh\,\III] all produce features coincident with the observed data. The lines that are prominent in our models, and potentially produce the observed emission features are presented in Table~\ref{tab:Kurucz candidates}. Details of each of these candidate species, and their viability, are discussed below. We note that for all candidate ions presented here, the inferred masses needed to match the observed feature strengths are typically higher than those supported by nucleosynthesis arguments.\footnote{For example, one can consult our composition profiles presented in \paperI, and, paired with the ejecta masses estimated for \gfo\ from lightcurve modelling \citep[$0.01 - 0.05$\,\msun; see e.g.][]{Cowperthwaite2017, Smartt2017, Tanvir2017, Coughlin2018, Waxman2018}, can estimate approximate element masses.}

\begin{itemize}

    \item \textbf{[Y\,$\boldsymbol{\textsc{ii}}$]:}
    This ion produces two emission features, centred at \mbox{$\sim 7800$} and 13750\,\AA. These are coincident with the emission features we observe at $\sim 0.79$\,\micron\ (the point at which we proposed the \SrII\ NIR triplet absorption rejoins the continuum; see Section~\ref{sec:Spectroscopic evolution of AT2017gfo}), and $\sim 1.40$\,\micron. It is possible that the $\sim 0.79$\,\micron\ region of the spectra has contribution from some emission feature, at least at late times. One issue with \YII\ as the main source of flux for this 1.40\,\micron\ feature is the larger than expected mass needed to power the observed emission feature. We require model masses, \mbox{$M_{\rm Y \textsc{ii}} \sim 8.0 \times 10^{-2}$}, $9.2 \times 10^{-3}$, and $4.2 \times 10^{-3}$\,\msun\ for our 2000, 3500 and 5000\,K models, respectively, to reproduce the measured flux of the observed 1.40\,\micron\ feature at +9.4\,d (see Section~\ref{tab:Gaussian fit parameters}). Note that higher temperature lowers the required mass, but increases the relative strength of the 7800\,\AA\ feature (which is not supported by the data, since the 1.40\,\micron\ feature is more prominent than the 0.79\,\micron\ feature). For \YII\ to be producing the $\sim 1.40$\,\micron\ feature, we need even higher $T$ (or some alternative non-thermal mechanism to populate the relevant levels).
    
    \item \textbf{[Zr\,$\boldsymbol{\textsc{ii}}$]:}
    This species produces an emission feature at 10800\,\AA, exactly coincident with one of our emission features. However, there is significant contribution from many weaker, blended lines. This emission feature becomes more prominent (relative to these other weaker lines) at higher $T$. We require large masses to reproduce the flux needed to power the 1.08\,\micron\ feature at +7.4\,d \mbox{($M_{\rm Zr \, \textsc{ii}} \sim 3.3 \times 10^{-2}$\,\msun} at 5000\,K).
    
    \item \textbf{Nb\,$\boldsymbol{\textsc{i}}$:}
    This ion produces a broad emission feature at 10720\,\AA, coincident with the 1.08\,\micron\ feature. However, it produces comparably strong emission features at 7800\,\AA\ (coincident with the 0.79\,\micron\ feature), and 9160 and 16760\,\AA, which do not match the observed spectra. Variations in temperature across the range explored does not improve the relative strengths, and so we disfavour this ion as a prominent source of emission in the \gfo\ spectra. This candidate ion has (intrinsically weak) E1 lines as the source of its most prominent emission in this regime. We note the presence of mid-infrared (MIR) lines that are coincident with the 3.6 and 4.5\,\micron\ \spitzer\ bands (35636 and 46410\,\AA; see Table~\ref{tab:Kurucz candidates}). We find the 3.6\,\micron\ line is stronger than the 4.5\,\micron\ line across all model temperatures, and so it seems unlikely there will be much Nb\,\I\ present (at least at the times of the \spitzer\ observations).
    
    \item \textbf{Mo\,$\boldsymbol{\textsc{i}}$:}
    This species produces a prominent feature (from E1 transitions) at 20800\,\AA, coincident with our blended 2.07\,\micron\ feature. This species also produces strong features at 5470 and 6700\,\AA, due to M1 and E2 transitions, respectively. These features are predicted to be stronger than our favourable 20800\,\AA\ feature. We do not see either of these features in the spectra, although they lie at wavelengths that suffer from line-blanketing (at least at early times). We need \mbox{$M_{\rm Mo \, \textsc{i}} \sim 6.1 \times 10^{-2}$\,\msun} (at $T = 5000$\,K) to match the flux producing the 2.07\,\micron\ feature at +7.4\,d.
    
    \item \textbf{[Rh\,$\boldsymbol{\textsc{ii}}$]:}
    This species produces a number of broad emission features, centred at $\sim 6730$, 8020, 10820, 17350, 21810 and 41650\,\AA. The emission feature at 1.08\,\micron\ is dominant at higher temperatures, and there is reasonable agreement between the model emission features (at $\sim 8020$ and 21810\,\AA) and the observed 0.79 and 2.07\,\micron\ features. The features at $\sim 6730$ and 17350\,\AA\ are also expected to be reasonably strong, but we see no evidence for them in the observed spectra. Therefore, we ultimately disfavour this ion as a significant contributor to the emission components observed in the late-phase spectra of \gfo. We note the presence of the MIR feature at 41650\,\AA, which lies within the wavelength range of the 4.5\,\micron\ \spitzer\ band, making it potentially detectable if present in significant quantities. We need masses, $M_{\rm Rh \, \textsc{ii}} \sim 6.0 \times 10^{-2}$ and 0.13\,\msun\ (at $T = 5000$\,K), to produce the flux needed to power the +7.4\,d 1.08\,\micron\ and +8.4\,d 2.135\,\micron\ features, respectively.

    \item \textbf{[Rh\,$\boldsymbol{\textsc{iii}}$]:}
    This ion produces a prominent M1 feature at 7690\,\AA, which is semi-coincident with the observed $\sim 0.79$\,\micron\ emission feature. We also find a strong emission feature at 46550\,\AA, which is coincident with the 4.5\,\micron\ \spitzer\ band, making this species detectable in the MIR.

\end{itemize}

\subsection{Discussion}

From this analysis, we find no convincing viable candidate species for any of the observed emission features, assuming they originate from an optically thin medium that is dominated by intrinsically weak lines (see Table~\ref{tab:Kurucz candidates} for a summary of all candidate transitions). From this analysis, the only possible identification comes from [\YII], although this ion has the issue of requiring more mass than is plausible for the \gfo\ ejecta. In fact, we find that for this approach, we need more mass for all potentially viable candidates than we expect is reasonable, based on our composition profiles from modelling the early phases, and the inferred ejecta mass for \gfo\ (see \paperI). The fact that the inferred masses for many of the ions discussed above are higher than what we can tolerate in the ejecta of \gfo\ is perhaps indicative of some error with our approach here; i.e. perhaps LTE level populations are not sufficiently accurate to reproduce reasonable estimates for the emission of the relevant lines. Alternatively, it may be evidence of intrinsically weak lines not being capable of reaching the observed luminosities of the features. If this is the case, then it indicates that intrinsically weak transitions cannot be responsible for the observed emission features in the \gfo\ spectra at late times. This would suggest the correct interpretation is that these features are produced by strong transitions (e.g. the lines presented in Section~\ref{sec:Search for potential candidate species (strong transitions)}). To quantify this issue, a more accurate calculation including non-thermal and non-LTE effects is necessary (such a study is not currently possible with the available data).

Ultimately, we lack extensive atomic data to explore the species heavier than $Z = 46$, and so we cannot constrain any more viable candidate ions than [\YII] from this analysis. We need to either rule out other species (for which we do not yet have complete atomic data), or identify other features that are also possibly produced by [\YII], which would help to corroborate its presence in the ejecta (although then we would have to reconcile the required mass with our expected composition).

\onecolumn
\footnotesize

\begin{ThreePartTable}
\renewcommand*{\arraystretch}{1.2}
\renewcommand\thetable{A1} 
\captionsetup{format=custom} 
\captionsetup{font=footnotesize}
    \begin{TableNotes}
        \footnotesize
        \item \textbf{Note.} $\lambda_{\rm air}$ values have been computed assuming the standard vacuum-to-air conversion from VALD3 \citep[see][]{Birch1994, Morton2000, VALD3}.
        \item[o] Denotes an odd parity.
        \item[a] This is a candidate transition for the bluer component of the feature at $\sim 2.07$\,\micron\ (2.059\,\micron; see Section~\ref{sec:Modelling the late-time emission features}).
        \item[b] This is a candidate transition for the redder component of the feature at $\sim 2.07$\,\micron\ (2.135\,\micron; see Section~\ref{sec:Modelling the late-time emission features}).
    \end{TableNotes}
    \begin{longtable}{lcccccrlccrlcc}
        \caption{
        \label{tab:NIST - Candidate transitions (E1)}
        Candidate E1 transitions for the $\sim 0.79$, 1.08, 1.23, 1.40, 1.58 and 2.07\,\micron\ features in the spectra of \gfo. Only transitions with a relative level population $\geq 0.1$ are included, for brevity. For each feature, two sets of transitions are given, separated by a vertical space. The upper set of transitions are those that belong to our most viable transitions, while the lower set are the potential candidate transitions. The species, wavelength (both in vacuum and air, for completeness), level energies, quantum J numbers, and the relative level population for the upper level are presented for each transition. The configurations and terms are displayed \citep[from the NIST ASD;][]{NIST2020} when they are easily expressible in the LS-coupling scheme.
        }   \\
        \hline
        \addlinespace[0.4ex]
        \hline
        \multirow{2}{*}{Species}     &\multirow{2}{*}{$\lambda_{\rm vac}$ (\AA)}      &\multirow{2}{*}{$\lambda_{\rm air}$ (\AA)}       &\multicolumn{2}{c}{Level energies (eV)}    &       &\multicolumn{3}{c}{Lower level}    &     &\multicolumn{3}{c}{Upper level}    &Relative level      \\
        \cline{4-5}
        \cline{7-9}
        \cline{11-13}
            &       &       &Lower      &Upper      &       &Configuration      &Term       &J      &    &Configuration      &Term      &J      &population    \\
        \hline
        \endfirsthead
        \caption[]{ \textit{(continued)}} \\
        \hline
        \addlinespace[0.4ex]
        \hline
        \multirow{2}{*}{Species}     &\multirow{2}{*}{$\lambda_{\rm vac}$ (\AA)}      &\multirow{2}{*}{$\lambda_{\rm air}$ (\AA)}       &\multicolumn{2}{c}{Level energies (eV)}    &       &\multicolumn{3}{c}{Lower level}    &     &\multicolumn{3}{c}{Upper level}    &Relative level      \\
        \cline{4-5}
        \cline{7-9}
        \cline{11-13}
            &       &       &Lower      &Upper      &       &Configuration      &Term       &J      &    &Configuration      &Term      &J      &population    \\
        \hline
        \endhead
        \hline
        \multicolumn{14}{r}{\footnotesize\itshape Continued on next page}
        \endfoot
        \endlastfoot
        %
        \multicolumn{14}{c}{0.79\,\micron\ feature}    \\
        \hline
        
        $^{88}$Ra\,\II 		 &8021.9	&8019.7         &1.704 		 &3.249 		&     &6d               &$^2$D     	 &\sfrac{5}{2} 		 &        &7p                	 &$^2$P$^{\textsc{o}}$   	 &\sfrac{3}{2} 			 &0.19 			 \\ 
        $^{89}$Ac\,\I 		 &8145.6	&8143.3         &0.000 		 &1.522 		&     &6d7s$^2$         &$^2$D  	 &\sfrac{3}{2} 		 &        &7s$^2$7p            	 &$^2$P$^{\textsc{o}}$   	 &\sfrac{3}{2} 	         &0.20 			 \\ 
        
        \addlinespace[2ex]
        
        $^{67}$Ho\,\I 		 &8104.4 	&8102.1         &0.000 		 &1.530 		&      &4f$^{11}$6s$^2$ 	                     &$^4$I$^{\textsc{o}}$ 	 &\sfrac{15}{2} 	&	 &$-$ 	                 &$-$ 	                  &\sfrac{15}{2} 	&0.13 			 \\ 
        $^{67}$Ho\,\I 		 &8100.7 	&8098.5         &0.000 		 &1.531 		&      &4f$^{11}$6s$^2$ 	                     &$^4$I$^{\textsc{o}}$ 	 &\sfrac{15}{2} 	&	 &$-$ 	                 &$-$ 	                  &\sfrac{13}{2} 	&0.12 			 \\ 
        $^{69}$Tm\,\I 		 &7622.2	&7620.1         &0.000 		 &1.627 		&      &4f$^{13}$($^2$F$^{\textsc{o}}$)6s$^2$ 	 &$^2$F$^{\textsc{o}}$ 	 &\sfrac{7}{2} 		&    &$-$ 	                 &$-$ 	                  &\sfrac{9}{2}     &1.0 			 \\ 
        $^{72}$Hf\,\I 		 &8206.8	&8204.6         &0.292 		 &1.803 		&      &5d$^2$6s$^2$                             &$^3$F     	         &3 		        &    &5d6s$^2$($^2$D)6p 	 &$^3$F$^{\textsc{o}}$    &3 	    	    &0.20 			 \\ 
        $^{72}$Hf\,\I 		 &8279.2    &8276.9         &0.292 		 &1.790 		&      &5d$^2$6s$^2$                             &$^3$F  	             &3 		        &    &5d6s$^2$($^2$D)6p 	 &$^3$F$^{\textsc{o}}$    &2 	    	    &0.15 			 \\ 
        $^{87}$Fr\,\I 		 &8171.7	&8169.4         &0.000 		 &1.517 		&      &7s                                       &$^2$S     	         &\sfrac{1}{2} 		&    &7p                  	 &$^2$P$^{\textsc{o}}$    &\sfrac{1}{2} 	&1.0 			 \\ 

        \hline
        \multicolumn{14}{c}{1.08\,\micron\ feature}    \\
        \hline \relax
        $^{38}$Sr\,\II 		 &10330 	&10327   	 &1.839          &3.040 	&	 &4p$^6$4d 	 &$^2$D 	 &\sfrac{5}{2} 	&     &4p$^6$5p 	 &$^2$P$^{\textsc{o}}$ 	 &\sfrac{3}{2}    &1.0 			 \\ 
        $^{38}$Sr\,\II 		 &10918 	&10915   	 &1.805          &2.940 	&	 &4p$^6$4d 	 &$^2$D 	 &\sfrac{3}{2} 	&     &4p$^6$5p 	 &$^2$P$^{\textsc{o}}$ 	 &\sfrac{1}{2}    &0.73 		 \\ 
        $^{58}$Ce\,\III 	 &10723 	&10720	     &0.000 		 &1.156 	&	 &4f$^2$ 	 &$^3$H 	 &4 		    &     &4f5d 	     &$^3$G$^{\textsc{o}}$ 	 &5               &0.16 		 \\ 
        $^{88}$Ra\,\II 		 &10791 	&10788 	     &1.498 		 &2.647 	&    &6d 	     &$^2$D 	 &\sfrac{3}{2} 	&     &7p 	         &$^2$P$^{\textsc{o}}$ 	 &\sfrac{1}{2}    &1.0 			 \\ 

        \addlinespace[2ex]

        $^{63}$Eu\,\II 		 &10907 	&10904   	 &2.108 		 &3.245 		&     &4f$^7$($^8$S$^{\textsc{o}}$)5d 	 &$^7$D$^{\textsc{o}}$ 	 &4 		      &   &$-$              &$-$                     &5 		         &0.48 			 \\ 
        $^{63}$Eu\,\II 		 &10739 	&10736   	 &2.090 		 &3.245 		&     &4f$^7$($^8$S$^{\textsc{o}}$)5d 	 &$^7$D$^{\textsc{o}}$ 	 &5 		      &   &$-$              &$-$                     &5 		         &0.48 			 \\ 
        $^{63}$Eu\,\II 		 &10312 	&10309   	 &2.125 		 &3.328 		&     &4f$^7$($^8$S$^{\textsc{o}}$)5d 	 &$^7$D$^{\textsc{o}}$ 	 &3 		      &   &$-$              &$-$                     &4 		         &0.28 			 \\ 
        $^{67}$Ho\,\I 		 &10932 	&10929    	 &0.000 		 &1.134 		&     &4f$^{11}$6s$^2$ 	                 &$^4$I$^{\textsc{o}}$ 	 &\sfrac{15}{2}   &	  &$-$ 	            &$-$ 	                 &\sfrac{13}{2} 	 &0.54 			 \\ 
        $^{69}$Tm\,\II 		 &11090 	&11087   	 &1.111 		 &2.229 		&     &$-$                               &$-$      	             &3 	          &   &$-$	            &$-$ 	                 &4 		         &0.42 			 \\ 
        $^{69}$Tm\,\II 		 &10714 	&10711   	 &1.111 		 &2.268 		&     &$-$                               &$-$                    &3 	          &   &$-$              &$-$                     &4 		         &0.36 			 \\ 
        $^{72}$Hf\,\II 		 &10904 	&10901   	 &2.343 		 &3.480 		&     &5d$^3$                            &$^4$F	                 &\sfrac{3}{2} 	  &   &5d6s($^3$D)6p 	&$^4$F$^{\textsc{o}}$ 	 &\sfrac{3}{2} 		 &1.0 			 \\ 
        $^{72}$Hf\,\II 		 &10787 	&10784   	 &2.496 		 &3.646 		&     &5d$^3$                            &$^4$F	                 &\sfrac{5}{2} 	  &   &5d6s($^3$D)6p 	&$^4$F$^{\textsc{o}}$ 	 &\sfrac{5}{2} 		 &0.79 			 \\ 

        \hline
        \multicolumn{14}{c}{1.23\,\micron\ feature}    \\
        \hline

        $^{58}$Ce\,\III 		 &12760 	&12757  	 &0.000 		 &0.972 	&	 &4f$^2$ 	 &$^3$H 	 &4 	&	 &4f5d 	 &$^3$G$^{\textsc{o}}$ 	 &4 	 &0.27 			 \\ 
        $^{58}$Ce\,\III 		 &12825 	&12822  	 &0.189 		 &1.156 	&	 &4f$^2$ 	 &$^3$H 	 &5 	&	 &4f5d 	 &$^3$G$^{\textsc{o}}$ 	 &5 	 &0.16 			 \\ 
                
        \addlinespace[2ex]
        
        $^{49}$In\,\I 		 &12916 	&12913    	 &3.022 		 &3.982 		&     &5s$^2$6s      	 &$^2$S              	 &\sfrac{1}{2} 		 &    &5s$^2$6p        	 &$^2$P$^{\textsc{o}}$           	 &\sfrac{3}{2}  	 &1.0 			 \\ 
        $^{67}$Ho\,\I 		 &11935 	&11931    	 &0.000 		 &1.039 		&     &4f$^{11}$6s$^2$ 	 &$^4$I$^{\textsc{o}}$ 	 &\sfrac{15}{2} 	 &    &$-$             	 &$-$                             	 &\sfrac{17}{2} 	 &1.0 			 \\ 
        $^{67}$Ho\,\I 		 &11866 	&11863    	 &0.000 		 &1.045 		&     &4f$^{11}$6s$^2$ 	 &$^4$I$^{\textsc{o}}$ 	 &\sfrac{15}{2} 	 &    &$-$             	 &$-$                           	 &\sfrac{15}{2} 	 &0.87 			 \\ 
        $^{68}$Er\,\I 		 &12992 	&12989    	 &0.000 		 &0.954 		&     &4f$^{12}$6s$^2$ 	 &$^3$H               	 &6            		 &    &$-$               &$-$                             	 &7                	 &0.90 			 \\ 
        $^{72}$Hf\,\II 		 &12604 	&12601   	 &2.496 		 &3.480 		&     &5d$^3$        	 &$^4$F                	 &\sfrac{5}{2} 		 &    &5d6s($^3$D)6p   	 &$^4$F$^{\textsc{o}}$            	 &\sfrac{3}{2}  	 &1.0 			 \\ 
        $^{72}$Hf\,\II 		 &12875 	&12871   	 &2.683 		 &3.646 		&     &5d$^3$        	 &$^4$F                	 &\sfrac{7}{2} 		 &    &5d6s($^3$D)6p   	 &$^4$F$^{\textsc{o}}$           	 &\sfrac{5}{2}   	 &0.79 			 \\ 
                
        \pagebreak
        \multicolumn{14}{c}{1.40\,\micron\ feature}    \\
        \hline \relax

        $^{57}$La\,\III 	 &13898 	&13894    	 &0.000 		 &0.892 	&   	 &5p$^6$5d 	 &$^2$D 	 &\sfrac{3}{2} 	 &	 &5p$^6$4f 	                       &$^2$F$^{\textsc{o}}$ 	 &\sfrac{5}{2} 		 &1.0 			 \\ 
        $^{57}$La\,\III 	 &14100 	&14096       &0.199 		 &1.078 	&   	 &5p$^6$5d 	 &$^2$D 	 &\sfrac{5}{2} 	 &	 &5p$^6$4f 	                       &$^2$F$^{\textsc{o}}$ 	 &\sfrac{7}{2} 		 &0.65 			 \\ 
        $^{58}$Ce\,\III 	 &14659 	&14655       &0.189 		 &1.035 	&   	 &4f$^2$ 	 &$^3$H 	 &5   	         &	 &4f5d 	                           &$^3$H$^{\textsc{o}}$ 	 &6 		         &0.31 			 \\ 
        $^{64}$Gd\,\III 	 &14337 	&14333       &0.295 		 &1.160 	&        &4f$^8$     &$^7$F 	 &6   	         &   &4f$^7$($^8$S$^{\textsc{o}}$)5d   &$^7$D$^{\textsc{o}}$     &5 	         	 &1.0 			 \\ 
        $^{89}$Ac\,\I 		 &13374 	&13370       &0.000 		 &0.927 	&        &6d7s$^2$ 	 &$^2$D      &\sfrac{3}{2} 	 &   &7s$^2$7p 	                       &$^2$P$^{\textsc{o}}$     &\sfrac{1}{2} 	     &1.0 			 \\ 

        \addlinespace[2ex]
        
        $^{49}$In\,\I 		 &13434 	&13430    	 &3.022 		 &3.945 	&	 &5s$^2$6s 	                          &$^2$S 	                 &\sfrac{1}{2} 		&      &5s$^2$6p 	      &$^2$P$^{\textsc{o}}$ 	 &\sfrac{1}{2} 	   &0.58 			 \\ 
        $^{63}$Eu\,\II 		 &13611 	&13607   	 &2.090 		 &3.001 	&    &4f$^7$($^8$S$^{\textsc{o}}$)5d      &$^7$D$^{\textsc{o}}$      &5             	&      &$-$               &$-$                       &4 		       &1.0 			 \\ 
        $^{63}$Eu\,\II 		 &14150 	&14146   	 &2.125 		 &3.001 	&    &4f$^7$($^8$S$^{\textsc{o}}$)5d      &$^7$D$^{\textsc{o}}$      &3              	&      &$-$               &$-$                       &4 		       &1.0 			 \\ 
        $^{63}$Eu\,\II 		 &13882 	&13878   	 &2.108 		 &3.001 	&    &4f$^7$($^8$S$^{\textsc{o}}$)5d      &$^7$D$^{\textsc{o}}$      &4              	&      &$-$               &$-$                       &4 		       &1.0 			 \\ 
        $^{67}$Ho\,\I 		 &14452 	&14448    	 &0.672 		 &1.530 	&	 &4f$^{11}$6s$^2$ 	                  &$^4$I$^{\textsc{o}}$ 	 &\sfrac{13}{2}     &      &$-$ 	          &$-$ 	                     &\sfrac{15}{2}    &0.13 			 \\ 
        $^{67}$Ho\,\I 		 &14441 	&14437    	 &0.672 		 &1.531 	&	 &4f$^{11}$6s$^2$ 	                  &$^4$I$^{\textsc{o}}$ 	 &\sfrac{13}{2} 	&      &$-$ 	          &$-$ 	                     &\sfrac{13}{2}    &0.12 			 \\ 
        $^{68}$Er\,\I 		 &13934 	&13931    	 &0.000 		 &0.890 	&    &4f$^{12}$6s$^2$ 	                  &$^3$H 	                 &6           	    &      &$-$               &$-$                       &6 		       &1.0 			 \\ 
        $^{68}$Er\,\II 		 &14653 	&14648   	 &0.000 		 &0.846 	&    &$-$                                 &$-$                       &\sfrac{13}{2} 	&      &4f$^{11}$6s$^2$   &$^4$I$^{\textsc{o}}$      &\sfrac{15}{2}    &1.0 			 \\ 
        $^{70}$Yb\,\I 		 &13888 	&13884    	 &2.143 		 &3.036 	&    &4f$^{14}$6s6p 	                  &$^3$P$^{\textsc{o}}$      &0            	    &      &4f$^{14}$5d6s 	  &$^3$D                     &1 		       &0.62 			 \\ 
        $^{71}$Lu\,\I 		 &13375 	&13372    	 &0.000 		 &0.927 	&    &5d6s$^2$ 	                          &$^2$D 	                 &\sfrac{3}{2} 	    &      &6s$^2$6p 	      &$^2$P$^{\textsc{o}}$      &\sfrac{3}{2} 	   &0.40 			 \\         
        $^{90}$Th\,\III      &13446 	&13442    	 &0.008 		 &0.930 	&	 &6d$^2$ 	                          &$^3$F 	                 &2 		        &      &$-$ 	          &$-$ 	                     &3		           &0.28 			 \\ 
                
        \hline
        \multicolumn{14}{c}{1.58\,\micron\ feature}    \\
        \hline \relax

        $^{58}$Ce\,\III 		 &15720 	&15716	 &0.000 		 &0.789 	&	 &4f$^2$ 	 &$^3$H 	 &4 	&	 &4f5d 	 &$^3$H$^{\textsc{o}}$ 	 &5 		 &0.68 			 \\ 
        $^{58}$Ce\,\III 		 &15961 	&15957	 &0.000 		 &0.777 	&	 &4f$^2$ 	 &$^3$H 	 &4 	&	 &4f5d 	 &$^3$G$^{\textsc{o}}$ 	 &3 		 &0.45 			 \\ 
        $^{58}$Ce\,\III 		 &15852 	&15848	 &0.189 		 &0.972 	&	 &4f$^2$ 	 &$^3$H 	 &5 	&	 &4f5d 	 &$^3$G$^{\textsc{o}}$ 	 &4 		 &0.27 			 \\ 
        $^{58}$Ce\,\III 		 &16133 	&16129	 &0.388 		 &1.156 	&	 &4f$^2$ 	 &$^3$H 	 &6 	&	 &4f5d 	 &$^3$G$^{\textsc{o}}$ 	 &5 		 &0.16 			 \\ 

        \addlinespace[2ex]

        $^{63}$Eu\,\II 		 &15322 	&15318	 &2.138 		 &2.948 	&	 &4f$^7$($^8$S$^{\textsc{o}}$)5d 	 &$^7$D$^{\textsc{o}}$ 	 &2 		 &     &$-$              &$-$       &3 		 &0.96 			 \\ 
        $^{63}$Eu\,\II 		 &15075 	&15071	 &2.125 		 &2.948 	&	 &4f$^7$($^8$S$^{\textsc{o}}$)5d 	 &$^7$D$^{\textsc{o}}$ 	 &3 		 &     &$-$              &$-$       &3 		 &0.96 			 \\ 
        $^{65}$Tb\,\II 		 &16068 	&16064	 &0.000 		 &0.772 	&	 &$-$                                &$-$                    &8 		 &     &$-$          	 &$-$       &7 		 &0.69 			 \\ 
        $^{65}$Tb\,\II 		 &16204 	&16199	 &0.126 		 &0.891 	&	 &$-$                                &$-$                    &7 		 &     &$-$          	 &$-$       &7 		 &0.43 			 \\ 
        $^{65}$Tb\,\II 		 &15967 	&15962	 &0.373 		 &1.150 	&	 &$-$                                &$-$                    &7 		 &     &$-$          	 &$-$       &8 		 &0.18 			 \\ 
        $^{70}$Yb\,\I 		 &15391 	&15387	 &2.231 		 &3.036 	&	 &4f$^{14}$6s6p 	                 &$^3$P$^{\textsc{o}}$   &1 		 &     &4f$^{14}$5d6s    &$^3$D     &1 		 &0.62 			 \\ 
        $^{90}$Th\,\III      &16064 	&16060	 &0.008 		 &0.780 	&	 &6d$^2$ 	                         &$^3$F                  &2 		 &     &$-$              &$-$       &2 		 &0.35 			 \\ 

        \hline
        \multicolumn{14}{c}{Blended 2.07\,\micron\ feature}    \\
        \hline \relax

        $^{58}$Ce\,\III$^{\rm a, b}$ 		 &20691 	&20686  	 &0.189 		 &0.789 	&	 &4f$^2$ 	 &$^3$H 	 &5 	&	 &4f5d 	 &$^3$H$^{\textsc{o}}$ 	 &5 		  &0.68 			 \\ 

        \addlinespace[2ex]

        $^{56}$Ba\,\I$^{\rm b}$ 		 &22318 	&22311	 &1.120 		 &1.676 	&	 &6s5d 	              &$^3$D 	              &1 		&            &6s6p 	                             &$^3$P$^{\textsc{o}}$ 	 &2 		 &1.0 			 \\ 
        $^{64}$Gd\,\III$^{\rm a, b}$     &21265 	&21260	 &0.622 		 &1.205 	&	 &4f$^8$ 	          &$^7$F 	              &4 		&            &4f$^7$($^8$S$^{\textsc{o}}$)5d 	 &$^7$D$^{\textsc{o}}$ 	 &4 		 &0.69 			 \\ 
        $^{64}$Gd\,\III$^{\rm a}$ 		 &20002 	&19996	 &0.622 		 &1.242 	&	 &4f$^8$ 	          &$^7$F 	              &4 		&            &4f$^7$($^8$S$^{\textsc{o}}$)5d 	 &$^7$D$^{\textsc{o}}$ 	 &3 		 &0.46 			 \\ 
        $^{70}$Yb\,\I$^{\rm a}$ 		 &19835 	&19830	 &2.444 		 &3.069 	&	 &4f$^{14}$6s6p 	  &$^3$P$^{\textsc{o}}$   &2 		&            &4f$^{14}$5d6s 	                 &$^3$D 	             &2 		 &0.92 			 \\ 
        $^{70}$Yb\,\I$^{\rm a, b}$ 		 &20926 	&20920	 &2.444 		 &3.036 	&	 &4f$^{14}$6s6p 	  &$^3$P$^{\textsc{o}}$   &2 		&            &4f$^{14}$5d6s 	                 &$^3$D 	             &1 		 &0.62 			 \\ 
        $^{70}$Yb\,\I$^{\rm b}$ 		 &22276 	&22270	 &2.875 		 &3.432 	&	 &$-$                 &$-$                    &2 		&            &4f$^{14}$5d6s 	                 &$^1$D 	             &2 		 &0.23 			 \\ 
        $^{70}$Yb\,\I$^{\rm b}$ 		 &22202 	&22196	 &3.133 		 &3.692 	&	 &4f$^{14}$5d6s 	  &$^3$D 	              &3 		&            &$-$                                &$-$                    &4 		 &0.15 			 \\ 
        $^{72}$Hf\,\I$^{\rm a, b}$ 		 &20533 	&20527	 &0.699 		 &1.303 	&	 &5d$^2$6s$^2$ 	      &$^1$D                  &2 		&            &5d6s$^2$($^2$D)6p 	             &$^1$D$^{\textsc{o}}$   &2 		 &1.0 			 \\ 
        $^{72}$Hf\,\I$^{\rm a}$ 		 &19865 	&19859	 &1.114 		 &1.738 	&	 &5d$^2$6s$^2$ 	      &$^3$P                  &2 		&            &5d6s$^2$($^2$D)6p 	             &$^3$D$^{\textsc{o}}$   &1 		 &0.11 			 \\ 
        $^{90}$Th\,\III$^{\rm a, b}$ 	 &20993 	&20987	 &0.008 		 &0.598 	&	 &6d$^2$ 	          &$^3$F 	              &2 		&            &$-$ 	                             &$-$ 	                 &3 		 &1.0 			 \\ 
        $^{90}$Th\,\III$^{\rm a}$        &20011 	&20005	 &0.008 		 &0.627 	&	 &6d$^2$ 	          &$^3$F 	              &2 		&            &5f6d 	                             &$^3$G$^{\textsc{o}}$ 	 &3 		 &0.89 			 \\ 
        $^{90}$Th\,\III$^{\rm a}$        &19948 	&19942	 &0.063 		 &0.685 	&	 &5f6d 	              &$^3$F$^{\textsc{o}}$   &2 		&            &6d7s 	                             &$^3$D 	             &1 		 &0.31 			 \\ 
        $^{90}$Th\,\III$^{\rm a, b}$ 	 &21510 	&21504	 &0.313 		 &0.890 	&	 &$-$ 	              &$-$   	              &3 		&            &6d7s 	                             &$^3$D 	             &2 		 &0.23 			 \\ 
        $^{90}$Th\,\III$^{\rm a, b}$ 	 &20306 	&20301	 &0.503 		 &1.113 	&	 &6d$^2$ 	          &$^3$F 	              &3 		&            &$-$ 	                             &$-$ 	                 &4 		 &0.18 			 \\ 

        \hline

        \insertTableNotes  
    \end{longtable}
\end{ThreePartTable}

\begin{ThreePartTable}
\renewcommand*{\arraystretch}{1.2}
\renewcommand\thetable{B1} 
\captionsetup{format=custom} 
\captionsetup{font=footnotesize}
    \begin{TableNotes}
        \footnotesize
        \item \textbf{Note.} $\lambda_{\rm air}$ values have been computed assuming the standard vacuum-to-air conversion from VALD3 \citep[see][]{Birch1994, Morton2000, VALD3}.
        \item[o] Denotes an odd parity.
        \item[a] This line falls in the overlap between our allowed wavelength ranges for both the 2.059 and 2.135\,\micron\ features, and so appears as a duplicate in the table.
        \item[b] Since we scaled the upper level populations to the lowest level that produced a line within $0.5 \leq \lambda \leq 2.5$\,\micron, it is possible for lines outside of this wavelength range (that originate from an upper level with lower energy) to produce a line with a relative level population $> 1$.
    \end{TableNotes}
    \begin{longtable}{lcccccrlccrlcc}
        \caption{
        \label{tab:Forbidden candidate transitions}
        Candidate M1 and E2 transitions for the \mbox{$\sim 0.79$, 1.08, 1.23, 1.40, 1.58, 2.059 and 2.135\,\micron} emission features in the late-phase spectra of \gfo. We also include the lines coincident with the 3.6 and 4.5\,\micron\ \spitzer\ bands that are discussed in the main text. Only lines with relative intensities $\geq 0.1$ have been included (for brevity), unless they are candidate transitions that have been discussed in the main text. For each feature, two sets of transitions are given, separated by a space (apart from the candidate transitions for the 1.58\,\micron\ feature, and the 3.6 and 4.5\,\micron\ \spitzer\ bands). The upper set of transitions are those that belong to our most viable transitions, while the lower set are the potential candidate transitions (in the case of the 1.58\,\micron\ feature these are all viable candidate transitions, whereas in the case of the 3.6 and 4.5\,\micron\ \spitzer\ bands, these are all potential candidate transitions). The species, wavelength (both in vacuum and air, for completeness), level energies, quantum J numbers, and the relative level population for the upper level are presented for each transition. The configurations and terms are displayed \citep[from the NIST ASD;][]{NIST2020} when they are easily expressible in the LS-coupling scheme.
        } \\
        \hline
        \addlinespace[0.4ex]
        \hline
        \multirow{2}{*}{Species}     &\multirow{2}{*}{$\lambda_{\rm vac}$ (\AA)}      &\multirow{2}{*}{$\lambda_{\rm air}$ (\AA)}      &\multicolumn{2}{c}{Level energies (eV)}    &       &\multicolumn{3}{c}{Lower level}    &     &\multicolumn{3}{c}{Upper level}    &Relative level      \\
        \cline{4-5}
        \cline{7-9}
        \cline{11-13}
            &       &       &Lower      &Upper      &       &Configuration      &Term       &J      &    &Configuration      &Term      &J      &population    \\
        \hline
        \endfirsthead
        \caption[]{ \textit{(continued)}} \\
        \hline
        \addlinespace[0.4ex]
        \hline
        \multirow{2}{*}{Species}     &\multirow{2}{*}{$\lambda_{\rm vac}$ (\AA)}      &\multirow{2}{*}{$\lambda_{\rm air}$ (\AA)}      &\multicolumn{2}{c}{Level energies (eV)}    &       &\multicolumn{3}{c}{Lower level}    &     &\multicolumn{3}{c}{Upper level}    &Relative level      \\
        \cline{4-5}
        \cline{7-9}
        \cline{11-13}
            &       &       &Lower      &Upper      &       &Configuration      &Term       &J      &    &Configuration      &Term      &J      &population    \\
        \hline
        \endhead
        \hline
        \multicolumn{14}{r}{\footnotesize\itshape Continued on next page}
        \endfoot
        \endlastfoot
        %
        \multicolumn{14}{c}{0.79\,\micron\ feature}    \\
        \hline

        [$^{53}$I\,\III] 		 &7943.9 		 &7941.8       &1.452 		 &3.013 		 &    &5s$^2$5p$^3$ 	 &$^2$D$^{\textsc{o}}$ 		 &\sfrac{3}{2} 		 &    &5s$^2$5p$^3$ 		 &$^2$P$^{\textsc{o}}$ 		 &\sfrac{1}{2} 	   &1.0 			 \\
        
        \addlinespace[2ex] \relax
        
        [$^{76}$Os\,\III] 		 &7977.3 		 &7975.1       &0.000 		 &1.554 		 &    &5d$^6$ 		     &$^5$D 		 &4 	        	 &    &$-$     	           	 &$-$    		 &4         	 &1.0 			 \\ \relax
        [$^{79}$Au\,\II] 		 &7857.8 		 &7855.7       &1.865 		 &3.443 		 &    &5d$^9$6s 		 &$^3$D     	 &3            		 &    &5d$^9$6s 	    	 &$^3$D 	   	 &1         	 &1.0 			 \\ 
                
        \hline
        \multicolumn{14}{c}{1.08\,\micron\ feature}    \\
        \hline

        [$^{53}$I\,\III] 		 &10641 		 &10638       &1.847 		 &3.013 	&	 &5s$^2$5p$^3$ 	   &$^2$D$^{\textsc{o}}$ 	 &\sfrac{5}{2} 		&     &5s$^2$5p$^3$ 	 &$^2$P$^{\textsc{o}}$ 	 &\sfrac{1}{2} 	     &1.0 			 \\ \relax
        [$^{79}$Au\,\I] 		 &10916 		 &10913       &0.000 		 &1.136 	&	 &5d$^{10}$6s 	   &$^2$S 	             	 &\sfrac{1}{2} 		&     &5d$^9$6s$^2$ 	 &$^2$D 	             &\sfrac{5}{2} 	     &1.0 			 \\ \relax
        [$^{89}$Ac\,\II] 		 &11004 		 &11001       &0.000 		 &1.127 	&	 &7s$^2$ 		   &$^1$S 	               	 &0                	&	  &6d7s 	       	 &$^1$D 	        	 &2            		 &1.0 			 \\
        
        \addlinespace[2ex] \relax
        
        [$^{43}$Tc\,\II] 		 &10922 	&10919       &0.429 	&1.564 	  &	   &4d$^6$ 		                             &$^5$D 	             &4 	         &     &4d$^5$($^6$S)5s 		          &$^5$S 	                &2 	             &1.0 			 \\ \relax
        [$^{45}$Rh\,\I] 		 &10845 	&10842       &0.000 	&1.143 	  &	   &4d$^8$($^3$F)5s 		                 &$^4$F 	             &\sfrac{9}{2} 	 &     &4d$^8$($^3$P)5s 		          &$^4$P 	                &\sfrac{5}{2} 	 &0.50 			 \\ \relax
        [$^{64}$Gd\,\III] 		 &10875 	&10872       &0.000 	&1.140 	  &	   &4f$^7$($^8$S$^{\textsc{o}}$)5d 	         &$^9$D$^{\textsc{o}}$ 	 &2 	         &     &4f$^7$($^8$S$^{\textsc{o}}$)6s    &$^9$S$^{\textsc{o}}$ 	&4 	             &1.0 			 \\ \relax
        [$^{70}$Yb\,\II] 		 &10936 	&10933       &2.656 	&3.789 	  &	   &4f$^{13}$($^2$F$^{\textsc{o}}$)6s$^2$ 	 &$^2$F$^{\textsc{o}}$ 	 &\sfrac{7}{2} 	 &     &$-$                               &$-$                      &\sfrac{11}{2}   &0.32 			 \\ \relax
        [$^{73}$Ta\,\I] 		 &10807 	&10804       &0.000 	&1.147    &    &5d$^3$6s$^2$ 		                     &$^4$F 		         &\sfrac{3}{2} 	 &	   &5d$^3$6s$^2$ 		              &$^4$P 		            &\sfrac{5}{2} 	 &0.33 			 \\ \relax
        [$^{78}$Pt\,\I] 		 &10761     &10758       &0.102 	&1.254 	  &	   &5d$^8$6s$^2$ 		                     &$^3$F 		         &4 	         &     &5d$^8$6s$^2$ 		              &$^3$F 	                &3 	             &0.26 			 \\
        
        \hline
        \multicolumn{14}{c}{1.23\,\micron\ feature}    \\
        \hline

        [$^{45}$Rh\,\II] 		 &12248 		 &12245       &0.000 		 &1.012 		 &    &4d$^8$         		 &$^3$F      &4 	     	 &    &4d$^8$   	       	 &$^3$P    &2 	 &1.0			 \\ \relax
        [$^{45}$Rh\,\II] 		 &12325 		 &12321       &0.298 		 &1.304 		 &    &4d$^8$         		 &$^3$F      &3       		 &    &4d$^8$   	       	 &$^3$P    &1    &0.19 			 \\ \relax
        [$^{52}$Te\,\III] 		 &12248 		 &12244       &0.000 		 &1.012 		 &    &5s$^2$5p$^2$    		 &$^3$P      &0 	     	 &    &5s$^2$5p$^2$    		 &$^3$P    &2 	 &0.32 			 \\ \relax
        [$^{77}$Ir\,\II] 		 &12215 		 &12211       &0.000 		 &1.015 		 &    &5d$^7$($^4$F)6s 		 &$^5$F      &5 	     	 &    &5d$^7$($^4$F)6s 		 &$^5$F    &3 	 &0.15 			 \\ 
        
        \addlinespace[2ex]

        [$^{52}$Te\,\II] 		 &12308 		&12305       &1.540 		 &2.547 		&    &5s$^2$5p$^3$  		 &$^2$D$^{\textsc{o}}$ 		 &\sfrac{5}{2} 	 &    &5s$^2$5p$^3$   		 &$^2$P$^{\textsc{o}}$ 		 &\sfrac{1}{2} 		 &1.0 			 \\ \relax
        [$^{54}$Xe\,\III] 		 &12300 		&12297       &0.000 		 &1.008 		&    &5s$^2$5p$^4$    		 &$^3$P              		 &2 	     	 &    &5s$^2$5p$^4$        	 &$^3$P                		 &0 	             &0.74 			 \\ \relax
        [$^{76}$Os\,\III] 		 &12567 		&12564       &0.568 		 &1.554 		&     &5d$^5$($^6$S)6s 		 &$^7$S 		             &3 		     &    &$-$ 		             &$-$ 		                 &4		             &1.0 			 \\ 

        \hline
        \multicolumn{14}{c}{1.40\,\micron\ feature}    \\
        \hline \relax
        
        [$^{39}$Y\,\II] 		 &13961 		&13957       &0.104 		 &0.992 	&	 &4d5s 		          &$^3$D 		 &1                	  &  	 &4d$^2$ 		          &$^3$F 		 &2 	          	 &0.79 			 \\ \relax
        [$^{53}$I\,\II] 		 &14111 		&14107       &0.000 		 &0.879 	&	 &5s$^2$5p$^4$ 		  &$^3$P 	     &2              	  &  	 &5s$^2$5p$^4$ 		      &$^3$P 		 &1 	           	 &1.0 			 \\ \relax
        [$^{66}$Dy\,\I] 		 &14183 		&14179       &0.000 		 &0.874 	&	 &4f$^{10}$6s$^2$ 	  &$^5$I 	     &8                	  &  	 &4f$^{10}$6s$^2$ 		  &$^5$I 		 &6          		 &0.21 			 \\ \relax
        [$^{77}$Ir\,\I] 		 &14071 		&14068       &0.000 		 &0.881 	&	 &5d$^7$6s$^2$ 		  &$^4$F 		 &\sfrac{9}{2} 	      &  	 &5d$^8$($^3$F)6s 		  &$^4$F 		 &\sfrac{7}{2} 		 &0.69 			 \\
        
        \addlinespace[2ex] \relax
        
        [$^{42}$Mo\,\III] 		 &14152 		&14148       &1.551 		 &2.427 	&	 &4d$^4$ 		     &$^3$P 	 &1 	              &  	 &4d$^4$ 		     &$^3$D 	 &2 		         &0.12 			      \\ \relax
        [$^{46}$Pd\,\III] 		 &14284 		&14280       &0.400 		 &1.268 	&	 &4d$^8$ 		     &$^3$F 	 &3 	              &  	 &4d$^8$ 		     &$^3$P 	 &2 		         &0.07 	  \\ \relax
        [$^{68}$Er\,\II] 		 &13987 		&13983       &0.000 		 &0.886 	&	 &$-$                &$-$      	 &\sfrac{13}{2} 	  &  	 &$-$                &$-$        &\sfrac{11}{2}		 &0.46 			      \\ \relax
        [$^{68}$Er\,\II] 		 &13898 		&13894       &0.000 		 &0.892 	&	 &$-$                &$-$      	 &\sfrac{13}{2} 	  &  	 &$-$                &$-$        &\sfrac{9}{2} 		 &0.37 			      \\ \relax
        [$^{73}$Ta\,\I] 		 &13806 		&13802       &0.249 		 &1.147 	&	 &5d$^3$6s$^2$ 		 &$^4$F 	 &\sfrac{5}{2} 		  &      &5d$^3$6s$^2$ 		 &$^4$P 	 &\sfrac{5}{2} 		 &0.33 			      \\ 

        \hline
        \multicolumn{14}{c}{1.58\,\micron\ feature}    \\
        \hline \relax

        [$^{53}$I\,\II] 		 &15509 		 &15505       &0.000 		 &0.799 	&	 &5s$^2$5p$^4$ 		    &$^3$P 		 &2            	&	 &5s$^2$5p$^4$ 		 &$^3$P 		 &0         		 &0.45 			 \\ \relax
        [$^{77}$Ir\,\I] 		 &15813 		 &15809       &0.000 		 &0.784 	&	 &5d$^7$6s$^2$ 		    &$^4$F 		 &\sfrac{9}{2} 	&	 &5d$^7$6s$^2$ 		 &$^4$F 		 &\sfrac{7}{2} 		 &1.0 			 \\ 

        \pagebreak
        \multicolumn{14}{c}{2.059\,\micron\ feature}    \\
        \hline \relax

        [$^{56}$Ba\,\II] 	     &20518 		 &20512       &0.000 		 &0.604 	&	 &6s 		         &$^2$S 	               	 &\sfrac{1}{2}      &    &5d 		         &$^2$D 		             &\sfrac{3}{2} 		 &0.98 			 \\ \relax
        [$^{65}$Tb\,\III] 		 &20760 		 &20755       &0.348 		 &0.945 	&	 &4f$^9$ 		     &$^6$H$^{\textsc{o}}$ 		 &\sfrac{13}{2}	    &    &4f$^9$ 		     &$^6$F$^{\textsc{o}}$ 		 &\sfrac{9}{2} 		 &0.21 			 \\ \relax
        [$^{77}$Ir\,\II] 		 &20886 		 &20880       &0.000 		 &0.594 	&	 &5d$^7$($^4$F)6s 	 &$^5$F 	              	 &5            	    &	 &5d$^7$($^4$F)6s	 &$^5$F 	            	 &4           		 &1.0 			 \\ \relax
        [$^{89}$Ac\,\I] 		 &20837 		 &20831       &0.927 		 &1.522 	&	 &7s$^2$7p 		     &$^2$P$^{\textsc{o}}$ 		 &\sfrac{1}{2} 	    &	 &7s$^2$7p 		     &$^2$P$^{\textsc{o}}$ 		 &\sfrac{3}{2} 		 &1.0 			 \\ 
        
        \addlinespace[2ex] \relax
        
        [$^{58}$Ce\,\III]$^{\rm a}$ 		 &20987 	&20982       	 &0.000 		 &0.591 	&	 &4f$^2$ 		     &$^3$H 		              &4 	           	&       &4f$^2$ 		         &$^3$F 		             &3 	           	 &0.87 			 \\        

        \hline
        \multicolumn{14}{c}{2.135\,\micron\ feature}    \\
        \hline \relax

        [$^{46}$Pd\,\III] 	     &21338 	&21332       &0.000 		 &0.581 	&	 &4d$^8$ 		    &$^3$F 	                 &4 	                &  	 &4d$^8$ 		         &$^3$F 		             &2          		 &1.0 			 \\ \relax
        [$^{47}$Ag\,\III]        &21696 	&21690       &0.000 		 &0.571 	&	 &4d$^9$ 		    &$^2$D 		             &\sfrac{5}{2}          &  	 &4d$^9$ 		         &$^2$D 		             &\sfrac{3}{2} 		 &1.0 			 \\ \relax
        [$^{52}$Te\,\I] 		 &21049 	&21044       &0.000 		 &0.589 	&	 &5p$^4$            &$^3$P 		             &2 	                &  	 &5p$^4$ 		         &$^3$P 		             &1 	           	 &1.0 			 \\ \relax
        [$^{52}$Te\,\I] 		 &21247 	&21241       &0.000 		 &0.584 	&	 &5p$^4$ 		    &$^3$P 		             &2 	                &  	 &5p$^4$ 		         &$^3$P 		             &0 	        	 &0.34 			 \\ \relax
        [$^{52}$Te\,\III] 	     &21050 	&21044       &0.000 		 &0.589 	&	 &5s$^2$5p$^2$ 	    &$^3$P 		             &0 	                &  	 &5s$^2$5p$^2$ 	         &$^3$P 		             &1 	         	 &1.0 			 \\ \relax
        [$^{65}$Tb\,\III] 		 &21121     &21116       &0.000         &0.587 	&	 &4f$^9$ 		    &$^6$H$^{\textsc{o}}$ 	 &\sfrac{15}{2}         &  	 &4f$^9$ 		         &$^6$H$^{\textsc{o}}$ 		 &\sfrac{11}{2}	     &1.0 	         \\ \relax
        [$^{89}$Ac\,\I] 		 &21458     &21452       &1.143 		 &1.721 	&	 &6d$^2$($^3$F)7s   &$^4$F 		             &\sfrac{3}{2}   	    &    &6d$^2$($^3$P)7s 		 &$^4$P 		             &\sfrac{5}{2} 		 &0.70 			 \\ 
        
        \addlinespace[2ex] \relax
        
        [$^{58}$Ce\,\III]$^{\rm a}$  &20987  &20982       &0.000  &0.591  &  &4f$^2$                                                   &$^3$H                     &4 	         &   &4f$^2$ 		                             &$^3$F 	            	 &3        		 &0.87  \\ \relax
        [$^{64}$Gd\,\II] 		     &21249  &21243       &0.079  &0.662  &  &4f$^7$($^8$S$^{\textsc{o}}$)5d($^9$D$^{\textsc{o}}$)6s   &$^{10}$D$^{\textsc{o}}$   &\sfrac{9}{2}   &   &4f$^7$($^8$S$^{\textsc{o}}$)5d$^2$($^3$F)	 &$^{10}$F$^{\textsc{o}}$	 &\sfrac{11}{2}	 &0.95  \\ \relax
        [$^{64}$Gd\,\II] 		     &21105  &21099       &0.144  &0.731  &  &4f$^7$($^8$S$^{\textsc{o}}$)5d($^9$D$^{\textsc{o}}$)6s   &$^{10}$D$^{\textsc{o}}$   &\sfrac{11}{2}  &   &4f$^7$($^8$S$^{\textsc{o}}$)5d$^2$($^3$F)	 &$^{10}$F$^{\textsc{o}}$    &\sfrac{13}{2}	 &0.85  \\ \relax
        [$^{64}$Gd\,\II] 		     &21414  &21408       &0.240  &0.819  &  &4f$^7$($^8$S$^{\textsc{o}}$)5d($^9$D$^{\textsc{o}}$)6s   &$^{10}$D$^{\textsc{o}}$   &\sfrac{13}{2}  &   &4f$^7$($^8$S$^{\textsc{o}}$)5d$^2$($^3$F)	 &$^{10}$F$^{\textsc{o}}$    &\sfrac{15}{2}	 &0.69  \\ \relax
        [$^{66}$Dy\,\II] 		     &21028  &21022       &0.000  &0.590  &  &$-$                                                      &$-$                       &\sfrac{17}{2}  &   &$-$                                        &$-$                        &\sfrac{13}{2}	 &0.72  \\ \relax
        [$^{68}$Er\,\II] 		     &21312  &21306       &0.055  &0.636  &  &$-$                                                      &$-$                       &\sfrac{11}{2}  &   &$-$                                        &$-$                        &\sfrac{9}{2} 	 &1.0 	\\ 
                
        \hline
        \multicolumn{14}{c}{3.6\,\micron\ \spitzer\ band}    \\
        \hline

        [$^{65}$Tb\,\III]       &35655       &35645       &0.000      &0.348      &        &4f$^9$             &6H$^{\textsc{o}}$    &\sfrac{15}{2}   &       &4f$^9$                 &$^6$H$^{\textsc{o}}$     &\sfrac{13}{2}        &2.9$^{\rm b}$                 \\ \relax
        [$^{89}$Ac\,\II]        &37218       &37208       &0.588      &0.921      &        &6d7s               &$^3$D                &1               &       &6d7s                   &$^3$D                    &3                    &3.1$^{\rm b}$       \\

        \hline
        \multicolumn{14}{c}{4.5\,\micron\ \spitzer\ band}    \\
        \hline

        [$^{52}$Te\,\II]        &45466        &45453       &1.267    &1.540      &    &5s$^2$5p$^3$      &$^2$D$^{\textsc{o}}$      &\sfrac{3}{2}       &      &5s$^2$5p$^3$       &$^2$D$^{\textsc{o}}$     &\sfrac{5}{2}      &148$^{\rm b}$   \\ \relax
        [$^{89}$Ac\,\I]         &44814        &44802       &0.000    &0.277      &    &6d7s$^2$          &$^2$D                     &\sfrac{3}{2}       &      &6d7s$^2$           &$^2$D                    &\sfrac{5}{2}      &185$^{\rm b}$   \\ \relax
        [$^{89}$Ac\,\II]        &46310        &46298       &0.653    &0.921      &    &6d7s              &$^3$D                     &2                  &      &6d7s               &$^3$D                    &3                 &3.1$^{\rm b}$   \\
        
        \hline

        \insertTableNotes  
    \end{longtable}
\end{ThreePartTable}

\begin{table}
    \renewcommand*{\arraystretch}{1.2}
    \renewcommand\thetable{C1} 
    \centering
    \caption{
        Candidate transitions from our line analysis based on Kurucz atomic data for the $\sim 0.79$, 1.08, 1.23, 1.40, 1.58, 2.059 and 2.135\,\micron\ emission features in the late-phase spectra of \gfo. We also shortlist the lines that are coincident with either of the 3.6 and 4.5\,\micron\ \spitzer\ bands. The relative intensities for all species have been computed with $T = 5000$\,K. Only lines with relative intensities $\geq 0.1$ have been included (for brevity), unless they are candidate transitions that have been discussed in the main text. The species, wavelength (both in air and vacuum, for completeness), level energies, $A$-value and transition type are presented for each transition.
    }
    \begin{threeparttable}
    \begin{tabular}{lccccccc}
    \hline
    \hline
    \multirow{2}{*}{Species}    &\multirow{2}{*}{$\lambda_{\rm air}$ (\AA)}    &\multirow{2}{*}{$\lambda_{\rm vac}$ (\AA)}    &\multicolumn{2}{c}{Level energies (eV)}    &$A$-value    &Transition    &Relative        \\
    \cline{4-5}
            &       &       &Lower      &Upper      &(s$^{-1}$)       &type       &intensity    \\
    \hline \relax
    
    [$^{39}$\YII]    &7904.0      &7906.2         &0.180      &1.748     &0.877        &E2     &1.00        \\ \relax
    [$^{39}$\YII]    &8070.9      &8073.2         &0.409      &1.944     &0.741        &E2     &0.94        \\ \relax
    [$^{39}$\YII]    &7659.0      &7661.1         &0.130      &1.748     &0.671        &E2     &0.79        \\ \relax
    [$^{39}$\YII]    &7586.3      &7588.4         &0.104      &1.738     &0.895        &E2     &0.65        \\ \relax
    [$^{39}$\YII]    &7954.2      &7956.4         &0.180      &1.738     &0.731        &E2     &0.51        \\ \relax
    [$^{39}$\YII]    &13707       &13711          &0.180      &1.084     &0.086        &E2     &0.47        \\ \relax
    [$^{39}$\YII]    &7787.0      &7789.2         &0.130      &1.721     &1.722        &E2     &0.42        \\ \relax
    [$^{39}$\YII]    &13727       &13731          &0.130      &1.033     &0.050        &E2     &0.24        \\ \relax
    [$^{39}$\YII]    &7540.6      &7542.7         &0.104      &1.748     &0.183        &E2     &0.22        \\ \relax
    [$^{39}$\YII]    &13957       &13961          &0.104      &0.992     &0.053        &E2     &0.20        \\ \relax
    [$^{39}$\YII]    &12986       &12990          &0.130      &1.084     &0.022        &E2     &0.13        \\ \relax
    [$^{39}$\YII]    &13351       &13355          &0.104      &1.033     &0.023        &E2     &0.12        \\ \relax
    [$^{39}$\YII]    &14368       &14372          &0.130      &0.992     &0.032        &E2     &0.12        \\ \relax
    [$^{39}$\YII]    &7706.1      &7708.2         &0.130      &1.738     &0.151        &E2     &0.11        \\ \relax
    [$^{39}$\YII]    &14535       &14539          &0.180      &1.033     &0.023        &E2     &0.11        \\

    \addlinespace[2ex] \relax
    
    [$^{40}$Zr\,\II]    &11563  &11566       &0.164  &1.236  &0.079  &E2 &1.00   \\ \relax
    [$^{40}$Zr\,\II]    &11134  &11137       &0.095  &1.208  &0.066  &E2 &0.62   \\ \relax
    [$^{40}$Zr\,\II]    &10604  &10607       &0.039  &1.208  &0.059  &E2 &0.58   \\ \relax
    [$^{40}$Zr\,\II]    &10465  &10468       &0.000  &1.184  &0.105  &E2 &0.55   \\ \relax
    [$^{40}$Zr\,\II]    &10860  &10863       &0.095  &1.236  &0.035  &E2 &0.47   \\ \relax
    [$^{40}$Zr\,\II]    &10821  &10824       &0.039  &1.184  &0.060  &E2 &0.31   \\ \relax
    [$^{40}$Zr\,\II]    &11699  &11702       &0.713  &1.773  &0.110  &E2 &0.26   \\ \relax
    [$^{40}$Zr\,\II]    &10261  &10264       &0.000  &1.208  &0.023  &E2 &0.23   \\ \relax
    [$^{40}$Zr\,\II]    &10208  &10211       &0.559  &1.773  &0.056  &E2 &0.15   \\ \relax
    [$^{40}$Zr\,\II]    &10356  &10358       &0.039  &1.236  &0.010  &E2 &0.14   \\ \relax
    [$^{40}$Zr\,\II]    &11132  &11135       &0.713  &1.827  &0.038  &E2 &0.13   \\ \relax
    [$^{40}$Zr\,\II]    &11204  &11207       &0.559  &1.665  &0.034  &M1 &0.11   \\

    \addlinespace[2ex]
    
    $^{41}$Nb\,\I\    &10716    &10719        &0.000    &1.157     &58.4     &E1     &1.00   \\
    $^{41}$Nb\,\I\    &7876.7   &7878.9       &0.000    &1.574     &75.9     &E1     &0.40   \\
    $^{41}$Nb\,\I\    &35636    &35646        &0.000    &0.348     &2.16     &E1     &0.07   \\
    $^{41}$Nb\,\I\    &46410    &46423        &0.000    &0.267     &1.87     &E1     &0.05   \\

    \addlinespace[2ex]
    
    $^{42}$Mo\,\I\ &21592  &21598        &3.208  &3.782  &20.4 &E1 &0.22   \\
    $^{42}$Mo\,\I\ &20444  &20449        &3.575  &4.181  &36.3 &E1 &0.12   \\
    $^{42}$Mo\,\I\ &20061  &20066        &3.560  &4.178  &68.2 &E1 &0.10   \\

    \addlinespace[2ex] \relax
    
    [$^{45}$Rh\,\II]    &41633     &41644        &0.000  &0.298  &0.360  &M1 &1.00   \\ \relax
    [$^{45}$Rh\,\II]    &10817     &10820        &0.298  &1.444  &0.946  &M1 &0.51   \\ \relax
    [$^{45}$Rh\,\II]    &8027.3    &8029.5       &0.298  &1.842  &0.229  &M1 &0.12   \\ \relax
    [$^{45}$Rh\,\II]    &21810     &21816        &0.444  &1.012  &0.136  &M1 &0.10   \\

    \addlinespace[2ex] \relax
    
    [$^{45}$Rh\,\III]   &7672.5   &7674.6    &0.000  &1.615  &1.71   &M1 &1.00   \\ \relax
    [$^{45}$Rh\,\III]   &46547    &46559     &0.000  &0.266  &0.331  &M1 &0.59   \\
    
    \hline
    \end{tabular}
    
    \begin{tablenotes}
        \footnotesize
        \item \textbf{Note.} $\lambda_{\rm vac}$ values have been computed assuming the standard air-to-vacuum conversion from VALD3 \citep[see][]{VALD3}.
    \end{tablenotes}
    
    \end{threeparttable}
    
    \label{tab:Kurucz candidates}
\end{table}


\bsp	
\label{lastpage}
\end{document}